\documentclass{cpamart1}
\usepackage{hyperref} 
\usepackage{amsmath}
\usepackage{xcolor}
\usepackage{mathtools}
\usepackage{soul}

\def\e{{e}}
\def\d{\textrm{d}}

\def\i{{i}}
\def\j{{j}}

\newcommand{\grad}{\nabla}

\newcommand{\Dopp}{\left( \partial_t + U \partial_x \right)}
\newcommand{\Doppz}{\left( \partial_t + U \partial_z \right)}

\newcommand{\Udim}{U_{s}}
\newcommand{\freq}{f_d}
\newcommand{\chord}{c}

\newcommand{\phasediff}{\varsigma}

\renewcommand{\Im}{\textrm{Im}}

\newcommand{\bv}{\boldsymbol{v}}
\newcommand{\bV}{\boldsymbol{V}}

\newcommand{\bL}{\boldsymbol{L}}
\newcommand{\bq}{\boldsymbol{q}}
\newcommand{\bu}{\boldsymbol{u}}

\newcommand{\bw}{\boldsymbol{w}}

\newcommand{\K}{\mathcal{K}}

\newcommand{\bJ}{\boldsymbol{J}}
\newcommand{\bI}{\boldsymbol{I}}
\newcommand{\bD}{\boldsymbol{D}}

\newcommand{\aPot}{f}
\newcommand{\APot}{F}
\newcommand{\baPot}{\boldsymbol{f}}

\newcommand{\map}{\mathcal{Z}}

\newlength\yshift
\newlength\xshift

\def\XXint#1#2#3{{\setbox0=\hbox{$#1{#2#3}{\int}$}
     \vcenter{\hbox{$#2#3$}}\kern-.5\wd0}}

\usepackage{subcaption}
\usepackage{lipsum}
\usepackage{amssymb}
\usepackage{graphicx}

\graphicspath{{images/}}
\usepackage{epstopdf}
\usepackage{tikz,pstool}
\usepackage{pgfplots}
\pgfplotsset{compat=1.6} 
\usetikzlibrary{plotmarks}

\usetikzlibrary{calc, 
				decorations.markings, 
				decorations.text,
				decorations.pathreplacing,
				shapes.multipart,
				positioning,
				arrows.meta,
				arrows,
				matrix,
				patterns,
				shadows,
				hobby}

\tikzset{
    side by side/.style 2 args={
        line width=2pt,
        #1,
        postaction={
            clip,postaction={draw,#2}
        }
    },
    circle node/.style={
        circle,
        draw,
        fill=white,
        minimum size=1.3cm
    }
}
\usetikzlibrary{fadings}
\tikzfading[name=fade out,
inner color=transparent!0,
outer color=transparent!100]

\tikzset{test/.style={
    postaction={
        decorate,
        decoration={
           markings,
            mark=at position \pgfdecoratedpathlength-0.5pt with {\arrow[blue,line width=#1] {}; },
            mark=between positions 0 and \pgfdecoratedpathlength-0pt step 0.5pt with {
                \pgfmathsetmacro\myval{multiply(divide(
                    \pgfkeysvalueof{/pgf/decoration/mark info/distance from start}, \pgfdecoratedpathlength),100)};
                \pgfsetfillcolor{white!\myval!black};
                \pgfpathcircle{\pgfpointorigin}{#1};
                \pgfusepath{fill};}
}}}}

\def\myCBar#1#2{
\begin{tikzpicture}[scale = 0.5]
\pgfplotscolorbardrawstandalone[ 
     colorbar style={
         height = 6cm,
 width = .5cm,
ylabel = {#2},
        ytick={-#1,-0.05,0,0.05,#1},
        yticklabels={$\leq-0.1$,$-0.05$,$0$,$0.05$,$\geq 0.1$},
ylabel shift = -10 pt,
           ylabel style={
            font = \Large}
},
colormap={mymap}{[1pt] rgb(0pt)=(0,0,1); rgb(127pt)=(1,1,1); rgb(128pt)=(1,1,1); rgb(255pt)=(1,0,0)},
colorbar,
    point meta min=-#1,
    point meta max=#1,
    of colormap/target pos min*=-.1, 
of colormap/target pos max*=.1, 
        ]
\end{tikzpicture}}

\newcommand{\TAv}{(T_1+T_0)/2}

\def\myCBarSum{
\begin{tikzpicture}[scale = 0.5]
\pgfplotscolorbardrawstandalone[ 
     colorbar style={
         height = 6cm,
 width = .5cm,
ylabel = {$\TAv$},
        ytick={0.9,0.95,1,1.05,1.1},
    yticklabels={$\leq0.9$,$0.95$,$1$,$1.05$,$\geq 1.1$},
ylabel shift = -10 pt,
           ylabel style={
            font = \Large}
},
colormap={mymap}{[1pt] rgb(0pt)=(0,0,1); rgb(127pt)=(1,1,1); rgb(128pt)=(1,1,1); rgb(255pt)=(1,0,0)},
colorbar,
    point meta min=.9,
    point meta max=1.1,
    of colormap/target pos min*=.9, 
of colormap/target pos max*=1.1, 
        ]
\end{tikzpicture}}

\def\myCBarTwo#1#2{
\begin{tikzpicture}[scale = 0.75]
\pgfplotscolorbardrawstandalone[ 
     colorbar style={
         height = 5cm,
 width = .2cm,
ylabel = {#2},
        ytick={-2,-1,0,1,2},
                    yticklabels={$\leq-2$,$-1$,$0$,$1$,$\geq2$},
               yticklabel style={font = \small},
ylabel shift = -10 pt,
           ylabel style={
            font = \Large}
},
colormap={mymap}{[1pt] rgb(0pt)=(0,0,1); rgb(127pt)=(1,1,1); rgb(128pt)=(1,1,1); rgb(255pt)=(1,0,0)},
colorbar,
    point meta min=-2,
    point meta max=2,
    of colormap/target pos min*=-2, 
of colormap/target pos max*=2, 
        ]
\end{tikzpicture}}

\def\myCBarThree{
\begin{tikzpicture}[scale = 0.5]
\pgfplotscolorbardrawstandalone[ 
     colorbar style={
         height = 6cm,
 width = .5cm,
ylabel = {$\TAv$},
        ytick={0.4,.7,1,1.3,1.6},
                yticklabels={$\leq0.4$,$0.7$,$1$,$1.3$,$\geq 1.6$},
ylabel shift = -10 pt,
           ylabel style={
            font = \Large}
},
colormap={mymap}{[1pt] rgb(0pt)=(0,0,1); rgb(127pt)=(1,1,1); rgb(128pt)=(1,1,1); rgb(255pt)=(1,0,0)},
colorbar,
    point meta min=0.4,
    point meta max=1.6,
    of colormap/target pos min*=.4, 
of colormap/target pos max*=1.6, 
        ]
\end{tikzpicture}}

\def\myCBarFour{
\begin{tikzpicture}[scale = 0.5]
\pgfplotscolorbardrawstandalone[ 
     colorbar style={
         height = 6cm,
 width = .5cm,
ylabel = {$T_1-T_0$},
        ytick={-1,-0.5,0,0.5,1},
        yticklabels={$\leq -1$,$-0.5$,$0$,$0.5$,$\geq 1$},
ylabel shift = -10 pt,
           ylabel style={
            font = \Large}
},
colormap={mymap}{[1pt] rgb(0pt)=(0,0,1); rgb(127pt)=(1,1,1); rgb(128pt)=(1,1,1); rgb(255pt)=(1,0,0)},
colorbar,
    point meta min=-1,
    point meta max=1,
    of colormap/target pos min*=-1, 
of colormap/target pos max*=1, 
        ]
\end{tikzpicture}}

\def\skCircDetColThick#1#2#3#4{
	\begin{scope}[shift={#1}]
		\draw[line width = \lw, #4,fill = confgrey] (0,0) circle (#2);
		\draw[fill=black] (0,0) circle (.01);
		\draw[latex-latex] (0,-.05)--(#2,-.05) node[midway, above] {$q_#3$};
		\node [left, below] {$\delta_#3$};
		\node[right = 3pt,above = 3pt] at ($#2*({cos(30)},{sin(30)})$) {$C_{#3}$};
	\end{scope}
}

\def\skCircBlankBThick#1#2#3{
	\begin{scope}[shift={#1}]
		\draw[line width = \lw, #3,fill = confgrey] (0,0) circle (#2);
	\end{scope}
}

\tikzset{
    mark position/.style args={#1(#2)}{
        postaction={
            decorate,
            decoration={
                markings,
                mark=at position #1 with \coordinate (#2);
            }
        }
    }
}

\def\myShiftUp#1{\raisebox{1ex}}
\def\myShiftDown#1{\raisebox{-2.5ex}}

\def\myPathTextAbove#1#2#3#4{
\draw [-LaTeX,thick,postaction={decorate,
               decoration={
                         raise=1ex,
                         text along path,
                         text align={center},
                         text={%
                                     |\color{black}| {#1} }
                              }
                        }
	  ] #2 to [bend left=#4] #3;
}

\usetikzlibrary{decorations.text,fit,chains,calc,shapes.geometric,intersections}

\def\myPathTextBelow#1#2#3#4{
\draw [-LaTeX,thick,postaction={decorate,
               decoration={
                         raise=-2ex,
                         text along path,
                         text align={center},
                         text={%
                                     |\color{black}| {#1} }
                              }
                        }
	  ] #2 to [bend left=#4] #3;
}

\newlength{\myww}
\newlength{\myhh}

\def\zeroRad{3}
\def\delIx{1}
\def\delIy{1}
\def\radI{1}
\def\delIIx{0}
\def\delIIy{-1.5}
\def\radII{.75}
\usepackage{amsfonts}
\def\radIII{.5}
\def\delIIIx{-1.5}
\def\delIIIy{0}
\def\lw{2}
\tikzset{->-/.style={decoration={
  markings,
  mark=at position #1 with {\arrow{latex}}},postaction={decorate}}}

\definecolor{themecolour}{rgb}{0.64, 0.76, 0.68}
\definecolor{mygray}{rgb}{0.7, 0.7, 0.7}
\definecolor{tablegray}{rgb}{0.75, 0.75, 0.75}
\definecolor{confcol}{RGB}{255, 255, 255}
\definecolor{camblue}{RGB}{108, 172, 228}
\definecolor{camred}{RGB}{213, 0, 50}
\definecolor{camnavy}{RGB}{0, 60, 113}
\definecolor{camgreen}{RGB}{114, 180, 49}

\definecolor{myblue}{rgb}{0 ,  0.4470 , 0.7410}
\definecolor{myorange}{rgb}{0.8500,    0.3250,    0.0980}
\definecolor{myyellow}{rgb}{0.9290,    0.6940,    0.1250}
\definecolor{mypurple}{rgb}{ 0.4940,    0.1840,    0.5560}
\definecolor{myred}{rgb}{     0.6350 ,   0.0780 ,   0.1840}
\definecolor{mygreen}{rgb}{         0.4660  ,  0.6740   , 0.1880}

\definecolor{branchCol}{rgb}{0, .8, .8}

\definecolor{prsared}{RGB}{ 219,25,73}
\definecolor{prsablue}{RGB}{ 4,146,210}

\definecolor{confblue}{rgb}{0.75, 0.72, 0.95}
\definecolor{confgreen}{rgb}{0.76, 1, 0.74}
\definecolor{confgrey}{rgb}{0.9, .9, 0.9}

\definecolor{cvblue}{rgb}{0.22,0.45,0.70}

\definecolor{matlab1}{rgb}{0,0.4470,0.7410}
\definecolor{matlab2}{rgb}{0.8500,0.3250,0.0980}
\definecolor{matlab3}{rgb}{0.9290,0.6940,0.0980}
\definecolor{matlab4}{rgb}{0.4940,0.1840,0.5560}
\definecolor{matlab5}{rgb}{0.4660,0.6740,0.1880}
\definecolor{matlab6}{rgb}{0.3010,0.7450,0.9330}
\definecolor{matlab7}{rgb}{0.6350,0.0780,0.1840}

\usepackage{times}

\theoremstyle{definition}

\theoremstyle{remark}

\usepackage{esint}

\newlength\fheight
\newlength\fwidth
\allowdisplaybreaks

\received{May 2021}
\volume{000}
\startingpage{1}
\title{Generalization of waving-plate theory to multiple interacting swimmers}
\author{Peter J. Baddoo}{Department of Mathematics, 
Massachusetts Institute of Technology, 
\mbox{77 Massachusetts Avenue,}
Cambridge, 
MA 02141, USA \vspace{0cm}}%

\author{Nicholas J. Moore}%
{Mathematics Department,
United States Naval Academy,
Chauvenet Hall,
\mbox{572C Holloway Road,}
Annapolis, MD 21402, USA \vspace{0cm}}%

\author{Anand U. Oza}%
{Department of Mathematical Sciences,
New Jersey Institute of Technology,
\mbox{University Heights,}
Cullimore Hall 510,
Newark, NJ 07102, USA \vspace{0cm}}%

\author{Darren G. Crowdy}%
{Department of Mathematics, 
Imperial College London, 
180 Queen’s Gate, 
\mbox{London SW7 2AZ}, UK \vspace{0cm}}

\authorheadline{Baddoo \emph{et al.}}%
\titleheadline{Generalization of waving-plate theory}%

\hypersetup{
     colorlinks = true,
     linkcolor = red!70!black,
     citecolor = blue!70!black,
     urlcolor = green!70!black
     }
\usepackage{overpic}

\begin{document}
\maketitle

\begin{abstract}
Early research in aerodynamics and biological propulsion was dramatically advanced by the analytical solutions of Theodorsen, von K\'{a}rm\'{a}n, Wu and others. While these classical solutions apply only to isolated swimmers, the flow interactions between multiple swimmers are relevant to many practical applications, including the schooling and flocking of animal collectives. In this work, we derive a class of solutions that describe the hydrodynamic interactions between an arbitrary number of swimmers in a two-dimensional inviscid fluid. Our approach is rooted in multiply-connected complex analysis and exploits several recent results. Specifically, the transcendental (Schottky--Klein) prime function serves as the basic building block to construct the appropriate conformal maps and leading-edge-suction functions, which allows us to solve the modified Schwarz problem that arises. As such, our solutions generalize classical thin aerofoil theory, specifically Wu's waving-plate analysis, to the case of multiple swimmers.
For the case of a pair of interacting swimmers, we develop an efficient numerical implementation that allows rapid computations of the forces on each swimmer. We investigate flow-mediated equilibria and find excellent agreement between our new solutions and previously reported experimental results. Our solutions recover and unify disparate results in the literature, thereby opening the door for future studies into the interactions between multiple swimmers.
\end{abstract}

\vspace{1.3cm}

\section{Introduction}
\label{Sec:Intro}

The question of how animals propel themselves through a fluid, whether swimming or flying, is one that has captured the imagination of the scientific community and the public alike. This question becomes especially subtle when considering groups of self-propelled bodies, for example schools or flocks.
Intriguingly, a number of observations~\cite{Partridge2, Herskin,Marras, AshrafPNAS,Li2020} suggest that schooling fish could benefit from hydrodynamic interactions by realizing significant energy savings. 
Practitioners have already begun incorporating some of these principles into the design of next-generation technologies, such as underwater vehicles that emulate fish-like swimming by periodically actuating a set of propulsors attached to the sides of the main body \cite{Triantafyllou_Review, Geder_2017}.

Experimental studies of interacting flapping foils in a water tank serve as valuable complements to observational studies of animal collectives, as the kinematics and flow conditions can be controlled. A natural starting point is the case of a pair of rigid foils that undergo two types of prescribed oscillatory motion at the same imposed frequency $\freq$ with a possible phase offset $\phasediff$ (Fig.~\ref{Fig:schematic}): a linear motion perpendicular to the swimming direction (heaving) or a rotational motion about a fixed pivot point (pitching). The line connecting the two foils may be along the swimming direction (in-line), perpendicular to it (side-by-side) or at an angle (staggered). Studies on flapping foils in an imposed flow, held a fixed distance apart in an in-line configuration, have shown that the thrust on the downstream foil is sensitive to both the separation distance and phase offset~\cite{Warkentin2007, Boschitsch2014, Gong2015,Lua2016}. Meanwhile, studies on pitching foils in a side-by-side configuration~\cite{Raspa2013, Dewey2014, Godoy2019} have suggested that the hydrodynamic thrust is maximized when the foils flap in antiphase, $\phasediff=\pi$. 


Recent experiments~\cite{ramananarivo2016flow,Newbolt2019} have demonstrated that pairs of {\em freely-swimming foils} in an in-line configuration, initially at an arbitrary distance apart, spontaneously assume ``schooling modes" in which the pair moves stably together at a fixed speed $\Udim$ while maintaining a constant separation distance $\ell$. These studies have highlighted the importance of the ``schooling number" $S = \ell \freq /\Udim$, a dimensionless quantity that represents the number of flapping wavelengths separating the swimmers. Specifically, integer values of $S$ correspond to spatially in-phase states in which the pair traces out the same path through space, while half-integer values indicate out-of-phase states. The experimental data across a range of flapping frequencies, amplitudes and phase offsets may be approximated using the simple relation $S\approx n + \phasediff/(2\pi)$ where $n\in\mathbb{Z}$. 

There are a number of numerical studies of pairs of flapping bodies in a viscous incompressible fluid, mostly in two dimensions (2D) and some in 3D. Studies on in-line flapping foils, held a fixed distance apart in an imposed flow, have shown that the thrust oscillates with both the distance between the foils $\ell$ and phase difference $\phasediff$~\cite{Akhtar2007,Deng2007,Broering2012b,Muscutt2017,Lin2019}, in qualitative agreement with experiments. Studies on flapping foils in side-by-side arrangements have demonstrated that the hydrodynamic thrust is relatively large when the foils flap in antiphase~\cite{Dong2007,Bao2017}. Methods that solve the Navier--Stokes equations directly, such as the immersed boundary method~\cite{Deng2007, Hoover2018, Lin2019} or marker-and-cell scheme~\cite{Alben2021_1,Alben2021_2}, require a fine computational grid near the bodies in order to resolve thin boundary layers and are thus limited to moderate Reynolds numbers, $\text{Re}=O(10^2)-O(10^3)$.
Accurate and tractable simulations in regimes relevant to many swimming animals and biomimetic vehicles (e.g. $\text{Re}=O(10^6)-O(10^7)$ for dolphins~\cite{Triantafyllou_Review}) have thus remained elusive. 

To overcome this limitation, some studies neglect the influence of viscosity and solve the Euler equations directly using vortex sheet methods~\cite{Alben2009,Fang2017}: the bodies are represented using bound vortex sheets, and free vortex sheets are shed from the trailing edges. The limitation of such methods is that the length of a vortex wake grows with time, and the simulation quickly becomes intractable owing to the large number of vortices. The far-field wake is thus approximated by a small number of point vortices~\cite{Alben2009,Fang2017}, or contributions to the wake are removed after some prescribed timescale~\cite{Huang2016}. 

Generally, a detailed parametric study of the system is prohibitive, owing to the computational cost of simulations and the large number of control parameters involved. Even for the relatively simple situation of a pair of rigid foils heaving in a viscous fluid, control parameters include the dimensionless amplitudes (Strouhal numbers), the reduced flapping frequencies, the  streamwise and lateral distances between foils, the phase offset, and the Reynolds number. If different foil motions are permitted (e.g.~pitching), or if the foil surfaces can deform due to elasticity~\cite{moore2015torsional, moore2017fast}, the parameter space becomes even larger. Clearly, analytical solutions, if feasible, would help conquer this vast parameter space.

Analytical progress is possible in the case of 2D flow over an infinitesimally-thin foil (flat plate) undergoing small-amplitude motions and shedding a \linebreak \mbox{non-deforming planar} wake from its trailing edge, a setting in which the Euler equations may be linearized. This so-called ``thin-airfoil theory" was pioneered in the early twentieth century by Theodorsen~\cite{Theodorsen1935} and von K\'{a}rm\'{a}n \& Sears~\cite{KarmanSears1938}, who obtained closed-form expressions for the lift and moment of pitching and heaving plates in a uniform background flow. Garrick~\cite{Garrick1936} extended the method outlined by von K\'{a}rm\'{a}n \& Burgers~\cite{KarmanBurgers1935} to derive expressions for the propulsive thrust on such plates. These results that were recently extended by Fernandez-Feria~\cite{Feria2016}, who considered the influence of the wake vorticity on the entire plate as opposed to the leading edge alone. 

Wu~\cite{wu1961swimming} proposed an alternative approach to explicitly calculate the hydrodynamic forces on a deformable plate undergoing time-harmonic undulations. Crucially, Wu worked directly with the pressure field rather than the velocity field, which considerably simplifies the analysis: the pressure is continuous everywhere in the domain except across the plate, while the tangential component of the velocity is discontinuous across the vortex wake. The small-amplitude theories of Theodorsen~\cite{Theodorsen1935}, von K\'{a}rm\'{a}n \& Sears~\cite{KarmanSears1938}, Garrick~\cite{Garrick1936}, and Wu have advanced our understanding of swimming animals~\cite{Smits2019} and biomimetic vehicles~\cite{Mueller2003}, whose propulsors often execute flapping motions with amplitude small relative to their body length.

While Wu's framework has been profitably extended to account for unsteady forward motion~\cite{Wu1971} and foil flexibility~\cite{moore2014analytical, moore2015torsional, moore2017fast}, it has not yet been extended to account for multiple flapping bodies. The problem of a pair of rigid foils heaving in an in-line configuration was treated in an approximate fashion by Ramananarivo {\it et al.}~\cite{ramananarivo2016flow} (Supplemental Material), who, following the approach of Wu \& Chwang~\cite{Wu1975}, first calculated the wavy flow generated by the leading foil, and then calculated the hydrodynamic thrust on the follower immersed in that background flow. Similarly, recent work has furnished approximate formulas for the forces and moments on a pair of rigid~\cite{Quesada2020} and deformable~\cite{Quesada2021} foils. These works neglect the contribution of the follower's vortex wake on the leader's wake for the sake of simplicity, an approximation expected to be valid only when the streamwise separation between the foils is large. In particular, such an approximation cannot be applied to the side-by-side configuration. 

Here, we present a natural extension of Wu's theory to account for multiple swimming bodies, with no such approximations made regarding the streamwise spacing between them. The newly developed theory leverages recent advances on conformal mappings of multiply-connected domains using the so-called Schottky-Klein prime function.
An outline of the paper is as follows.
In \S\ref{Sec:Modelling}, we present the governing equations for a collection of plates undergoing oscillatory kinematics in a uniform background flow, following Wu's original framework~\cite{wu1961swimming}. In \S\ref{Sec:Conformal}, we introduce the prime function and its relevant properties. In \S\ref{Sec:Solution} we demonstrate how this function can be used to solve the governing equations for an arbitrary number of plates, and explicate solutions for the special case of two plates. 
In \S\ref{sec:results}, we explore the physical consequences of our solutions for pairs of heaving, pitching and waving plates, plotting the pressure fields and deducing the dependence of the time-averaged thrust on both the distance between the plates and the flapping phase difference. The results are compared to existing experimental and numerical results in the literature. Conclusions and future directions are discussed in \S\ref{Sec:Conclusions}. 
\section{Modelling}
\label{Sec:Modelling}

\begin{figure}
	\resizebox{\linewidth}{!}{
	\begin{tikzpicture}[remember picture]
\def\airfoilLength{5}
	\node at (-4,3.5) {};
	\node at (-4,-3.5) {};
	\begin{scope}[shift = {(-2,1)}]
		\node (0,0) [below left = 7pt and -1cm] {\small $x^-_{0}+\i y^-_{0}$};
		\node[rotate = 10] at (\airfoilLength/1.8,0.8) {\small $h_{0}(x,t)$};
		\foreach \y in {1,...,8}
		{%
		\draw[ black!40!white, domain=0:\airfoilLength,samples = 30]   plot (\x,{.1*\x*sin(80*\x+\y*180/8)});
	}%
		\draw[ ultra thick, domain=0:\airfoilLength,samples = 30]   plot (\x,{.1*\x*sin(80*\x+4/5*180)});
\def\cl{.5}

\draw[blue, ultra thick, domain=0:.2,shift = {(\airfoilLength,0)},path fading = west,samples =10]   plot (\x,{-.1*sin(180*\x)});
\draw[blue, ultra thick, domain=.2:2*\cl,shift = {(\airfoilLength,0)},path fading = east,samples =10]   plot (\x,{-.1*sin(180*\x)});
\draw[red, ultra thick, domain=2*\cl:3*\cl,shift = {(\airfoilLength,0)},path fading = west,samples =10]   plot (\x,{-.1*sin(180*\x)});
\draw[red, ultra thick, domain=3*\cl:4*\cl,shift = {(\airfoilLength,0)},path fading = east,samples =10]   plot (\x,{-.1*sin(180*\x)});
\draw[blue, ultra thick, domain=4*\cl:5*\cl,shift = {(\airfoilLength,0)},path fading = west,samples =10]   plot (\x,{-.1*sin(180*\x)});
\draw[blue, ultra thick, domain=5*\cl:6*\cl,shift = {(\airfoilLength,0)},path fading = east,samples =10]   plot (\x,{-.1*sin(180*\x)});
\draw[red, ultra thick, domain=6*\cl:7*\cl,shift = {(\airfoilLength,0)},path fading = west,samples =10]   plot (\x,{-.1*sin(180*\x)});
\draw[red, ultra thick, domain=7*\cl:8*\cl,shift = {(\airfoilLength,0)},path fading = east,samples =10]   plot (\x,{-.1*sin(180*\x)});
\draw[blue, ultra thick, domain=8*\cl:9*\cl,shift = {(\airfoilLength,0)},path fading = west,samples =10]   plot (\x,{-.1*sin(180*\x)});
\draw[blue, ultra thick, domain=9*\cl:10*\cl,shift = {(\airfoilLength,0)},path fading = east,samples =10]   plot (\x,{-.1*sin(180*\x)});
\draw[red, ultra thick, domain=10*\cl:11*\cl,shift = {(\airfoilLength,0)},path fading = west,samples =10]   plot (\x,{-.1*sin(180*\x)});
\draw[red, ultra thick, domain=11*\cl:12*\cl,shift = {(\airfoilLength,0)},path fading = east,samples =10]   plot (\x,{-.1*sin(180*\x)});
\draw[blue, ultra thick, domain=12*\cl:13*\cl,shift = {(\airfoilLength,0)},path fading = west,samples =10]   plot (\x,{-.1*sin(180*\x)});
\draw[blue, ultra thick, domain=13*\cl:14*\cl,shift = {(\airfoilLength,0)},path fading = east,samples =10]   plot (\x,{-.1*sin(180*\x)});
\draw[red, ultra thick, domain=14*\cl:15*\cl,shift = {(\airfoilLength,0)},path fading = west,samples =10]   plot (\x,{-.1*sin(180*\x)});
\draw[red, ultra thick, domain=15*\cl:16*\cl,shift = {(\airfoilLength,0)},path fading = east,samples =10]   plot (\x,{-.1*sin(180*\x)});
\draw[blue, ultra thick, domain=16*\cl:17*\cl,shift = {(\airfoilLength,0)},path fading = west,samples =10]   plot (\x,{-.1*sin(180*\x)});
\draw[blue, ultra thick, domain=17*\cl:18*\cl,shift = {(\airfoilLength,0)},path fading = east,samples =10]   plot (\x,{-.1*sin(180*\x)});
\draw[red, ultra thick, domain=18*\cl:19*\cl,shift = {(\airfoilLength,0)},path fading = west,samples =10]   plot (\x,{-.1*sin(180*\x)});
\draw[red, ultra thick, domain=19*\cl:20*\cl,shift = {(\airfoilLength,0)},path fading = east,samples =10]   plot (\x,{-.1*sin(180*\x)});
\draw[blue, ultra thick, domain=20*\cl:21*\cl,shift = {(\airfoilLength,0)},path fading = west,samples =10]   plot (\x,{-.1*sin(180*\x)});
\draw[blue, ultra thick, domain=21*\cl:22*\cl,shift = {(\airfoilLength,0)},path fading = east,samples =10]   plot (\x,{-.1*sin(180*\x)});
\end{scope}

	\begin{scope}[shift = {(4,-2)}]
		\node (0,0) [below left = 7pt and -1cm] {\small $x^-_{1}+\i y^-_{1}$};
		\node[rotate = 8] at (\airfoilLength/2,.8) {\small $h_{1}(x,t)$};
		\foreach \y in {-4,...,4}
		{%
			\draw[black!40!white]  (0,0)--($({\airfoilLength*cos(\y*2)},{\airfoilLength*sin(\y*2)})$);
	}%
	\draw[ ultra thick]  (0,0)--(\airfoilLength,0) ;
\def\cl{.5}
\foreach \y in {0,2}
{%
\draw[red, ultra thick, domain=0:.2,shift = {(\airfoilLength,0)},path fading = west,samples =10]   plot (\x,{-.1*sin(180*\x)});
\draw[red, ultra thick, domain=.2:2*\cl,shift = {(\airfoilLength,0)},path fading = east,samples =10]   plot (\x,{-.1*sin(180*\x)});
\draw[blue, ultra thick, domain=2*\cl:3*\cl,shift = {(\airfoilLength,0)},path fading = west,samples =10]   plot (\x,{-.1*sin(180*\x)});
\draw[blue, ultra thick, domain=3*\cl:4*\cl,shift = {(\airfoilLength,0)},path fading = east,samples =10]   plot (\x,{-.1*sin(180*\x)});
\draw[red, ultra thick, domain=4*\cl:5*\cl,shift = {(\airfoilLength,0)},path fading = west,samples =10]   plot (\x,{-.1*sin(180*\x)});
\draw[red, ultra thick, domain=5*\cl:6*\cl,shift = {(\airfoilLength,0)},path fading = east,samples =10]   plot (\x,{-.1*sin(180*\x)});
\draw[blue, ultra thick, domain=6*\cl:7*\cl,shift = {(\airfoilLength,0)},path fading = west,samples =10]   plot (\x,{-.1*sin(180*\x)});
\draw[blue, ultra thick, domain=7*\cl:8*\cl,shift = {(\airfoilLength,0)},path fading = east,samples =10]   plot (\x,{-.1*sin(180*\x)});
\draw[red, ultra thick, domain=8*\cl:9*\cl,shift = {(\airfoilLength,0)},path fading = west,samples =10]   plot (\x,{-.1*sin(180*\x)});
\draw[red, ultra thick, domain=9*\cl:10*\cl,shift = {(\airfoilLength,0)},path fading = east,samples =10]   plot (\x,{-.1*sin(180*\x)});
}%
\end{scope}
\begin{scope}[shift = {(0,-5)}]
		\node (0,0) [below left = 12pt and -.5cm] {\small $x^-_{2}+\i y^-_{2}$};
		\node at (\airfoilLength/2,.75) {\small $h_{2}(x,t)$};
		\foreach \y in {-4,...,4}
		{%
			\draw[ black!40!white]  (0,{sin(6*\y)})--(\airfoilLength,{sin(6*\y)});
	}%
	\draw[ ultra thick] (0,0)--(\airfoilLength,0);
\def\cl{.5}

\draw[blue, ultra thick, domain=0:.2,shift = {(\airfoilLength,0)},path fading = west,samples =10]   plot (\x,{-.1*sin(180*\x)});
\draw[blue, ultra thick, domain=.2:2*\cl,shift = {(\airfoilLength,0)},path fading = east,samples =10]   plot (\x,{-.1*sin(180*\x)});
\draw[red, ultra thick, domain=2*\cl:3*\cl,shift = {(\airfoilLength,0)},path fading = west,samples =10]   plot (\x,{-.1*sin(180*\x)});
\draw[red, ultra thick, domain=3*\cl:4*\cl,shift = {(\airfoilLength,0)},path fading = east,samples =10]   plot (\x,{-.1*sin(180*\x)});
\draw[blue, ultra thick, domain=4*\cl:5*\cl,shift = {(\airfoilLength,0)},path fading = west,samples =10]   plot (\x,{-.1*sin(180*\x)});
\draw[blue, ultra thick, domain=5*\cl:6*\cl,shift = {(\airfoilLength,0)},path fading = east,samples =10]   plot (\x,{-.1*sin(180*\x)});
\draw[red, ultra thick, domain=6*\cl:7*\cl,shift = {(\airfoilLength,0)},path fading = west,samples =10]   plot (\x,{-.1*sin(180*\x)});
\draw[red, ultra thick, domain=7*\cl:8*\cl,shift = {(\airfoilLength,0)},path fading = east,samples =10]   plot (\x,{-.1*sin(180*\x)});
\draw[blue, ultra thick, domain=8*\cl:9*\cl,shift = {(\airfoilLength,0)},path fading = west,samples =10]   plot (\x,{-.1*sin(180*\x)});
\draw[blue, ultra thick, domain=9*\cl:10*\cl,shift = {(\airfoilLength,0)},path fading = east,samples =10]   plot (\x,{-.1*sin(180*\x)});
\draw[red, ultra thick, domain=10*\cl:11*\cl,shift = {(\airfoilLength,0)},path fading = west,samples =10]   plot (\x,{-.1*sin(180*\x)});
\draw[red, ultra thick, domain=11*\cl:12*\cl,shift = {(\airfoilLength,0)},path fading = east,samples =10]   plot (\x,{-.1*sin(180*\x)});
\draw[blue, ultra thick, domain=12*\cl:13*\cl,shift = {(\airfoilLength,0)},path fading = west,samples =10]   plot (\x,{-.1*sin(180*\x)});
\draw[blue, ultra thick, domain=13*\cl:14*\cl,shift = {(\airfoilLength,0)},path fading = east,samples =10]   plot (\x,{-.1*sin(180*\x)});
\draw[red, ultra thick, domain=14*\cl:15*\cl,shift = {(\airfoilLength,0)},path fading = west,samples =10]   plot (\x,{-.1*sin(180*\x)});
\draw[red, ultra thick, domain=15*\cl:16*\cl,shift = {(\airfoilLength,0)},path fading = east,samples =10]   plot (\x,{-.1*sin(180*\x)});
\draw[blue, ultra thick, domain=16*\cl:17*\cl,shift = {(\airfoilLength,0)},path fading = west,samples =10]   plot (\x,{-.1*sin(180*\x)});
\draw[blue, ultra thick, domain=17*\cl:18*\cl,shift = {(\airfoilLength,0)},path fading = east,samples =10]   plot (\x,{-.1*sin(180*\x)});
\end{scope}
\draw [thick, -Latex] (-3.5,-2)--(-1.5,-2) node[midway, above=4pt] {$U$};
\draw [thick, -Latex] (-3.5,-2.2)--(-1.5,-2.2);

\end{tikzpicture}
				}
	\caption{Schematic: a group of foils executing various flapping motions in a steady background flow of speed $U$.
	The foils have mean profile $h_{m}(x,t)$ and leading edges $x_{m}^- + \i y_{m}^-$. The motions depicted include undulation ($h_0$), pitching about the leading edge ($h_1$), and heaving ($h_2$). Each foil sheds a vortex sheet from its trailing edge, the strength of which is determined by the Kutta condition. This figure illustrates a staggered configuration, although any configuration is permitted.
}
	\label{Fig:schematic}
\end{figure}
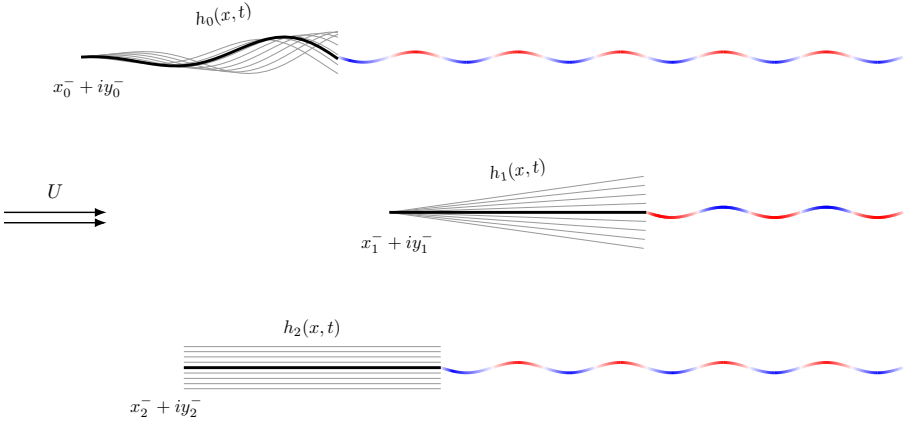

The basic hydrodynamic model employed here follows closely that of Wu \cite{wu1961swimming} and Moore \cite{moore2014analytical, moore2017fast}, with the appropriate modifications needed to model multiple swimmers. Consider a group of $M+1$ foils swimming together at speed $\Udim$ through an otherwise quiescent fluid. The foils may have different chord lengths, with $\chord$ chosen as a characteristic value (e.g.~the mean). In the present paper, the foils all swim in the same direction and at the same speed. The general framework, however, permits small deviations in swimming speed and direction, an extension that will be explored in future work.
The swimmers' propulsion arises from a flapping motion driven at a characteristic frequency $\freq$. We employ the same non-dimensionalization as in earlier work \cite{moore2014analytical, moore2017fast}, with distances scaled on $\chord/2$, time on $1/\freq$, and velocity on $\chord \freq/2$. With the appropriate Gallilean transformation, the swimmers can be considered stationary and embedded in a flow of dimensionless free-stream value $U = 2 \Udim/(\chord \freq)$, as illustrated in figure \ref{Fig:schematic}.

Each plate executes transverse motions with its mean surface expressed as $h_{m}(x,t)$, where the subscript denotes the $m$-th plate for $m = 0, \dots, M$. The key assumption underlying this theory is that the characteristic amplitude of the transverse motion, $A_s$, is small compared to the body lengthscale $\chord$. In the dimensionless variables, both $h_{m}$ and $\partial_x h_{m}$ are $\mathcal{O}(\epsilon)$, where $\epsilon = A_s/\chord \ll 1$. 
We associate the physical domain with a complex variable $z = x + \i y$.
The edges of the swimmers are located at $z^\pm_{m} = x^\pm_{m} + \i y^\pm_{m}$
 where $\pm$ corresponds to the trailing and leading edge, respectively. While the fluid velocity is permitted to have an integrable singularity at each leading edge, the Kutta condition requires that the velocity remain finite at each trailing edge. Each plate sheds a vortex sheet that extends from its trailing edge indefinitely, $z \to +\infty$, the strength of which is prescribed by the Kutta condition. These sheets represent the vorticity that is accumulated and eventually shed from the viscous boundary layer surrounding each foil. Apart from this vortex-shedding effect, the fluid is assumed otherwise inviscid.

Upon linearizing the Euler equation in small-amplitude motion, the incompressible (divergence-free) velocity field $\bu = (U + u, v)$ satisfies
\begin{align}
	\Dopp \bu = \nabla \phi
	\label{Eq:Euler}
\end{align}
where $\phi(z,t) = (p_\infty - p)/\rho$, $\rho$ is fluid density, and $p$ is the pressure
with far-field value $p_\infty$. Taking the divergence of \eqref{Eq:Euler} and enforcing incompressibility of $\bu$ implies that $\phi$ is a harmonic function throughout the domain,
\begin{align}
	\nabla^2 \phi = 0.
	\label{Eq:presAn}
\end{align}
Thus, $\phi$ admits a harmonic conjugate which gives rise to an analytic function
\begin{align}
	\aPot(z,t) = \phi(z,t) + \i \psi(z,t).
	\label{Eq:gDef}
\end{align}
In the literature, $\aPot$ is known as the complex acceleration potential, due to the fact that $\grad \phi$ yields the fluid acceleration as expressed in \eqref{Eq:Euler}. Alternatively, $\phi$ can be regarded simply as a negative, normalized pressure field.
In terms of the complex velocity field $w(z,t) = u(z,t) - \i v(z,t)$, \eqref{Eq:Euler} implies the relationship
\begin{align}
	\partial_z \aPot = \Doppz w.
	\label{Eq:velPres}
\end{align}

An important distinction exists regarding the domains over which $\aPot$ and $w$ enjoy analyticity. The acceleration potential, $\aPot$, is analytic everywhere except on the plates themselves. In particular, $\aPot$ is analytic across each of the semi-infinite vortex sheets, whereas $w$ possesses branch cuts along these sheets. Thus, working with $\aPot$ as the primary state variable considerably simplifies the analysis as the vortex sheets do not require special attention. These vortex sheets are certainly present in the model (indeed, without them, there would be no thrust), but they need not be computed explicitly. If desired, they can be computed as a post-processing step.

On each plate, the fluid velocity must satisfy a no-flux condition, which, upon linearization, reads
\begin{align}
	v_{m}(x,t) = \Dopp h_{m} (x,t) 
	\qquad \qquad \textrm{for } x_m^-< x < x_m^+,
	\label{Eq:bc}
\end{align}
where $v_{m}$ is the vertical velocity on the $m$-th plate.
Combining
\eqref{Eq:velPres} and \eqref{Eq:bc}
furnishes the condition 
\begin{align}
	\partial_x \psi(x+\i y_m, t) = -\Dopp^2 h_{m}(x,t)
	\quad \qquad \textrm{for } x_m^-< x < x_m^+. 
	\label{Eq:convDeriv}
\end{align}
We antidifferentiate the boundary data \eqref{Eq:convDeriv} to obtain the functions
\begin{align}
	\Psi_m(x,t) = -\int_{x_m^-}^{x} 
    \Dopp^2 h_{m}(x^\prime, t) \d {x}^\prime
    \quad \qquad \textrm{for } x_m^-< x < x_m^+.
\label{Eq:PsiDefn}
\end{align}
Thus, on each boundary component, $\psi$ matches $\Psi_m$ up to an integration constant. These constants, which may be different on each boundary component, will later be determined by a compatibility condition.

In addition, the velocity must vanish far upstream and the pressure must tend to the reference value in the far-field, giving conditions:
\begin{align}
	& |w(z,t)|\rightarrow 0 \qquad\textrm{ as } \quad x \rightarrow -\infty,
	\label{Eq:velFF} \\
	& \phi(z,t)\rightarrow 0 \qquad \textrm{ as } \quad |z| \rightarrow \infty.
	\label{Eq:presFF}
\end{align}

Due to geometric considerations, the tangential fluid velocity possesses a square-root singularity at the leading edge $z_m^-$ of each plate~\cite{Katz2009, wu1961swimming}. Moreover, the velocity at the trailing-edge $z_m^+$ must remain finite due to the Kutta condition. We thus obtain two endpoint conditions on each plate:
\begin{align}
\label{Eq:SqSing}
& w(z,t)= O\left(\frac{1}{\sqrt{z-z_m^-}}\right)\qquad\textrm{ as }\quad z\rightarrow z_m^- \\
\label{Eq:Kutta}
& |w(z_m^+,t)| < \infty.
\end{align}

In summary, we seek a complex acceleration potential $f(z,t)$, analytic in the domain $D_z$ exterior to the $M+1$ plates, whose imaginary part matches~\eqref{Eq:PsiDefn} up to a constant of integration on each plate and whose real part satisfies the far-field condition~\eqref{Eq:presFF}. Further, the complex velocity $w(z,t)$, related to $\aPot(z,t)$ through~\eqref{Eq:velPres}, must satisfy  boundary condition~\eqref{Eq:bc}, far-field condition~\eqref{Eq:velFF}, and endpoint conditions~\eqref{Eq:SqSing}--\eqref{Eq:Kutta}.

\subsection{Harmonic motions}
\label{Sec:harmonic}

With the model established for general motions, we now consider the special case of harmonic motions. That is, each plate oscillates harmonically in time, with its own amplitude and phase.
Owing to linearity, it suffices to consider the case of a single oscillation frequency $\freq$; the case of multiple oscillation frequencies can be obtained by Fourier superposition, as shown in Appendix \ref{Ap:Fourier}.
Scaling time on $1/\freq$, the motion of the $m$-{th} foil is expressed as
\begin{align}
	h_{m}(x,t) = h_{m}^c(x) \cos(2 \pi t) +  h_{m}^s(x) \sin(2 \pi t).
\end{align}
where $t$ is the dimensionless time.

We similarly decompose $\aPot(z,t)$ and $w(z,t)$ and introduce the vector quantities
\begin{align}
	\baPot = \begin{bmatrix} \aPot^c(z)\\ \aPot^s(z) \end{bmatrix},\qquad
	\bw = \begin{bmatrix} w^c(z)\\ w^s(z) \end{bmatrix}.
\end{align}
In this notation, \eqref{Eq:velPres} becomes
\begin{align}
	\partial_z \baPot(z) = 2 \pi \bJ \bw(z)  + U \partial_z \bw(z)
	\label{Eq:pde1}
\end{align}
where
\begin{align}
	\bJ =
	\begin{bmatrix}
		0& 1\\
		-1 & 0
	\end{bmatrix}.
\end{align}
We remark that $\bJ$ can be viewed as a matrix representation of the imaginary unit since $\bJ^2 = -\bI$, where $\bI$ is the identity matrix. The eigendecomposition of $\bJ$ gives the convenient formula
\begin{align}
	\e^{\sigma \bJ z} =
	\begin{bmatrix}
		\cos(\sigma z) & \sin(\sigma z)\\
		-\sin(\sigma z) & \cos(\sigma z)
	\end{bmatrix}
	\label{Eq:eigen}
\end{align}
which is directly analogous to Euler's formula for $e^{\sigma \j z}$. Some previous studies~\cite{wu1961swimming, moore2014analytical, moore2017fast} used an imaginary unit $\j = \sqrt{-1}$, distinct from the unit $\i$ so that $\i \times \j \ne -1$. We instead use the above matrix representation to avoid the (potentially significant) confusion of having two distinct imaginary units, which would be compounded by the task of working in multiply-connected domains in the complex plane.

Momentarily treating $\baPot(z)$ as given, \eqref{Eq:pde1} can be viewed as a first-order ODE system for $\bw(z)$, whose solution is given by
\begin{align}
	\bw(z) = \frac{1}{U} \e^{-\sigma\bJ z}
	\int_{-\infty}^z \e^{\sigma \bJ z^\prime}
	\partial_z \baPot(z^\prime) \d z^\prime \label{Eq:matSol}
\end{align}
where $\sigma = 2 \pi/ U$ is called the reduced frequency. In terms of physical parameters, $\sigma = \pi \freq \chord / \Udim$, so the reduced frequency measures the ratio of the flapping period to the  time required for the free-stream flow to travel across a typical plate.

The boundary data \eqref{Eq:PsiDefn} can be written as
\begin{align}
	\boldsymbol{\Psi}_{m}
	=
	\begin{bmatrix}
		\Psi^c_m\\
		\Psi^s_m
	\end{bmatrix}=
	- U^2 \int_{x_m^-}^{x}
	\left( \sigma \bJ + \bI\frac{\d}{\d x^\prime}  \right)^2 \boldsymbol{h}_m(x^\prime)
	\d x^\prime \quad \qquad \textrm{for } x_m^-< x < x_m^+,
\label{Eq:vectorBC}
\end{align}
where $\boldsymbol{h}_m = \left[h^c_m\quad h^s_m \right]^T$.

\subsection{The solution for a single plate}
\label{Sec:singlePlate}

Here, we briefly rederive Wu's single-plate result~\cite{wu1961swimming} in a way that will naturally generalize to multiple plates. After non-dimensionalizing, the single plate occupies the slit $z \in [-1,1]$. Following Wu, we associate the physical $z$-plane with a parametric $\zeta$-plane via the Joukowski map
\begin{equation}
\map(\zeta) = {1 \over 2} \left ( \frac{1}{\zeta} + \zeta \right ) \, .
\label{Jouk}
\end{equation}
This function conformally maps the unit disk in the $\zeta$-plane to the exterior of the plate in the $z$-plane. The unit circle $C_0$ maps to the plate, with the trailing and leading edges corresponding to $\zeta_0^\pm = \pm 1$. We let $\APot(\zeta, t) \equiv \aPot(z, t)$ represent the complex acceleration potential in the $\zeta$-plane.
The behavior of the map~\eqref{Jouk} near $\zeta=\zeta_0^-=-1$, combined with~\eqref{Eq:velPres} and~\eqref{Eq:SqSing}, 
implies a simple pole in the complex potential $\APot(\zeta,t)$:
\begin{align}
    \APot(\zeta,t) \sim \frac{\i a_0(t)}{\zeta + 1} \qquad \textrm{as } \zeta \rightarrow -1.
\label{Eq:Fsing}
\end{align}
Since the velocity of the plate is finite, so too is $\text{Im}[F(\zeta,t)]$ on $C_0$ by the boundary condition~\eqref{Eq:convDeriv}, which implies that $a_0$ is real. Note that $F(\zeta,t)$ must be finite as $\zeta\rightarrow \zeta_0^+=+1$ by the Kutta condition~\eqref{Eq:Kutta}. 

At this point, we depart from Wu's derivation and introduce notation that may seem extraneous for the single-plate problem, but that will generalize nicely to the multiply-connected setting. In particular, the function
\begin{align}
\hat{\K}(\zeta,\alpha) = -\frac{\i \alpha}{\zeta-\alpha} 
\end{align}
with $\alpha=\zeta_0^-=-1$, has the form required to capture the singularity in \eqref{Eq:Fsing}. We therefore resolve $\APot(\zeta,t)$ into singular and regular components:
\begin{align}
    \APot(\zeta,t) = a_0(t)\hat{\K}(\zeta, -1) + L(\zeta,t) , \qquad a_0 \in \mathbb{R} ,
    \label{Eq:g1}
\end{align}
where $L(\zeta,t)$, the regular part, is analytic inside the unit disc and {\em finite} on the boundary $C_0$. Importantly, the function $\hat{\K}(\zeta, -1)$ takes constant imaginary part on the boundary:  $\Im[\hat{\K}(\zeta,-1)]=1/2$ for $\zeta \in C_0$. Therefore, taking the imaginary part of \eqref{Eq:g1}, evaluating on the boundary and applying \eqref{Eq:convDeriv}--\eqref{Eq:PsiDefn} gives
\begin{align}
    \Im[L(\zeta, t)] = \Psi_0(\zeta, t),\qquad \textrm{for } \zeta \in C_0. 
\end{align}
where we have fixed a constant of integration without losing generality. 

Thus, $L(\zeta,t)$ is a (single-valued) analytic function that is finite on the boundary and whose imaginary part takes prescribed values on $C_0$. Mathematically, this is known as a (modified) \emph{Schwarz problem} \cite{Crowdy2008}. In the simply-connected case, the solution is given by the Poisson formula \cite{Fokas2003,CrowdyBook}:
\begin{equation}
L(\zeta, t)  =
 {1 \over 2 \pi}
\oint_{C_0} 
\Psi_0(\zeta^\prime,t) \left [{\zeta'+ \zeta \over \zeta'-\zeta} \right ] {\mathrm{d} \zeta' \over \zeta'} + D(t), \qquad D(t) \in \mathbb{R}.
\label{Eq:Poisson}
\end{equation}
The (time-dependent) constant $D(t)$ is chosen to satisfy the far-field condition \eqref{Eq:presFF}.
Equation \eqref{Eq:Poisson} can be discretized by a quadrature method of the reader's choice, and we therefore consider the regular part of the solution to be completely determined.

All that remains is to determine the coefficient $a_0(t)$ of the singular part in~\eqref{Eq:g1}. It suffices to evaluate the no-flux condition at a single point on the plate. For example, if the plate executes harmonic motions, we may take the imaginary part of the contour integral~\eqref{Eq:matSol} and apply the no-flux condition~\eqref{Eq:bc} to yield an expression for $a_0$ in terms of Bessel functions.
The details of how to derive these Bessel functions, and further connections to Wu's solution are explicated in Appendix \ref{Sec:wuComp}.

In summary, we deployed four steps in the solution of Wu's problem:
\begin{enumerate}
\item Conformally map the circular domain to the physical domain;
\item Construct a function, $\hat{\K}(\zeta,\alpha)$, that has the appropriate leading-edge singularity and, simultaneously, assumes constant imaginary part on the boundary;
\item Solve the Schwarz problem to determine the regular part of the solution;
\item Calculate the coefficient $a_0(t)$ by evaluating the no-flux condition at a single point on the boundary.
\end{enumerate}
In \S\ref{Sec:Conformal} and \S\ref{Sec:Solution}, we will explain how to generalise each of these steps to multiple plates.

We remark that Wu did not explicitly use the Poisson formula \eqref{Eq:Poisson} to solve for the acceleration potential, but instead matched coefficients in a series expansion \cite{wu1961swimming}. The advantage of the Poisson-formula representation, however, is that, once the requisite mathematical objects are introduced, it easily generalizes to any connectivity (\S \ref{Sec:Schwarz}). Further, Wu's series expansion can be recovered directly from the Poisson formula above; indeed, a series expansion could be viewed as a convenient quadrature method for \eqref{Eq:Poisson}. We provide some relevant details in Appendix \ref{Sec:wuComp}.
 \section{Conformal maps and the prime function}
\label{Sec:Conformal}
\begin{figure}[t!]
\centering
	\resizebox{\linewidth}{!}{
	\begin{tikzpicture}[remember picture]
	\begin{scope}[scale = 0.7,shift = {(-7,0)}, every node/.style={scale=.6}]
	\path [path fading = circle with fuzzy edge 15 percent, fill=confgrey, even odd rule] (0,0) circle (\zeroRad) (0,0) circle (1.3*\zeroRad);
\draw[line width = \lw, blue,fill= confcol] (0,0) circle (\zeroRad);

\node[right = 8pt, above = 8pt] at ($1.2*\zeroRad*({cos(-30)},{sin(-30)})$) {$C_0$};

\node at (-1.25,1.75) {\Large $D_\zeta$};
\skCircDetColThick{(\delIx,\delIy)}{\radI}{1}{red}
\skCircDetColThick{(\delIIx,\delIIy)}{\radII}{2}{green}
\skCircDetColThick{(\delIIIx,\delIIIy)}{\radIII}{3}{magenta}
\fill[fill =cyan] (-1.5,-1.5) circle (2pt) node[below]  {\Large$\beta$};
\end{scope}
\myPathTextAbove{$z = \map(\zeta)$}{(-2.6,2)}{(-.6,2)}{20}
	\begin{scope}[scale = 0.7,shift = {(1,0)}]
\draw[line width = \lw, red,shift = {(-1.5,-.5)}] (0,0)--(2.5,0);
\draw[line width = \lw, blue,shift = {(2,-1.5)}] (0,0)--(2,0);
\draw[line width = \lw, green,shift = {(2,.5)}] (0,0)--(2,0);
\draw[line width = \lw, magenta,shift = {(-1,1.5)}] (0,0)--(1.5,0);
	\node at (2.5,1.5) {\large $D_z$};
\draw[-Latex,thick] (-2,-2)--(-.5,-2);
\draw[-Latex,thick] (-2,-2.3)--(-.5,-2.3) node[below, midway] {$U$};
\end{scope}	
	\end{tikzpicture}
	}
\caption{The conformal map. We map a canonical circular domain $D_\zeta$ of connectivity $M+1$ to the unbounded fluid region $D_z$ exterior to $M+1$ swimming plates. 
The circular domain $D_\zeta$ is
characterized by the centres $\lbrace \delta_m | m=1, \dots, M \rbrace$ and 
radii $\lbrace q_m | m=1, \dots, M \rbrace$ of $M$ discs with boundaries $C_m$ excised from a unit disc with boundary $C_0$. 
\label{NewFig1}}
\end{figure}
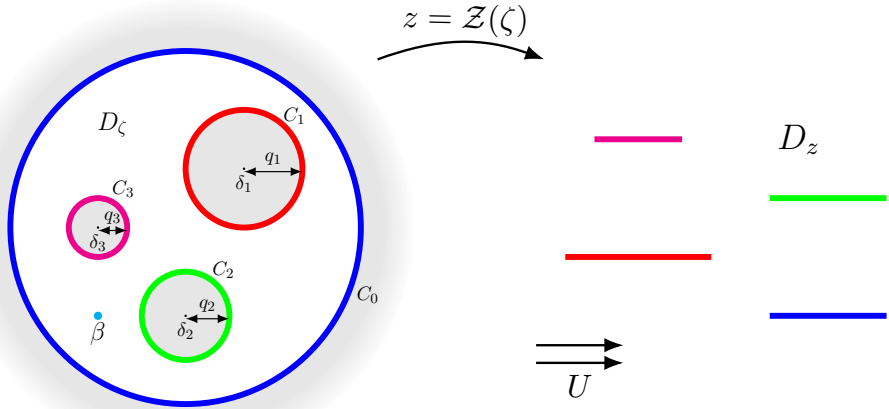

We now introduce the main mathematical machinery that will be used to generalize Wu's solution to multiple swimmers. For a single swimmer, the familiar Joukowski map \eqref{Jouk} associates the physical domain with the unit disc in the $\zeta$-plane. For multiple swimmers, the unit disc is no longer appropriate, but an extension of the Riemann mapping theorem guarantees the existence of a conformal mapping from a multiply connected circular
domain $D_\zeta$ of the kind shown 
in Figure \ref{NewFig1} to the unbounded fluid region exterior to $M+1$ swimming plates.
The domain $D_\zeta$ is the unit $\zeta$ disc with $M$ smaller circular discs excised, the smaller discs having centres $\lbrace \delta_m | m=1, \dots, M \rbrace$ and radii $\lbrace q_m | m=1, \dots, M \rbrace$. The outer boundary of $D_\zeta$, the unit circle, will be denoted by $C_0$,
its interior circular boundaries by $\lbrace C_m | m=1, \dots, M \rbrace$. 
Here we exploit this generalized mapping theorem and carry out an 
analysis that is a direct theoretical extension
of Wu's single plate analysis to multiple swimming plates.

The appropriate mathematical machinery exists:
a general framework for solving problems in multiply connected planar domains
has
been described in a recent monograph \cite{CrowdyBook} and is
based on consideration of a so-called {\em prime function} (also known as the ``Schottky--Klein prime function'' \cite{Baker1897}),
denoted here by
\begin{equation}
    \omega(\zeta,\alpha),
    \label{SKP}
\end{equation}
that can naturally be associated
with any given domain $D_\zeta$.
 The notation (\ref{SKP}) is convenient, but note that it hides the dependence of the prime function on the  parameters
$\lbrace \delta_m, q_m \rbrace$ characterizing the domain $D_\zeta$. 
The prime function is a classical special function, but its significance for solving problems in multiply connected domains -- such as the swimming plate problem
of interest here -- had been missed until recently. 
The prime function for a 
disc with no holes is the simple function
\begin{equation}
\omega(\zeta,\alpha) = \zeta-\alpha.
\end{equation}
Invoking this important fact in the general formulation to follow will retrieve Wu's results
for a single swimming plate.

The treatment  in \cite{CrowdyBook} demonstrates that the theory of the prime function associated with $D_\zeta$
is connected
to potential theory in that domain; indeed, the prime function can be defined as a double limit
of a so-called third kind differential built from a suitable potential theoretic Green's function 
defined in $D_\zeta$.
While the Green's function is the more familiar object to applied scientists, it turns out that
the lesser-known prime function is the more versatile object for representing the solution of many different
applied problems involving multiply connected domains (see~\cite{CrowdyBook} Part 2). Numerical codes are now available \cite{ACCA} for efficient and accurate
calculation of the prime function associated with a given circular domain $D_\zeta$ \cite{Crowdy2016a}.

One of 
the main arguments \cite{CrowdyBook} for the importance, and theoretical significance, of the prime function is that many results familiar in a simply connected setting -- here, Wu's single-plate analysis --
can be generalized in a natural way to the multiply connected setting.
Presenting the specific details of this generalization for Wu's swimming plate problem is the aim of this paper.

\subsection{Modified Green's functions and the prime function}

\label{Sec:genresults}
With the prime function $\omega(.,.)$ associated with a multiply connected circular domain $D_\zeta$ understood to
be a well-defined and computable function, we now record all the requisite results  needed
to generalize Wu's single-plate analysis to multiple swimming plates.
These results are reported here without proof;
 full details can be found in \cite{CrowdyBook}.

An important observation for present purposes concerns the {\em modified Green's function} ${G}_0(\zeta,\alpha)$ 
in $D_\zeta$ that satisfies
\begin{equation}
\nabla^2 G_0(\zeta,\alpha) = - \delta(\zeta-\alpha),  \qquad \zeta, \alpha \in D_\zeta
\end{equation}
and assumes constant values on each boundary component of $D_\zeta$.
It can be
expressed in terms of the prime function associated with $D_\zeta$ as
\begin{equation}
G_0(\zeta,\alpha) = {\rm Im}[{\mathcal G}_0(\zeta,\alpha)], \qquad 
{\mathcal G}_0(\zeta,\alpha) = {1 \over 2 \pi {\rm i}} \log \left ( {\omega(\zeta,\alpha) \over |\alpha| \omega(\zeta,1/\overline{\alpha})} \right ).
\label{eq:dgc0}
\end{equation}
Since  $D_\zeta$ is $(M+1)$-connected, there are $M$ additional modified Green's functions conveniently
expressed as
\begin{equation}
G_m(\zeta,\alpha) = {\rm Im}[{\mathcal G}_m(\zeta,\alpha)], \qquad 
{\mathcal G}_m(\zeta,\alpha) = {1 \over 2 \pi {\rm i}} \log \left ( {\omega(\zeta,\alpha) \over |\alpha| \omega(\zeta,\theta_m(1/\overline{\alpha}))} \right ), 
\label{MoreModG}
\end{equation}
where 
\begin{equation}
\theta_m(\zeta) = \delta_m + {q_m^2 \zeta \over 1- \overline{\delta_m} \zeta}, \qquad m=1, \dots, M
\label{thetadef}
\end{equation}
are M\"obius mappings associated with each boundary circle.
The modified Green's functions satisfy
\begin{equation}
\nabla^2 G_m(\zeta,\alpha) = - \delta(\zeta-\alpha), \qquad \zeta, \alpha \in D_\zeta, \qquad m=1, \dots, M
\end{equation}
and all assume constant values on the boundary components of $D_\zeta$. The difference between the various
modified Green's functions
lies in whether
their analytic extensions $\lbrace {\mathcal G}_m(\zeta,\alpha) \rbrace$ are single-valued around the boundary circles: the function
${\mathcal G}_m(\zeta,\alpha)$ is single-valued around all boundaries of $D_\zeta$ except for $C_m$ \cite{CrowdyBook}.
Crowdy \cite{CrowdyFOILS} first introduced the
functions (\ref{MoreModG}) in the context of calculating the lift forces on a set of
aerofoils in a steady uniform flow (see
also Sec. 4.10 of \cite{CrowdyBook}).

These modified Green's functions enable us to solve the swimming plate problem for multiple interacting plates.
In particular, we use the modified Green's functions to generalise
the first three steps outlined at the end of \S\ref{Sec:singlePlate}:
constructing the conformal map, constructing the leading-edge singularity functions, and
solving the modified Schwarz problem.

Before proceeding
it is useful to introduce the function
\begin{equation}
{\mathcal K}(\zeta,\alpha) \equiv \alpha {\partial \over \partial \alpha} \log \omega(\zeta,\alpha),
\label{Kdef}
\end{equation}
which has a singularity at $\alpha$ of the form
\begin{align}
    \mathcal{K}(\zeta, \alpha) \sim -\frac{\alpha}{\zeta - \alpha}.
    \label{Eq:Kasymp}
\end{align}
{This function is important in the general framework \cite{CrowdyBook} and also plays a prominent role in our analysis.
Indeed, the complete solution to our problem can be phrased exclusively in terms of the function $\K$.}

\subsection{Parallel slit maps}\label{Sec:Parallel}

First, it can be shown \cite{Crowdy2006,CrowdyBook} that within a class of so-called {\em parallel slit maps} 
is the function given by
\begin{equation}
- 2 \pi {\rm i} \left [
{\partial \over \partial \alpha} - {\partial \over \partial \overline{\alpha}} \right ] {\mathcal G}_m(\zeta,\alpha), \qquad m=0,1, \dots, M.
\label{eq:dgc1}
\end{equation}
The point $\alpha$, assumed to be strictly inside $D_\zeta$, is mapped to infinity in an image plane.
Moreover, all circles $\lbrace C_m | m=0,1, \dots, M \rbrace$ are transplanted to slits
of
finite length
all parallel to the real
axis -- hence the designation ``parallel slit maps''.
These properties follow directly from those of the modified Green's functions
and the observation that
\begin{equation}
{\partial \over \partial \alpha} - {\partial \over \partial \overline{\alpha}} = {\partial \over \partial ({\rm i} \alpha_y)},
\end{equation}
where $\alpha = \alpha_x + {\rm i} \alpha_y$. Since $\mathcal{G}_m$ has constant imaginary part when $\zeta\in\partial D_{\zeta}$, so too does~\eqref{eq:dgc1}, as it is the parametric derivative of $-2\pi\mathrm{i}\mathcal{G}_m$ with respect to the purely imaginary parameter $\mathrm{i}\alpha_y$.  
It is also easily verified that the map
(\ref{eq:dgc1}) 
has a simple pole with unit residue at $\zeta=\alpha$.
Consequently, the map transplants $D_\zeta$ to the unbounded region exterior to $M+1$
slits all parallel to the real axis in the image plane; see \cite{CrowdyBook} for more details.

For $m=0$, for example, \eqref{Kdef} can be used to write~\eqref{eq:dgc1} as
\begin{equation}
\varphi_0(\zeta,\alpha) = 
- {1 \over \alpha} {\mathcal K}(\zeta,\alpha) + {1 \over \overline{\alpha}} {\mathcal K}(\zeta, 1/\overline{\alpha}),
\end{equation}
up to terms independent of $\zeta$. The mapping $\varphi_0(\zeta,\alpha)$ has a constant imaginary part when $\zeta$ lies on any boundary circle of $D_\zeta$. Thus, any function 
\begin{equation}
\map(\zeta) =
A \varphi_0(\zeta,\beta) + B, \qquad A \in \mathbb{R}^+, B \in \mathbb{C},
\label{phimap}
\end{equation}
serves as a mapping from $D_\zeta$
to $M+1$ swimming plates parallel to the real axis with $\beta$ now used to denote the preimage of the point at infinity for consistency (see figure \ref{NewFig1}).
Given the locations of the end points of $M+1$ swimming plates, namely,
\begin{equation}
[z_{m}^-, z_m^+], \qquad m = 0,1, \dots, M,
\end{equation}
the parameters $A, B, \lbrace \delta_m, q_m | m=1, \dots, M \rbrace$ can be determined by solving a parameter problem. 
Specifically, we denote the preimages of $z_m^{\pm}$ as $\zeta_m^{\pm}$, each pair being on one of the boundary circles.
Then, in order that the images are the end points of a slit, necessary conditions are
\begin{equation}
    {\map}(\zeta_{m}^\pm) = z_{m}^\pm,
    \qquad {\mathrm{d} {\map} \over \mathrm{d} \zeta} \biggl |_{\zeta_{m}^\pm} = 0,\qquad m = 0,1, \dots, M.\\
\label{zeroderiv}
\end{equation}
Since the functional form of ${\map}(\zeta)$
is known from (\ref{phimap}) up to a finite set of parameters, the set (\ref{zeroderiv}) is a finite system of nonlinear equations for those parameters.
In the problem considered here, where the arrangement of swimming plates is fixed at the outset, these parameters can be solved for as a set-up step in the calculation. The number of parameters matches the number of constraints, as shown in Appendix~\ref{Sec:dof}.

Once the parameters are found,
the mapping (\ref{phimap}) is precisely the function we need to generalize the
classical Joukowski map (\ref{Jouk}) used by Wu.
To see this explicitly, note that, for a single plate ($M=0$),
\begin{equation}
\omega(\zeta,\alpha) = \zeta-\alpha, \qquad {\mathcal K}(\zeta,\alpha) = - {\alpha \over \zeta-\alpha}
\label{mzero}
\end{equation}
and (\ref{phimap}) becomes
\begin{equation}
\map(\zeta)
= 
A\left [ {1 \over \zeta-\beta} - {1 \over \zeta-1/\overline{\beta}}\right ] + B
= 
A\left [ {1 \over \zeta-\beta} + {1 \over \overline{\beta}} {1 \over (1-\overline{\beta} \zeta) } \right ] + B.
\label{phimap2}
\end{equation}
Setting $A=1/2$ and $B=-A/\overline{\beta}$, and then letting $\beta \to 0$ retrieves the Joukowski map \eqref{Jouk}.

\subsection{Leading-edge singularity functions}
\label{Sec:Kfun} 

Geometrically, as $\alpha$ tends to a boundary point of $D_\zeta$, the functions (\ref{eq:dgc1}) give a class of so-called parallel slit mappings that transplant the boundary circle on which $\alpha$ sits to an infinite line parallel to the finite-length slit images of all other circles. 
Such functions are not of interest to us as conformal mappings, but they are useful for describing the pressure singularities present at the leading edge of each wing.

Suppose, for example, that $\alpha$ tends to $C_0$, i.e.,
$\alpha \to e^{{\rm i} \vartheta}$ for some $0 \le \vartheta < 2\pi$.
Then $\alpha \to 1/\overline{\alpha}$ and
\begin{equation}
\varphi_0(\zeta,\alpha) \to
- \e^{-{\rm i} \vartheta}{\mathcal K}(\zeta,\alpha) +\e^{{\rm i} \vartheta} {\mathcal K}(\zeta, \alpha) = 2 {\rm i} {\mathcal K}(\zeta, \alpha) \sin \vartheta.
\label{eq:dgc2}
\end{equation}
Recall that, for $\zeta \ne \alpha$, $\varphi_0(\zeta,\alpha)$ has constant imaginary part on all boundary circles of $D_\zeta$. Therefore, \eqref{eq:dgc2} shows 
that ${\rm i} {\mathcal K}(\zeta,\alpha)$ has  constant imaginary part on the $M$ boundary circles
$\lbrace C_m | m=1, \dots, M \rbrace$ and, away from the singularity at $\alpha$, also on $C_0$.

If, on the other hand, $\alpha \to \delta_m+ q_m \e^{{\rm i} \vartheta}$ for some $m=1, \dots, M$,
an analogous analysis of the limit of (\ref{eq:dgc1}) 
allows us to deduce that the function
\begin{equation}
\varphi_m(\zeta,\alpha)= -{1 \over \alpha} {\mathcal K}(\zeta,\alpha) 
\left [1 + {\partial \over \partial \overline{\alpha}} \theta_m(1/\overline{\alpha}) \right ]
\end{equation}
has constant imaginary part on each boundary circle;
here we have used the fact that
if $\alpha$ lies on $C_m$ then $\theta_m(1/\overline{\alpha}) =\alpha$.
But, from (\ref{thetadef}),
\begin{equation}
{\partial \over \partial \overline{\alpha}} \theta_m(1/\overline{\alpha}) = - \e^{2 {\rm i} \vartheta}
\end{equation}
which implies that
\begin{equation}
\varphi_m(\zeta,\alpha)={2 \sin \vartheta \over q_m} \left [{\rm i} \left ({\alpha-\delta_m \over \alpha} \right ) {\mathcal K}(\zeta,\alpha) \right ],
\label{gencase}
\end{equation}
which shows that, for $\zeta \ne \alpha$, the function in square brackets has constant imaginary part on {each} boundary circle (different constant values on different circles). 
Formula (\ref{gencase}) reduces to (\ref{eq:dgc2}) when $q_m=1, \delta_m=0$.

In summary, if we define $\delta_0=0$, we can say that when $\alpha$ tends to {\em any} boundary circle $C_m$,
the function
\begin{equation}
\hat{\K}(\zeta, \alpha) \equiv {\rm i} \left ({\alpha-\delta_m \over \alpha} \right ) {\mathcal K}(\zeta,\alpha), 
 \qquad \qquad m=0,1, \dots, M
\label{Eq:hatKdef}
\end{equation}
has a singular real part with a simple pole at $\zeta=\alpha$ and (finite) constant imaginary part when $\zeta\in\partial D_{\zeta}$ (different constant values on different boundary circles). This fact will be useful 
in resolving the leading-edge pressure singularities.

For the case of a single plate, $M=0$, using \eqref{mzero} with $\delta_0=0$ and $\alpha=-1$, the function \eqref{Eq:hatKdef} becomes
\begin{equation}
\hat{\K}(\zeta, -1) =
{{\rm i} \over \zeta+1}
\label{sing3}
\end{equation}
which is precisely the function used by Wu \cite{wu1961swimming} to account for a pressure singularity on the swimming plate, as described in \S\ref{Sec:singlePlate}.

\subsection{Solution of the modified Schwarz problem in $D_\zeta$} 
\label{Sec:Schwarz}

Determining the complex acceleration potential can be viewed as a {\em modified Schwarz problem} \cite{Crowdy2008, CrowdyBook}, that is, the problem of finding a single-valued analytic function with prescribed real (or imaginary) part on the boundary. In multiply connected domains, the solution to such a problem can be written explicitly in terms of the prime function \cite{Crowdy2008, CrowdyBook}. In the present context, we seek an analytic function $\APot(\zeta) = \phi+\i \psi$ whose imaginary part is known up to a constant on all boundary components. Eventually, $\APot$ will represent the regular part of the complex acceleration potential (specifically, $L(\zeta,t)$ in~\eqref{Eq:schwarzSol}),
but for the purposes of this section we use $\APot$ simply to refer to an analytic function. Multiplication by $-\i$ allows us to map to the standard modified Schwarz problem:
\begin{equation}
{\rm Re}[-\i\APot(\zeta)] = 
{\rm Im}[\APot(\zeta)]=
\psi = \left \lbrace
\begin{array}{ll}
\Psi_0, & {\rm on}~C_0, \\
\Psi_m + d_m, & {\rm on}~C_m, ~ m=1, \dots, M,
\end{array} \right .
\end{equation}
where $\lbrace \Psi_m | m=0,1, \dots, M \rbrace$ are considered known (via \eqref{Eq:PsiDefn} in the present context), and the $M$ constants $\lbrace d_m | m =1, \dots, M \rbrace$ must  satisfy a set of $M$ {\em compatibility conditions} for a single-valued solution to exist \cite{Crowdy2008, CrowdyBook}. Given that these conditions are met, the solution is
\begin{equation}
\begin{split}
\APot(\zeta)  &= {1 \over 2 \pi} 
\int_{\partial D_\zeta} \psi(\zeta^\prime) 
\left [\mathrm{d} \log \omega(\zeta',\zeta) + \mathrm{d} \log \overline{\omega}(\overline{\zeta'},1/{\zeta}) 
\right ]
+ D \\
&= {1 \over 2 \pi} 
\int_{\partial D_\zeta} \psi(\zeta^\prime) 
\left [{\mathcal K}(\zeta,\zeta') {\mathrm{d} \zeta'
\over \zeta'} + \overline{{\mathcal K}}(1/{\zeta},\overline{\zeta'}){\mathrm{d} \overline{\zeta'}
\over \overline{\zeta'}} 
\right ]
+ D, \qquad D \in \mathbb{R},
\end{split}
\label{eq:dgc3}
\end{equation}
where $\overline{g}(\zeta,\alpha)\equiv \overline{g(\overline{\zeta},\overline{\alpha})}$ denotes the Schwarz conjugate of $g$. Here, integration over the multi-component boundary $\partial D_\zeta$ is defined as
\begin{equation}
\int_{\partial D_\zeta} = \oint_{C_0} - \sum_{m=1}^M \oint_{C_m}
\end{equation}
where $\oint$ denotes integration over a closed contour in the counter-clockwise direction.
The integral formula (\ref{eq:dgc3}) in terms of the prime function was first written down  in \cite{Crowdy2008}; see also Chapter 13 of \cite{CrowdyBook}. 
In the single-plate case $M=0$ and on use of (\ref{mzero}), equation  (\ref{eq:dgc3}) reduces to the Poisson formula \eqref{Eq:Poisson} and thus recovers Wu's single-plate solution.
 \section{Solution}
\label{Sec:Solution}

Having introduced the necessary mathematical objects, we now solve the \linebreak \mbox{multiple-swimmer problem} introduced in \S\ref{Sec:Modelling}.
Specifically, we seek an analytic function $\APot(\zeta,t)$ (the complex acceleration potential) whose imaginary part \linebreak matches~\eqref{Eq:PsiDefn} up to a constant of integration on each boundary component of $\partial D_\zeta$ and whose real part satisfies the far-field condition \eqref{Eq:presFF}. Further, the complex velocity, related to the acceleration potential through~\eqref{Eq:velPres}, must satisfy the boundary condition~\eqref{Eq:bc}, far-field condition~\eqref{Eq:velFF}, and edge conditions~\eqref{Eq:SqSing}--\eqref{Eq:Kutta}. We present the solution decomposition (\S\ref{Sec:Decomp}) and the regular part of the solution (\S\ref{Sec:Regular}) for general motions. We then present the singular part of the solution (\S\ref{Sec:suctions}) for time-harmonic motions (\S\ref{Sec:harmonic}), 
for which~\eqref{Eq:vectorBC} sets the boundary conditions on the acceleration potential and \eqref{Eq:matSol} relates it to the velocity field.

\subsection{Solution decomposition}\label{Sec:Decomp}

First, by the same reasoning used to obtain \eqref{Eq:Fsing}, the complex acceleration potential $\APot(\zeta,t)$ must possess a simple pole at the pre-image of each leading edge \mbox{$\zeta = \zeta_m^-$} for \mbox{$m = 0, \dots, M$}. The function $\hat{K}(\zeta, \zeta_m^-)$ defined in \eqref{Eq:hatKdef} also has a simple pole at $\zeta_m^-$.
We can therefore extract the singularities from $\APot$ as follows:
\begin{align}
	\APot(\zeta,t) = \sum_{m=0}^M a_{m}(t) \hat{\K}\left(\zeta,\zeta^-_{m}\right) + L(\zeta,t) +  D(t).
	\label{Eq:singularityExt}
\end{align}
What remains, $L(\zeta,t)$, is an analytic function that is {\em finite} on all boundary components. We thus refer to $L(\zeta,t)$ as the regular part of the solution, and $D=D(t)$ is an overall constant that enforces the far-field condition \eqref{Eq:presFF}. The coefficients $a_m(t)$ physically represent the strength of the leading-edge suction forces on each plate. Importantly, since both $F$ and $\hat{\K}$ have finite imaginary part but singular real part on the boundary circles, the coefficients $a_m$ must be real.

\subsection{Regular part of the solution}\label{Sec:Regular}

We showed in \S\ref{Sec:Kfun} that each function $\hat{\K}\left(\zeta,\zeta^-_{m}\right)$ assumes a constant imaginary part on each boundary component of $D_{\zeta}$. Therefore, evaluating the imaginary part of \eqref{Eq:singularityExt} on the boundary gives
\begin{align}
	 \Im \left[ L(\zeta,t)  \right] 
	 = \Psi_{m}(\zeta,t) + d_{m}(t) , \qquad\qquad \zeta \in C_{m},\,  m = 0,\dots, M
\label{Eq:schwarz3}
\end{align}
where $\Psi_{m}$ is boundary data that is known through \eqref{Eq:PsiDefn}, and the (time-dependent) constants $d_m$ arise from two sources: first, from anti-differentiating \eqref{Eq:convDeriv} to obtain $\Psi_m$, and second from the evaluation of $\hat{\K}\left(\zeta,\zeta^-_{m}\right)$ on each boundary. In our formulation, it is unnecessary to distinguish between these two contributions, so we combine them into a single term $d_m$. One of these can be absorbed into the overall constant $D$, and so, without loss of generality, we take $d_0=0$.

Now, \eqref{Eq:schwarz3} takes the form of a modified Schwarz problem as discussed in \S\ref{Sec:Schwarz}. An integral formula for the solution  therefore follows from (\ref{eq:dgc3}):
\begin{align}
	L(\zeta,t) &=	\frac{1}{2 \pi}  \int_{\partial D_\zeta}\left(\Psi_{m}(\zeta^\prime,t) + d_{m}(t) \right)
	\left [{\mathcal K}(\zeta,\zeta') {\mathrm{d} \zeta'
\over \zeta'} + \overline{{\mathcal K}}(1/{\zeta},\overline{\zeta'}){\mathrm{d} \overline{\zeta'}
\over \overline{\zeta'}} 
\right ],
	\label{Eq:schwarzSol}
\end{align}
where the (time-dependent) constants $d_m(t)$ can be determined by enforcing compatibility conditions.
The coefficients $a_m(t)$ of the singular terms are not determined by this formula and must be treated separately.

We note that an analogous procedure applies in the case of harmonic motions (\S\ref{Sec:harmonic}), with $\Psi_m$
replaced by its vectorised equivalent \eqref{Eq:vectorBC}.
We then express the solution \eqref{Eq:schwarzSol} as $\bL = [L^c(\zeta)\,\,\,\, L^s(\zeta)]^T$.
Closed-form expressions for $\bL$ -- i.e. without integrals -- are available for the simple case of in-phase heaving motions, as shown in Appendix \ref{Ap:exact}.

\subsection{Singular part of the solution}
\label{Sec:suctions}

With the regular part of the solution given by \eqref{Eq:schwarzSol}, all that remains is to determine the singular part. For simplicity, we consider the case of harmonic motions.
By decomposition \eqref{Eq:singularityExt}, the singular part is fully specified by the $2(M+1)$ real constants $\boldsymbol{a}_{m}= \left[a_m^c \quad a_m^s \right]^T$ for $m = 0, \dots, M$.

To determine these constants, we will construct a linear system of $2(M+1)$ equations.
To this end,  we sample the plate boundaries at $M+1$ points labelled $z_m=x_m +\i y_m$, with one point on each plate. Substituting~\eqref{Eq:singularityExt} into~\eqref{Eq:matSol} and enforcing the no-flux boundary condition~\eqref{Eq:bc} at $z=z_m$ yields the equation 
\begin{align}
	\Im \left[
	 -\sum_{n=0}^M 
		\frac{1}{U}
	\int_{-\infty}^{z_{m}}
	\e^{\sigma \bJ (z - z_{m})}
\boldsymbol{a}_{n}
	\frac{\d 
	\hat{\K}}{\d z}(\zeta(z), \zeta^-_{n})
\,\d z \right]
=\bv(z_{m})+\bq(z_{m})
\label{Eq:sucInts}
\end{align}
for $m=0,\dots,M$, where
\begin{align}
	\bv(z_m) &= U\left(\sigma\boldsymbol{J}+ \boldsymbol{I}\frac{\mathrm{d}}{\mathrm{d}x}\right)\boldsymbol{h}_m(x_m)
\label{Eq:qDef}
\end{align}
and
\begin{align}
	\bq(z_{m}) &= \Im \left[\frac{1}{U} 
		\int_{-\infty}^{z_{m}}
		\e^{\sigma \bJ (z - z_{m})}
		\frac{\d \bL}{\d z}(\zeta(z))\,
\d z \right].
\label{Eq:qDef}
\end{align}
Note that $\bv(z_m)$ is known because the plates' kinematics are prescribed,
and $\bq(z_m)$ is already known since it depends only on the regular part $\bL$ of the solution. Therefore, the right side of \eqref{Eq:sucInts} is known.

As established in \S\ref{Sec:Decomp}, the coefficients $\boldsymbol{a}_{n}$ are real and can thus be moved outside the brackets in \eqref{Eq:sucInts}.
Therefore, we obtain the linear system
\begin{align}
\boldsymbol{D}\boldsymbol{a} = \boldsymbol{b},
\label{Eq:linear}
\end{align}
where
\begin{align}
    \boldsymbol{a} = 
    \begin{bmatrix}
      \boldsymbol{a}_{0}\\
       \vdots\\
      \boldsymbol{a}_{M}
    \end{bmatrix},\qquad
    \boldsymbol{b} = 
    	    \begin{bmatrix}
      \bv(z_{0})+\bq(z_{0})\\
       \vdots\\
     \bv(z_{M})+\bq(z_{M})
    \end{bmatrix},
\end{align}
and
\begin{align}
\boldsymbol{D}_{m,n}
	&=-\frac{1}{U}
	\Im \left[
	 	\e^{-\sigma \bJ  z_{m}}
	\int_{-\infty}^{z_{m}}
	\e^{\sigma \bJ z}
	\frac{\d 
	\hat{\K}}{\d z}(\zeta, \zeta_{n}^-)
\d z \right]\label{Eq:DMatrix}
\end{align}
is the $2 \times 2$ block of $\bD$ in position $(m,n)$. Solving the linear system \eqref{Eq:linear} yields the values of the $2(M+1)$ constants $\boldsymbol{a}$.

In the simply-connected case, the integrals in \eqref{Eq:sucInts} and \eqref{Eq:qDef} can be expressed in terms of Bessel functions, but analytical expressions are unavailable in the general multiply-connected case. We therefore evaluate these integrals numerically using a rapidly-converging quadrature scheme detailed in Appendix \ref{Ap:integrals}. 
Taken together, Eqs.~\eqref{Eq:singularityExt},~\eqref{Eq:schwarzSol}, and~\eqref{Eq:linear} comprise the solution for the acceleration potential $\APot(\zeta,t)$.

\subsection{Two swimming plates}\label{Sec:TwoSwim}

The case of two swimming plates, corresponding to $M=1$, is of special interest for two reasons. First, the interesting physics of interaction is likely to be well captured by just two interacting plates.
The second reason is that the general
formulation presented thus far affords significant mathematical simplifications in the case $M=1$
because the preimage domain $D_\zeta$ can be taken to be the concentric annulus
$\rho < |\zeta| < 1$, for which
\begin{equation}
    \delta_1 = 0, \qquad q_1 = \rho < 1.
\end{equation}
On a practical level this means that
use of numerical codes to compute the prime function can be avoided, and the following (suitably truncated) infinite product and sum representations of the requisite functions can be used directly. We emphasize that our general formulation, and the formulas derived in \S\ref{Sec:Decomp}-\ref{Sec:suctions}, 
pertain to {\em any number} of swimming plates. We plan to use our general formulation to investigate the case of three or more swimming 
plates in future work.

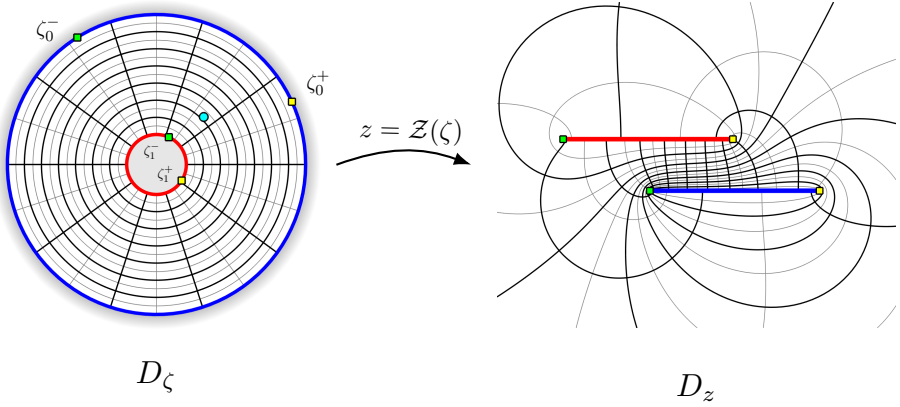
\begin{figure}[t]
	\centering
	\resizebox{\linewidth}{!}{
	\begin{tikzpicture}[scale=1.2, transform shape]
		    \node (image) at (-3,0) 
	    {\includegraphics[width=.3\linewidth]{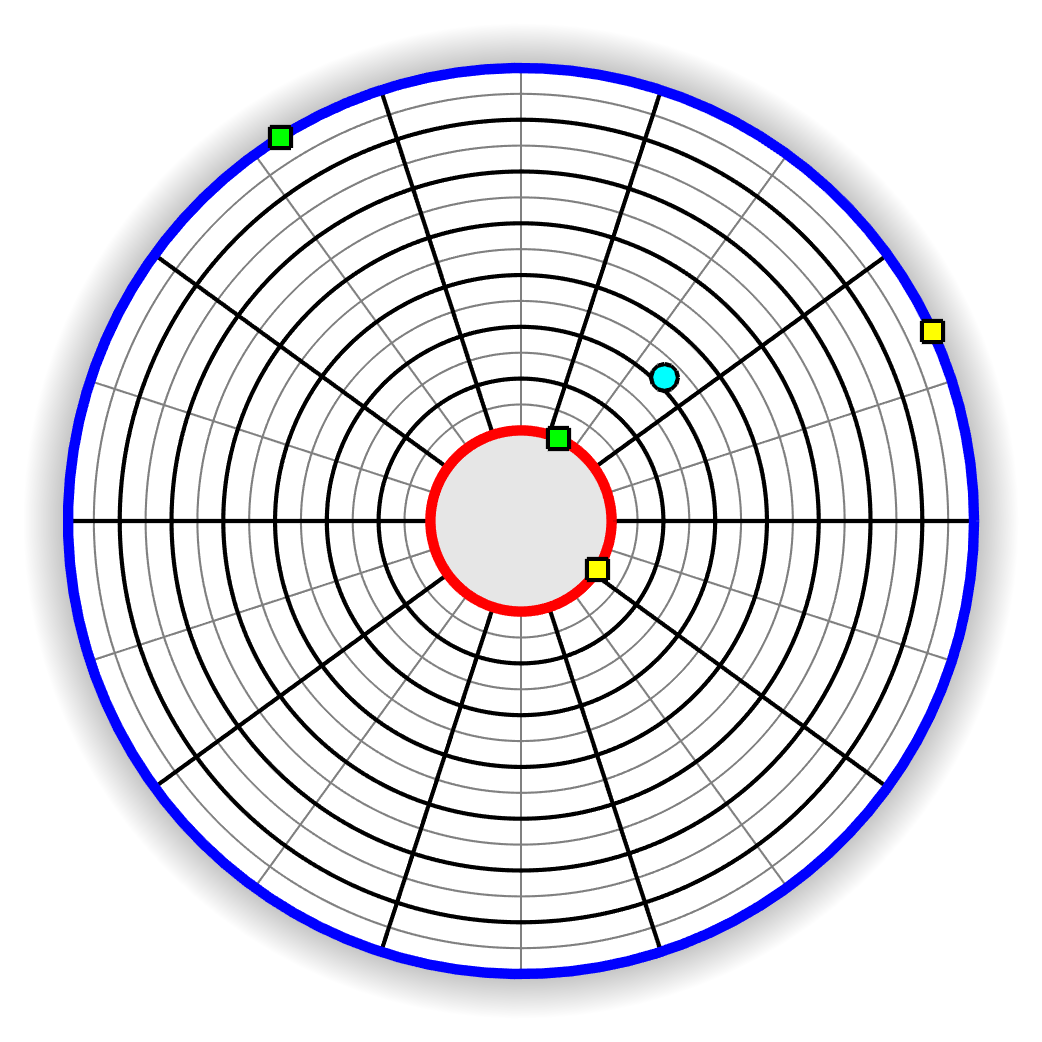}};
    \node[below = 0cm of image] 
    {\small{$D_\zeta$}};
\node[scale = 0.6] at (-4.2,1.5) {$\zeta_0^-$};
\node[scale = 0.6] at (-1.2,.9) { $\zeta_0^+$};
\node[scale = 0.4] at (-3.05,.15) { $\zeta_1^-$};
\node[scale = 0.4] at (-2.9,-.1) { $\zeta_1^+$};

	    \myPathTextAbove{\small{ $z = \map(\zeta)$}}{(-1,0)}{(.5,0)}{20}
	    \node (image) at (3,0) {\includegraphics[width=.35\linewidth]{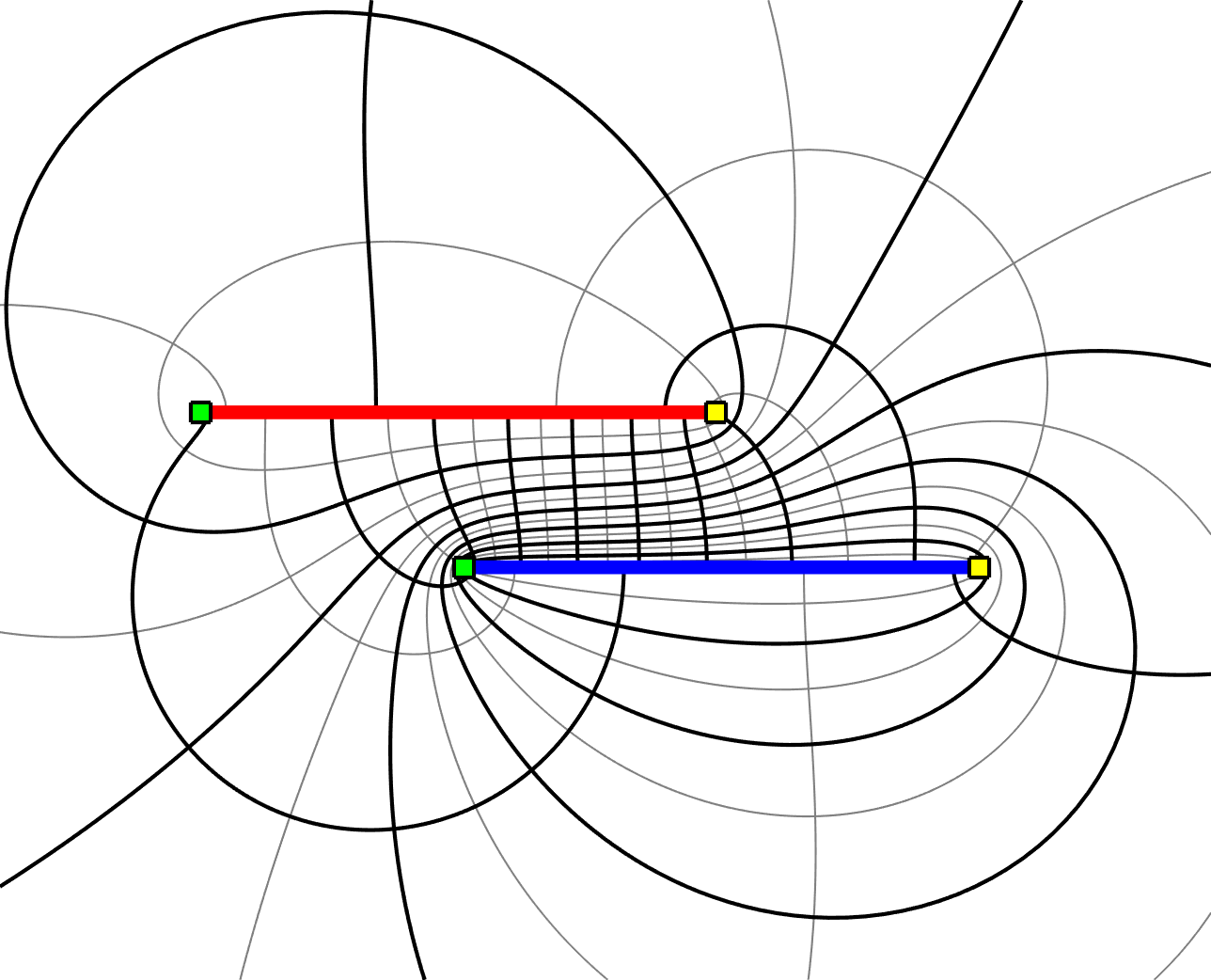}};
	    \node[below = .25cm of image] {\small{$D_z$}};
	\end{tikzpicture}
}
	\caption{An example of a parallel slit map $\map(\zeta)$ between doubly connected domains. The annulus is mapped
	to the exterior of two finite, parallel slits.
	The blue (red) circle is mapped to the blue (red) slit.
	The cyan dot in the $D_\zeta$ plane represents the preimage of infinity, $\beta$.
	The preimages of the edges of the slits are denoted by the green ($\zeta_i^-$) and yellow ($\zeta_i^+$) circles for the leading and trailing edges, respectively.
	In this plot, the conformal modulus is taken to be $\rho = 0.05$ and $\beta = \sqrt{\rho} \e^{\i \pi/4}$.}
	\label{Fig:conformal}
\end{figure}

It can be shown
-- see Chapters 5 and 14 of \cite{CrowdyBook} -- 
that for the concentric annulus $\rho < |\zeta| < 1$,
\begin{equation}
\omega(\zeta,\alpha) = - {\alpha \over C^2} P(\zeta/\alpha), \qquad C = \prod_{n=1}^\infty (1-\rho^{2n}),\label{Eq:Omega2}
\end{equation}
where
\begin{equation}
P(\zeta) = (1-\zeta) \prod_{n=1}^\infty (1-\rho^{2n}\zeta) (1-\rho^{2n}/\zeta)\label{Eq:P2}
\end{equation}
is a (uniformly) convergent product for $0 \le \rho < |\zeta| < 1$. This
function satisfies the functional
relations \cite{CrowdyBook}:
\begin{equation}
    P(\rho^2 \zeta) = -\zeta^{-1}P(\zeta) = P(1/\zeta).
    \label{Pprop}
    \end{equation}
Substituting~\eqref{Eq:Omega2} and~\eqref{Eq:P2} into the definition of $\mathcal{K}$ given in~\eqref{Kdef}, we obtain
\begin{equation}
{\mathcal K}(\zeta,\alpha) = {1 \over 1-\zeta/\alpha} + \sum_{n=1}^\infty \left \lbrace
{\rho^{2n} \zeta/\alpha \over 1-\rho^{2n} \zeta/\alpha } -
{\rho^{2n} \alpha/\zeta \over 1-\rho^{2n} \alpha/\zeta } \right \rbrace.
\label{Kdef2}
\end{equation}
Figure~\ref{Fig:conformal} shows the corresponding conformal map $\map(\zeta)$, as given by~\eqref{phimap}, for a typical set of parameters. We note that choosing $|\beta|=\sqrt{\rho}$ leads to a pair of plates of equal length. For a given configuration of plates, it is necessary to determine $\rho$ and $\arg(\beta)$ by solving the parameter problem described in \S\ref{Sec:Parallel}. Alternatively, a range of configurations may be produced by sweeping across values of these parameters, as we do in \S\ref{SSec:Staggered}. We note that the special case of real (imaginary) $\beta$ produces the in-line (stacked) configuration considered in \S\ref{sec:results}.

Equation~\eqref{Pprop} can be used to write~\eqref{Eq:schwarzSol} as
\begin{equation}
\begin{split}
L(\zeta,t) &=
 {1 \over 2 \pi } \oint_{C_0} \Psi_0(\zeta^\prime,t) 
 \left [2 {\mathcal K}(\zeta,\zeta') -1 \right ] 
 {\mathrm{d}\zeta' \over \zeta'} 
 \\ &-
{1 \over 2 \pi } 
\oint_{C_1} (\Psi_1(\zeta^\prime,t)+ d_1(t))
[2 {\mathcal K}(\zeta,\zeta') ] {\mathrm{d}\zeta' \over \zeta'} 
+ D(t),
\end{split}
   \label{eq:dgc5}
\end{equation}
where $D(t) \in \mathbb{R}$.
This is an integral formula for the solution of the modified Schwarz problem in the concentric annulus
\cite{CrowdyBook}.
The single compatibility condition which ensures that the acceleration potential is single-valued (as discussed in~\cite{Crowdy2008}) is
\begin{equation}
\oint_{C_0} \Psi_0(\zeta,t) {\mathrm{d} \zeta \over \zeta} = \oint_{C_1} \left [\Psi_1(\zeta,t)  + d_1(t) \right ] {\mathrm{d} \zeta \over \zeta},
\end{equation}
which affords an explicit expression for $d_1(t)$ in terms of $\Psi_0$ and $\Psi_1$.
For the two-plate case it turns out that
use of the integral formula
(\ref{eq:dgc5}) can be circumvented, since a
computationally efficient method 
based on Laurent series can be used instead. That method, which we use here,
 is described in appendix D of \cite{Crowdy2018}.

The results in this section may readily be connected to Wu's single-plate theory~\cite{wu1961swimming} by taking the limit $\rho\to 0$. Specifically, this limit reduces the annulus to the unit disc,
and formula~\eqref{eq:dgc5} reduces to the Poisson integral formula \eqref{Eq:Poisson}. 
A further simplification of the general theory is that, since $\delta_0 = \delta_1 = 0$,~\eqref{Kdef2} may be used to show that~\eqref{Eq:hatKdef} reduces to the singularity function~\eqref{sing3} used by Wu.

 \section{Results for pairs of swimmers}%
\label{sec:results}

Using the solution derived in \S\ref{Sec:TwoSwim}, we now present the pressure fields (\S\ref{SSec:Pressure}) and hydrodynamic thrust (\S\ref{SSec:Inline} and \S\ref{SSec:Staggered}) associated with pairs of swimmers, and give physical interpretations of our results. 

\subsection{Pressure fields}
\label{SSec:Pressure}

Our formulation casts the entire hydrodynamic problem in terms of the complex acceleration potential, $\aPot(z,t) = \phi(z,t) + \i \psi(z,t)$, whose real part $\phi(z,t)$ is simply the (negative and normalized) pressure field. Other quantities, such as velocity and vorticity, can be extracted from $\aPot(z,t)$ if desired, but they are not needed to determine the hydrodynamic forces on swimmers. Hence, in this section, we focus on the pressure field as the main physical diagnostic for gaining insights into fluid-mediated interactions between two swimmers.

\begin{figure}[t]
	\begin{subfigure}[h]{.3\linewidth}
	\begin{center}
	\begin{tikzpicture}
	\node[inner sep=0pt, draw = black!30!white,ultra thick] (0,0)
    {\includegraphics[width=\linewidth]{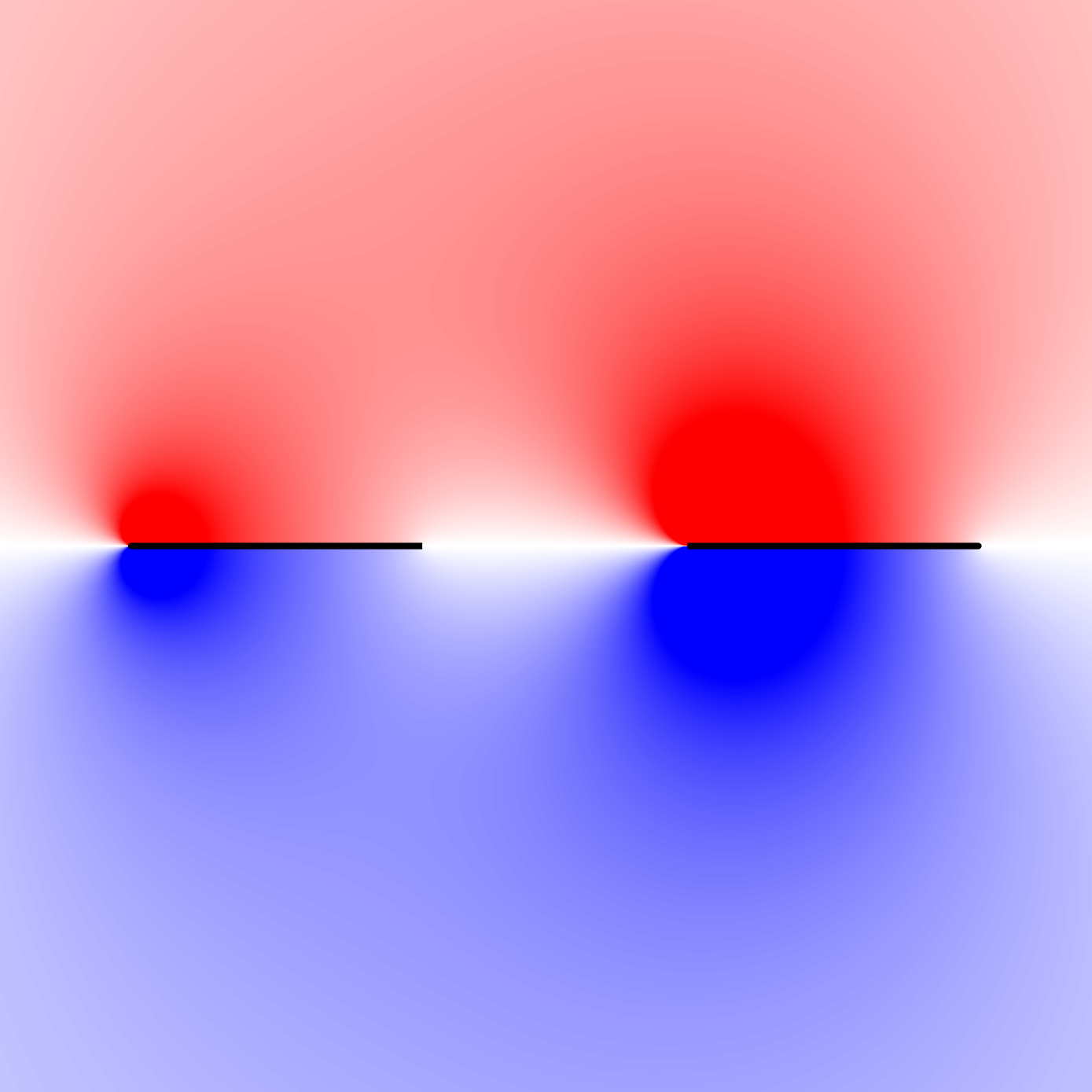}};
    \draw[-Latex,gray!70!black] (-1.6,1.3)--(-.5,1.3) node[black, midway, above] {\small $U$};
     \draw[-Latex,gray!70!black] (-1.6,1.1)--(-.5,1.1);
    \end{tikzpicture}
	\end{center}
	\caption{Inline}
	\label{fig:HeaveInPhase1}
	\end{subfigure}
	\hfill
	\begin{subfigure}[h]{.3\linewidth}
	\begin{center}
	\begin{tikzpicture}
	\node[inner sep=0pt, draw = black!30!white,ultra thick] (0,0)
    {\includegraphics[width=\linewidth]{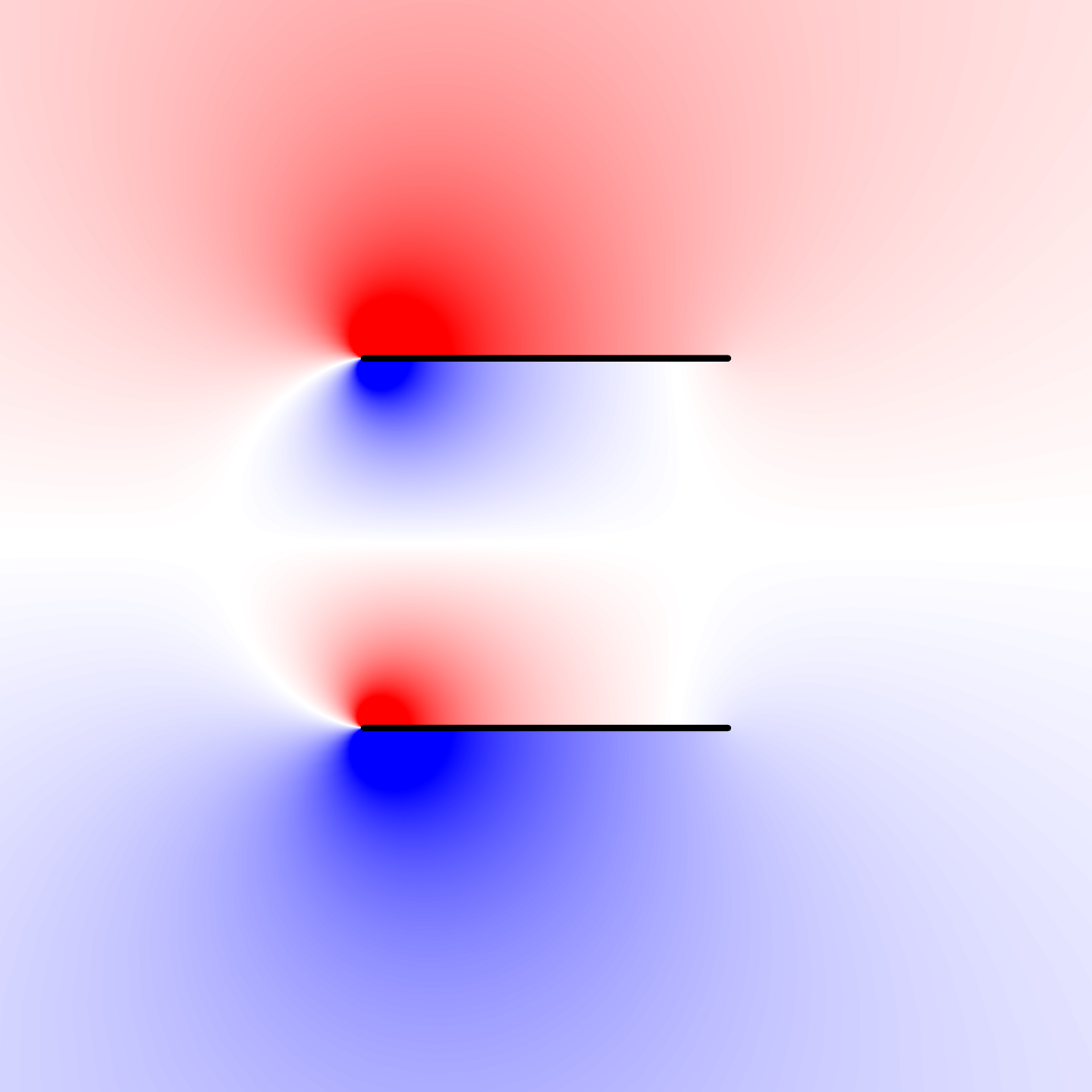}};
    \end{tikzpicture}
	\end{center}
	\caption{Side-by-side}
	\label{fig:HeaveInPhase2}
	\end{subfigure}
	\hfill
	\begin{subfigure}[h]{.3\linewidth}
	\begin{tikzpicture}
	\node[inner sep=0pt, draw = black!30!white,ultra thick] (0,0)
    {\includegraphics[width=\linewidth]{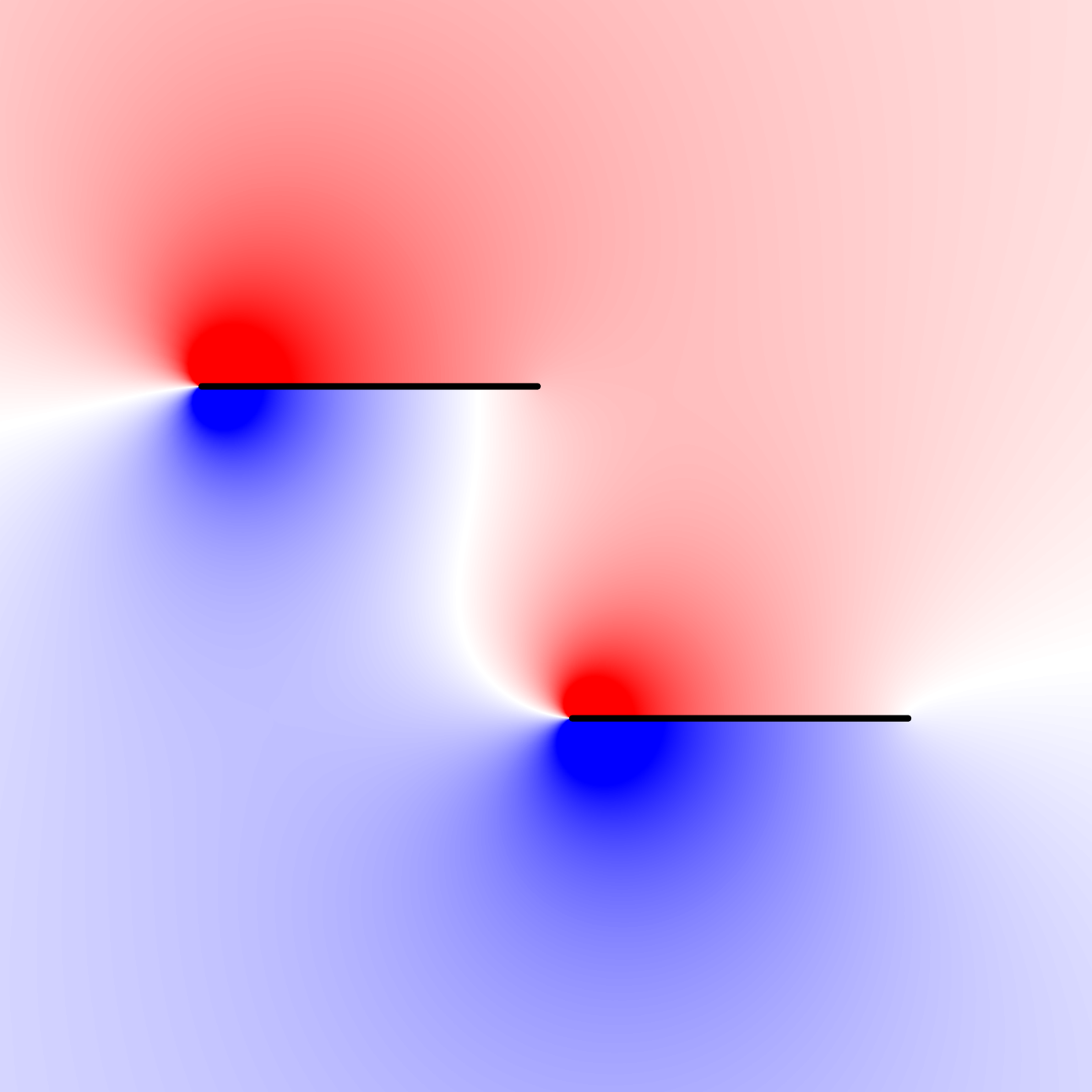}};
    \end{tikzpicture}
	\caption{Diagonal}
	\label{fig:HeaveInPhase3}
	\end{subfigure}
	\caption{Instantaneous pressure fields for three different foil configurations. The foils are heaving in phase at a reduced frequency of $\sigma = 1$.  At the instant shown, both foils are in the upward phase of their motion, experiencing high pressure (red) along the top surface and low pressure (blue) along the bottom.}
	\label{Fig:pressureIn}
	\end{figure}
	
	\begin{figure}[t]
	\begin{subfigure}[h]{.3\linewidth}
	\begin{center}
	\begin{tikzpicture}
	\node[inner sep=0pt, draw = black!30!white,ultra thick] (0,0)
    {\includegraphics[width=\linewidth]{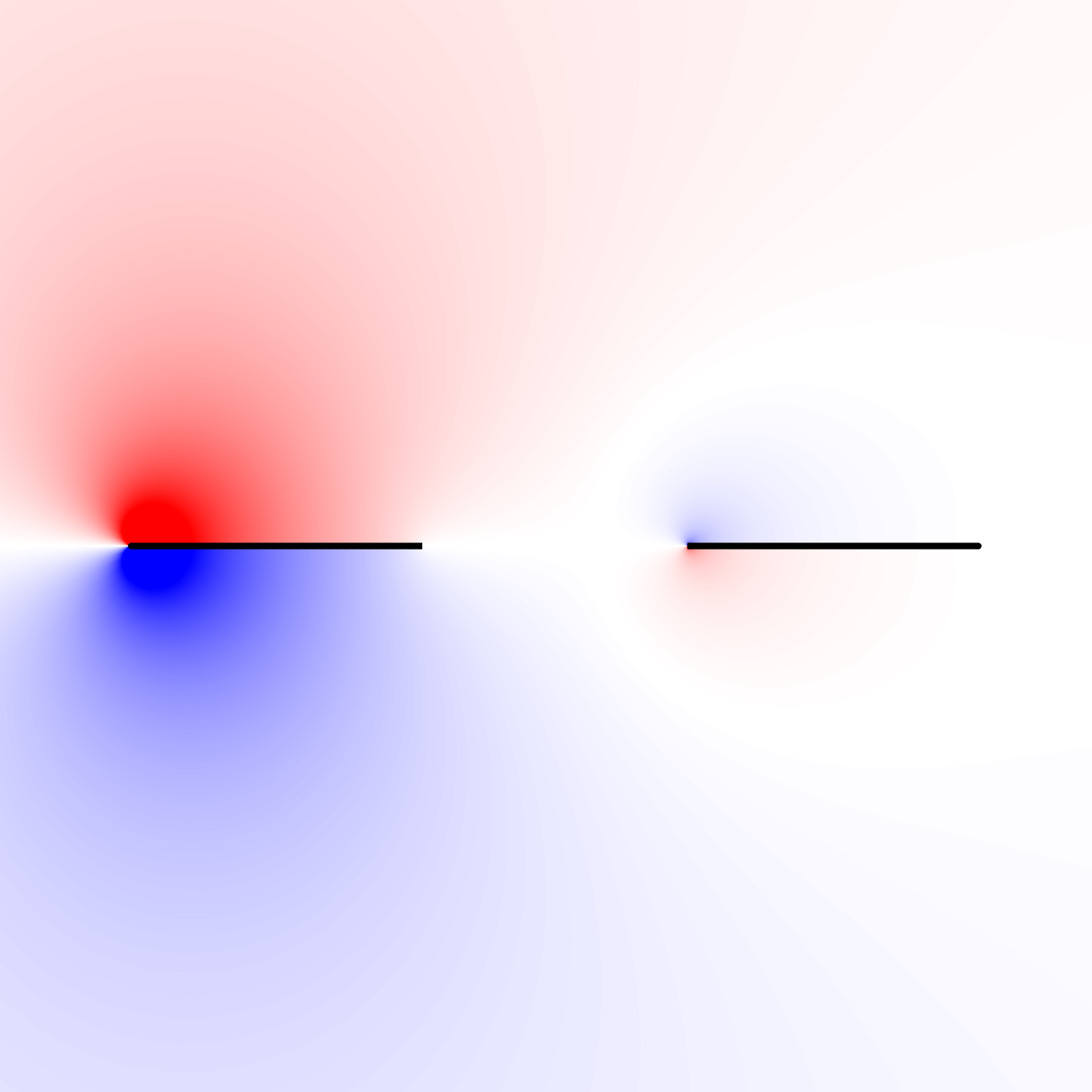}};
    \draw[-Latex,gray!70!black] (-1.6,1.3)--(-.5,1.3) node[black, midway, above] {\small $U$};
     \draw[-Latex,gray!70!black] (-1.6,1.1)--(-.5,1.1);
    \end{tikzpicture}
	\end{center}
	\caption{Inline}
	\label{fig:HeaveOutPhase1}
	\end{subfigure}
	\hfill
	\begin{subfigure}[h]{.3\linewidth}
	\begin{center}
	\begin{tikzpicture}
	\node[inner sep=0pt, draw = black!30!white,ultra thick] (0,0)
    {\includegraphics[width=\linewidth]{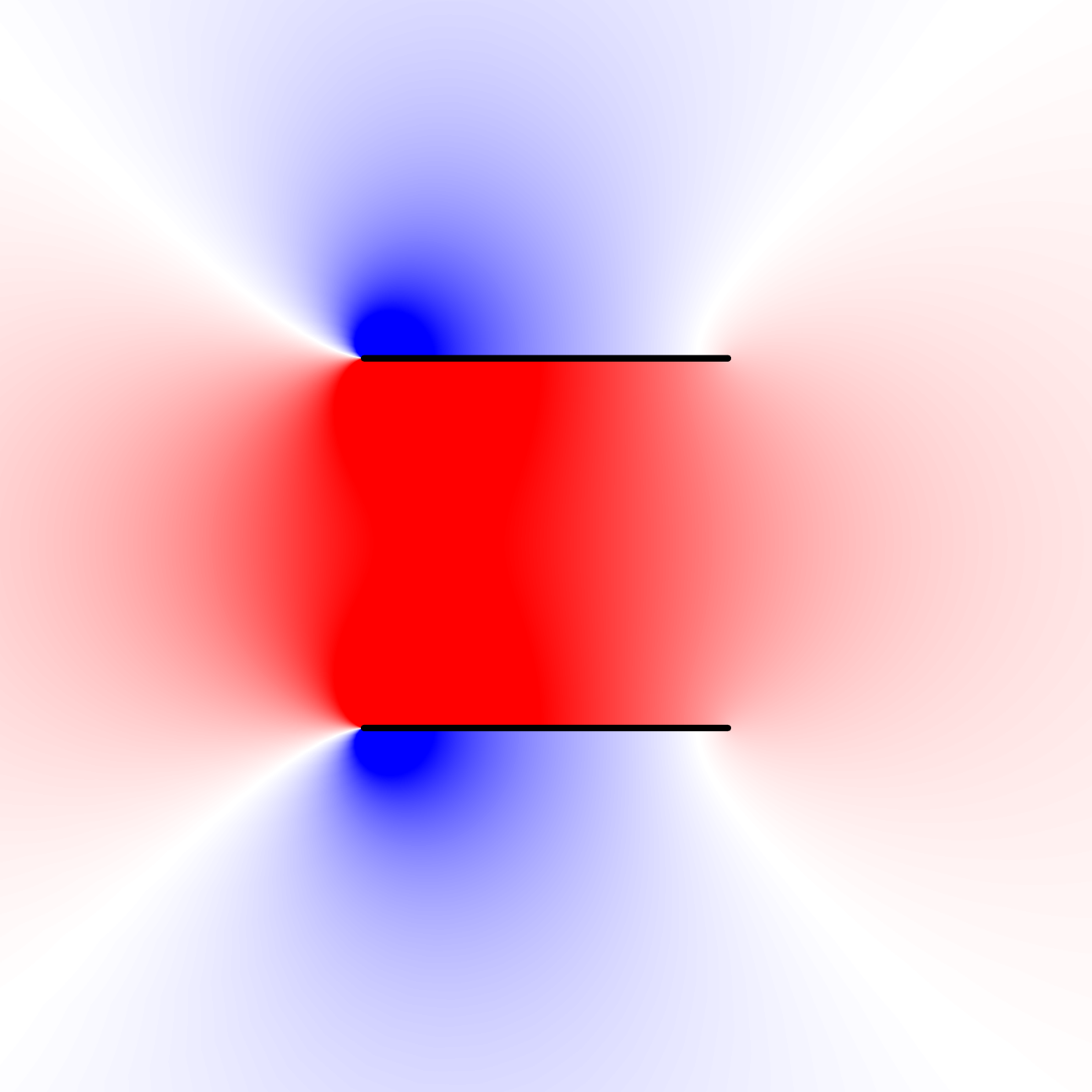}};
    \end{tikzpicture}
	\end{center}
	\caption{Side-by-side}
	\label{fig:HeaveOutPhase2}
	\end{subfigure}
	\hfill
	\begin{subfigure}[h]{.3\linewidth}
	\begin{center}
	\begin{tikzpicture}
	\node[inner sep=0pt, draw = black!30!white,ultra thick] (0,0)
    {\includegraphics[width=\linewidth]{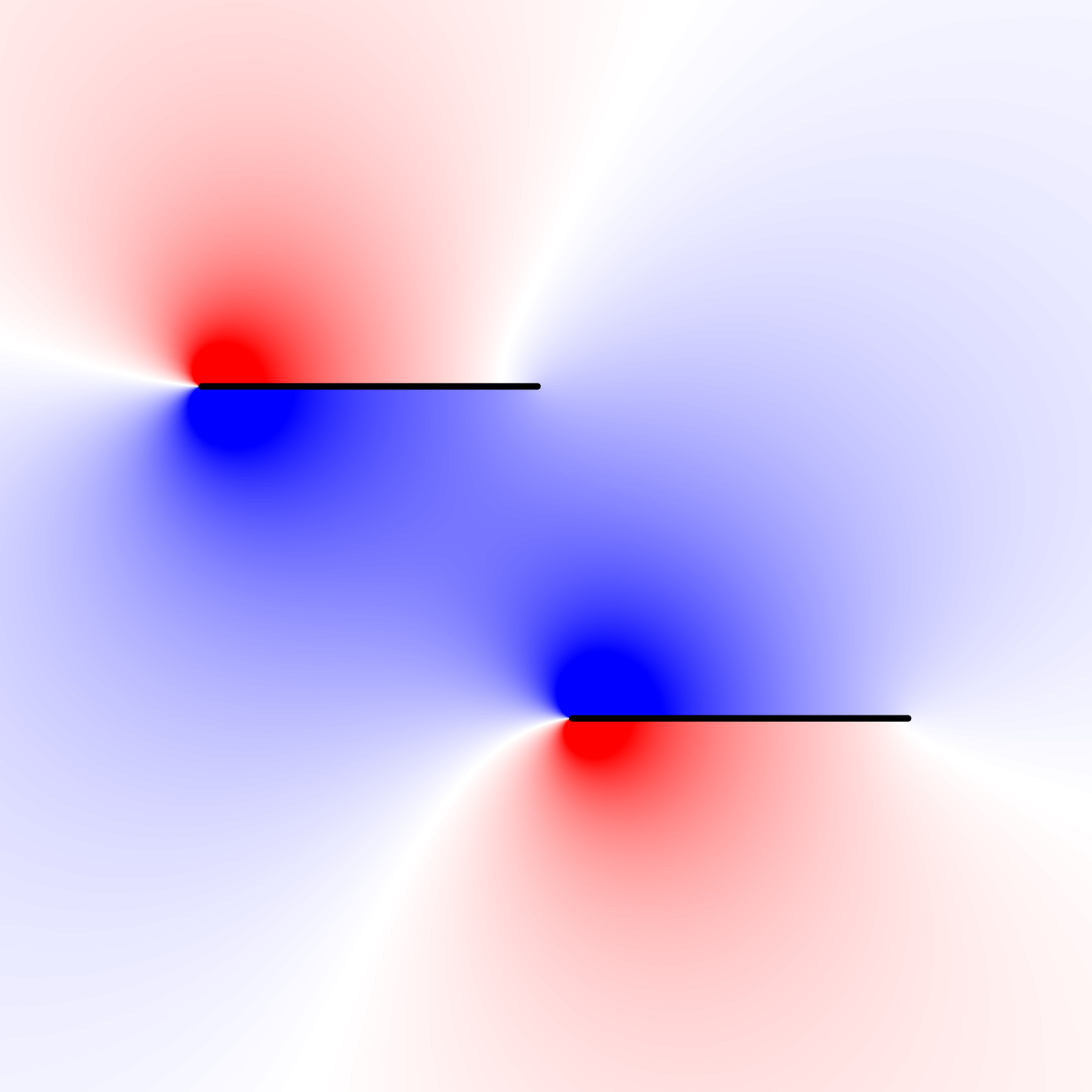}};
    \end{tikzpicture}	\end{center}
	\caption{Diagonal}
	\label{fig:HeaveOutPhase3}
	\end{subfigure}
	\caption{Instantaneous pressure fields for three different foil configurations. The foils are heaving in anti-phase at a reduced frequency of $\sigma = 1$.
	}
	\label{Fig:pressureOut}
\end{figure}

Figures \ref{fig:HeaveInPhase1}--\ref{fig:HeaveInPhase3} show instantaneous pressure fields surrounding two foils that heave in phase. The figures show a few different configurations: (A) inline, (B) side-by-side, and (C) a diagonal configuration. Figures \ref{fig:HeaveOutPhase1}--\ref{fig:HeaveOutPhase3} show similar plots but for foils that heave in anti-phase. Distinct features in the surrounding pressure fields are apparent. 

First consider the inline and in-phase case shown in Fig.~\ref{fig:HeaveInPhase1}. As indicated in the figure, the free-stream flow is left to right, so we refer to the foil on the left as the leader and the right as the follower.
At the instant shown, both foils are in the upward phase of their motion and thus experience high pressure (red) along the top surface and low pressure (blue) along the bottom. The interaction with the leader's wake evidently increases the pressure differential across the follower, thereby enhancing its own lift and thrust forces. Whether the follower's hydrodynamic forces are strengthened or weakened by the leader's wake depends on many details, such as the separation distance and relative flapping phase, as will be explored more thoroughly in what follows.

Figures~\ref{fig:HeaveInPhase2} and \ref{fig:HeaveInPhase3} reveal features of the surrounding pressure fields for the side-by-side and diagonal arrangements. We note that in-phase heaving foils in the side-by-side arrangement (Fig.~\ref{fig:HeaveInPhase2}) effectively move the fluid in between them as a cohesive unit. This coherent motion reduces the pressure differential due to less relative motion between foil and fluid which, as will be shown in \S\ref{SSec:Staggered}, ultimately reduces the thrust that is generated.

The case of anti-phase heaving, shown in Fig.~\ref{fig:HeaveOutPhase1}--\ref{fig:HeaveOutPhase3}, is perhaps more subtle. For the inline configuration (Fig.~\ref{fig:HeaveOutPhase1}), the leader is heaving upwards and the follower downwards at the instant shown.
Hence, the leader (follower) experiences high pressure along its top (bottom) surface, with the pressure differential across the follower being significantly weakened by the interaction. 
The side-by-side case (Fig.~\ref{fig:HeaveOutPhase2}) is of particular interest as it is equivalent to the ground-effect problem of a plate swimming above an impenetrable wall \cite{Baddoo2020Rapids}. At the instant shown, both foils are heaving towards the center, thus squeezing the fluid in between and creating a region of high pressure. As we will see in \S\ref{SSec:Staggered}, this squeezing action significantly enhances the thrust that is generated. The staggered case (Fig.~\ref{fig:HeaveOutPhase3}) shows a combination of the features from the inline and side-by-side arrangements. At the instant shown, the staggered foils are heaving away from one another.

We remark that the singular term in \eqref{Eq:singularityExt} can produce large pressure values in the vicinity of each foil's leading edge, as are visible in a few of these figures. How visible the effect is depends on details like which phase is shown. These singularities also contribute to the net thrust through the so-called leading-edge suction, as detailed in Appendix \ref{Ap:forces}. 

\subsection{Hydrodynamic thrust for the inline configuration}\label{SSec:Inline}

\begin{figure}[t]
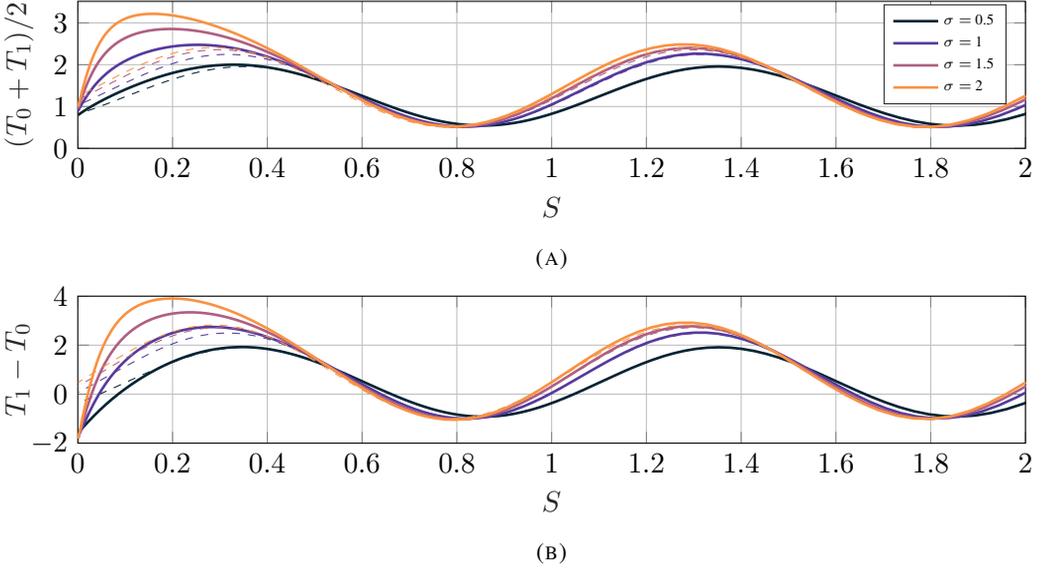

	\setlength{\fheight}{2.1cm}
	\setlength{\fwidth}{\linewidth}
	\begin{subfigure}{\linewidth}
	{\centering
	 \input{images/inline-thrust-sum.tex}
	 \caption{}
	 \label{fig:ThrustA}
	}
	\end{subfigure}

	\begin{subfigure}{\linewidth}
	{\centering
	 \input{images/inline-thrust-diff.tex}
	 \caption{}
	 \label{fig:ThrustB}
	}
	\end{subfigure}
	\caption{The normalised average (a) and difference (b) of thrusts for a pair of inline foils that are heaving in-phase. The full solutions are plotted in thick lines whereas the asymptotic solution of \cite{ramananarivo2016flow} is plotted in dashed lines.}
	\label{fig:Thrust}
\end{figure}

We now further quantify the inline, in-phase configuration 
by showing in Fig.~\ref{fig:Thrust} the thrust produced by the foils, time-averaged over the flapping period. The thrust is calculated using the formulas derived in Appendix \ref{Ap:forces}. We introduce the so-called schooling number to quantify the horizontal spacing $\ell$ between the foils \cite{ramananarivo2016flow, OzaPRX2019}: $S = \ell \freq / U_{\infty}=\ell\sigma/\pi c$, where $\ell$ is the (dimensional) distance between the trailing edge of the leader and the leading edge of the follower. Thus, the schooling number represents the spacing between the foils normalized by the wake wavelength.

Figure \ref{fig:ThrustA} shows the average thrust $(T_1+T_0)/2$ produced by the leader ($T_0$) and follower ($T_1$) as a function of the schooling number $S$. Here, $(T_1+T_0)/2$ is normalized by the thrust produced by an isolated foil.
The figure reveals that $(T_1 + T_0)/2 > 1$ for most schooling numbers, meaning that the two foils generate greater thrust when flapping together than they would in isolation. There is also a smaller range (roughly $0.65<S<1.0$, the precise values depending on $\sigma$) for which the interaction leads to reduced thrust. For $S$ outside of that range,  the thrust enhancement can be substantial. In the most extreme case, $S=0.15$ and $\sigma=2$, the interacting foils generate more than triple the thrust they would if they were performing the same motions in isolation. The greatest thrust enhancement occurs for closely-spaced foils, roughly $S=0.1-0.3$, and for relatively large $\sigma$. The peak in thrust enhancement is seen to reoccur periodically, roughly $S=1.3, 2.3, 3.3,\dots$, due to the periodicity of the wake.

Figure \ref{fig:ThrustB} shows the thrust difference, $T_1 - T_0$, normalized by the thrust produced by a single foil performing the same motions. For most schooling numbers, $T_1-T_0 > 0$, indicating that the follower produces greater thrust than the leader. In these cases, the follower reaps a net benefit from interacting with the leader's wake. There is also a small range (roughly $S=0.65-1.0$) for which $T_1 - T_0 < 0$, and the leader's wake
effectively exerts a drag on the follower. In both Figs.~\ref{fig:ThrustA} and \ref{fig:ThrustB}, we also plot the asymptotic solution of \cite{ramananarivo2016flow}, which is valid for large $S$. It is evident that, for all values of $\sigma$ considered, the asymptotic solution agrees well with our computation for large $S$, thus validating our results.

Importantly, Fig.~\ref{fig:ThrustB} offers some insight into the dynamics of self-propelled swimmers. For example, if the follower were to begin at a spacing of $S=1.5$ behind the leader, it would generate greater thrust and thus begin to catch up. The follower would continue to close the distance until the two thrusts become equal, $T_1=T_0$, which is seen to occur at approximately $S=0.95-1.05$ with only weak dependence on $\sigma$. Hence, our computation predicts an equilibrium spacing of $S\approx 1$ between the free swimmers. 

\begin{figure}[t]
	\centering
	\begin{subfigure}[h]{0.4\textwidth}
    \setlength{\fwidth}{4.8cm}
    \setlength{\yshift}{0cm}%
 \begin{tikzpicture}[%
trim axis left, trim axis right
]
 \begin{axis}[
  set layers,
ylabel shift = \yshift,
xlabel shift = \xshift,
 width=\fwidth,
 at={(0,0)},
scale only axis,
enlargelimits=false,
axis on top,
xlabel={$S$}, 
ytick={0,2,4,6,8},
yticklabels={0,2,4,6,8},
tick label style={font=\tiny},
ylabel={$U$},
legend style={fill opacity=0.9, nodes={scale=0.7, transform shape}},
legend pos = north east,
]
    \addlegendimage{red,mark=triangle*,color=black,fill=black!40!white,only marks};
        \addlegendimage{red,mark=*,color=black,fill=black!40!white,only marks};
            \addlegendimage{red,mark=square*,color=black,fill=black!40!white,only marks};
    \addlegendentry{2 Hz};
    \addlegendentry{3 Hz};
    \addlegendentry{4 Hz};
\addplot graphics[xmin=0,xmax=3.5,ymin=0,ymax=8] {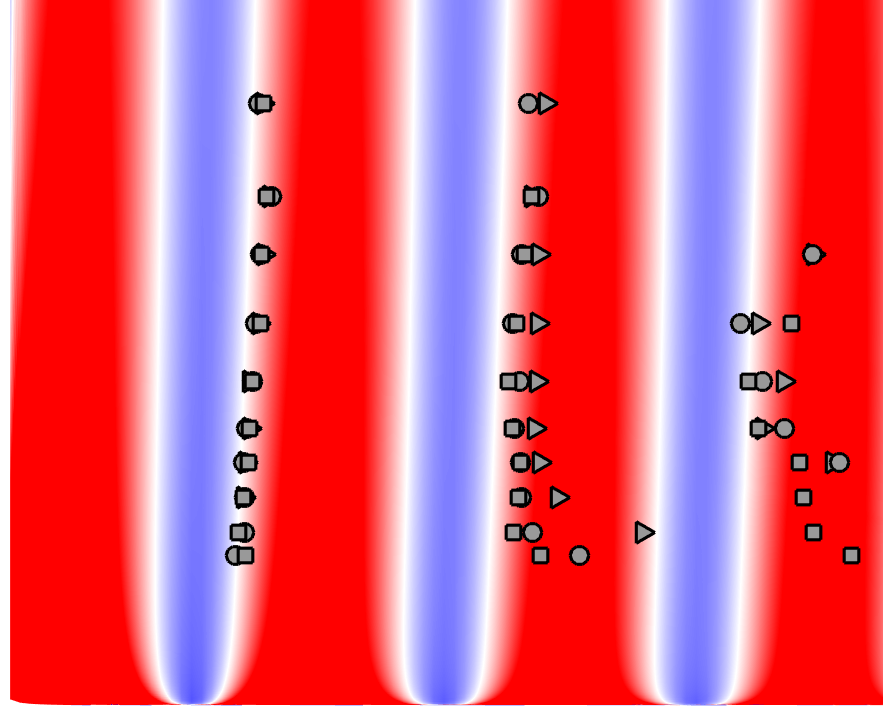};
    \end{axis}
    \end{tikzpicture}%
\caption{}
\label{fig:exptA}
	 \end{subfigure}
	 \hfill
	 \begin{subfigure}[h]{0.4\textwidth}
    \centering
    \setlength{\fwidth}{4.8cm}
    \setlength{\yshift}{-.4cm}
 \begin{tikzpicture}[%
trim axis left, trim axis right
]
 \begin{axis}[
  set layers,
ylabel shift = \yshift,
xlabel shift = \xshift,
 width=\fwidth,
 at={(0,0)},
scale only axis,
enlargelimits=false,
axis on top,
xlabel={$S$}, 
ytick={0,1.571
,3.141,4.7124,6.283},
yticklabels={0,$\pi/2$,$\pi$,$3\pi/2$,$2\pi$},
tick label style={font=\tiny},
ylabel={$\phasediff$},
legend style={fill opacity=0.9, nodes={scale=0.7, transform shape},cells={align=right}},
legend pos = north east,
]
    \addlegendimage{red,mark=triangle*,color=black,fill=black!40!white,only marks};
    \addlegendentry{2 Hz};
\addplot graphics[xmin=0,xmax=3.5,ymin=0,ymax=6.2832] {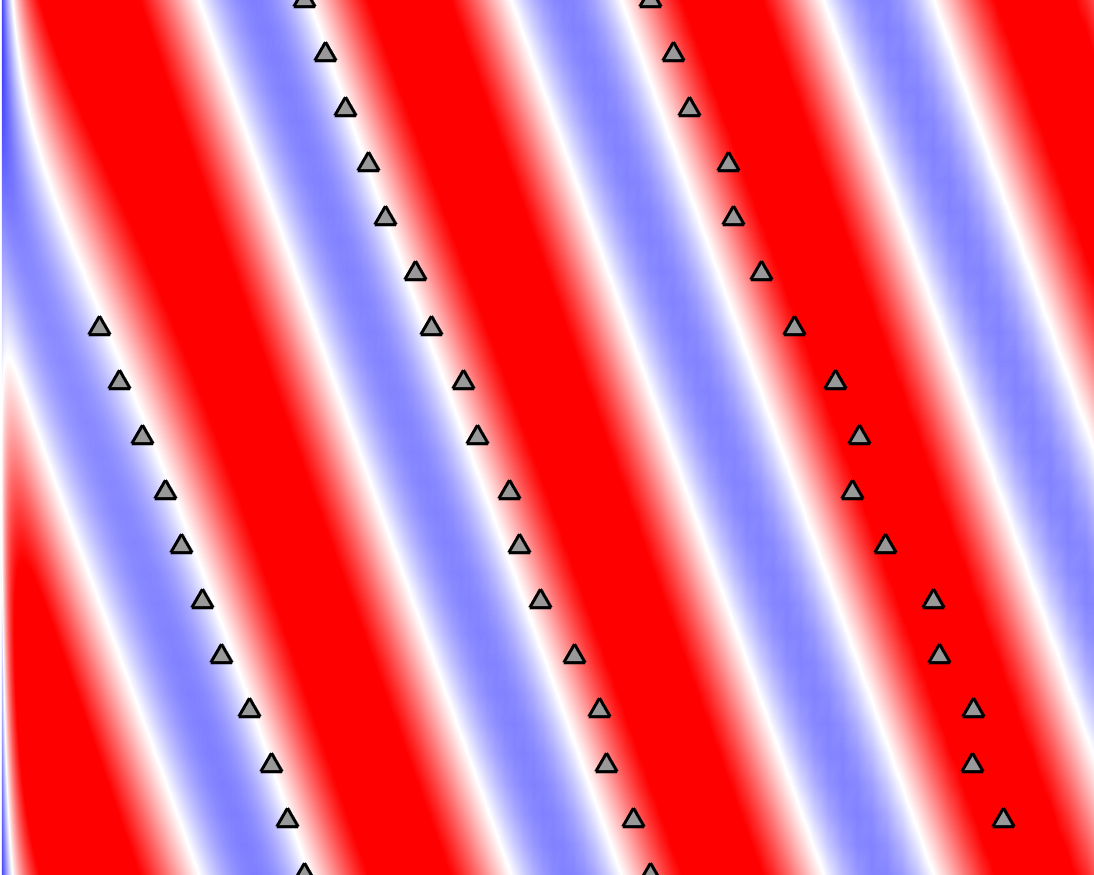};
    \end{axis}
    \end{tikzpicture}%
\caption{}
\label{fig:exptB}
	 \end{subfigure}
	\begin{subfigure}[h]{.1\textwidth}
	\begin{center}
	\vspace{-1.4cm}
	
\myCBarTwo{1}{$T_1-T_0$}
\end{center}
	\end{subfigure}
	 \caption{(A) The thrust difference $T_1-T_0$ for a pair of inline foils that are heaving in-phase. Horizontal cuts of the plot (fixed $U = 2\pi/\sigma$) correspond to the curves shown in Fig.~\ref{fig:ThrustB}. The points indicate experimental data on pairs of freely swimming flapping foils in a water tank~\cite{ramananarivo2016flow}. Different symbols correspond to the flapping frequencies indicated in the legend, and the flapping amplitude ranges from $0.05c$ to $0.4c$. (B) The thrust difference $T_1-T_0$ for a pair of inline foils heaving with the indicated phase difference $\phasediff$. The points indicate experimental data~\cite{Newbolt2019} for foils flapping with frequency 2 Hz and amplitude $0.4c$. The value of $\sigma$ corresponds to the midpoint of the range of horizontal velocities $U_{\infty}=18-28$ cm/s reported in experiments.}
	 \label{fig:expt}
\end{figure}

The results in Fig.~\ref{fig:ThrustB} are extended in Fig.~\ref{fig:exptA} to a larger range of schooling numbers $S$ and velocities $U = 2\pi/\sigma$. The white regions of the plots indicate parameters for which the leader and follower produce equal thrust, $T_1=T_0$, while blue (red) indicates the leader (follower) producing greater thrust.
By the foregoing arguments, our theory predicts stable equilibrium spacings to occur at $S\approx n$ for $n\in\mathbb{N}$. This prediction exhibits excellent agreement with experimental data (points in Fig.~\ref{fig:exptA}) on the equilibrium configurations adopted by freely swimming heaving foils in an in-line configuration~\cite{ramananarivo2016flow}. Specifically, the data points for a range of flapping frequencies and amplitudes lie near the stable zero-contours of the thrust difference, $T_1-T_0$, that were calculated by the theory. The same is true for Fig.~\ref{fig:exptB}, wherein we explore the influence of the flapping phase difference $\phasediff$ and compare with experimental data~\cite{Newbolt2019}. The theory and experiments agree well throughout the parameter space, especially for smaller schooling numbers $S < 2.5$ where the agreement is quantitative. For larger $S$, the theory still predicts the qualitative trend of the data, but the agreement is less accurate, likely due to the increased importance of viscous effects far downstream of the leader's wake. Indeed, experimental measurements~\cite{ramananarivo2016flow, Newbolt2019} have shown  the strength of the wake to decay exponentially over a timescale of a few seconds in water, and these effects are not accounted for in the present inviscid theory.

\subsection{Thrust on staggered configurations}\label{SSec:Staggered}

\begin{figure}
	\begin{subfigure}[h]{.28\textwidth} \begin{center} 
	\includegraphics[width = \linewidth]{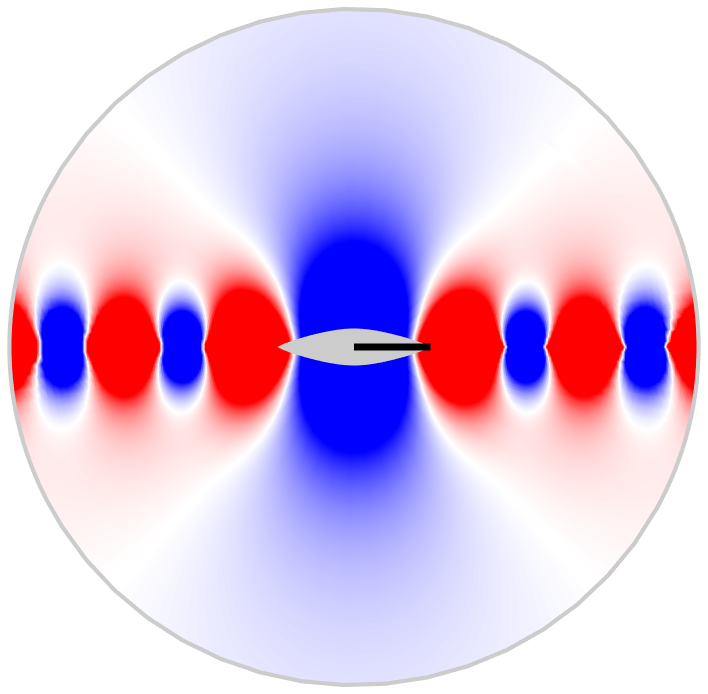} \end{center}
	\caption{$\phasediff = 0$} 
	\label{Fig:sumThrustHeave0}
	\end{subfigure} \hfill
	\begin{subfigure}[h]{.28\textwidth} \begin{center}
	\includegraphics[width = \linewidth]{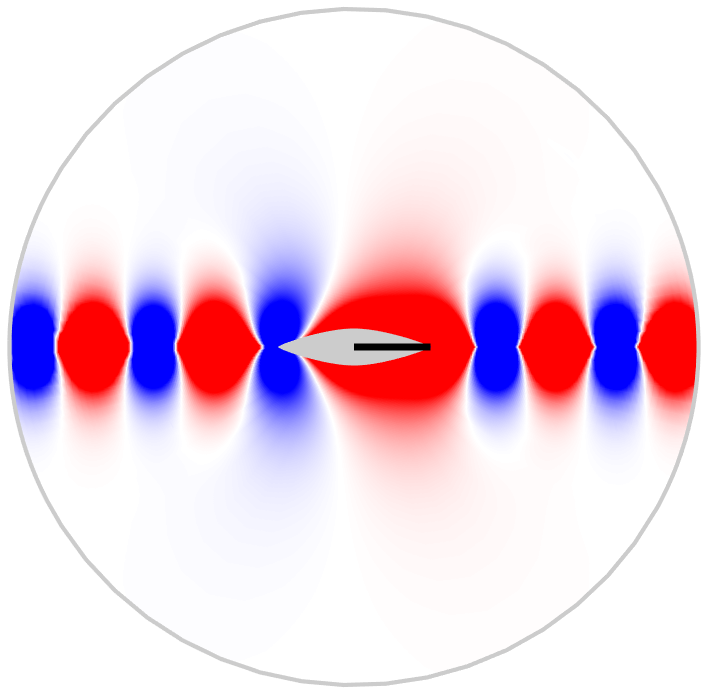} \end{center}
	\caption{$\phasediff = \pi/2$} 
	\label{Fig:sumThrustHeave90}
	\end{subfigure} \hfill
	\begin{subfigure}[h]{.28\textwidth} \begin{center}
	\includegraphics[width = \linewidth]{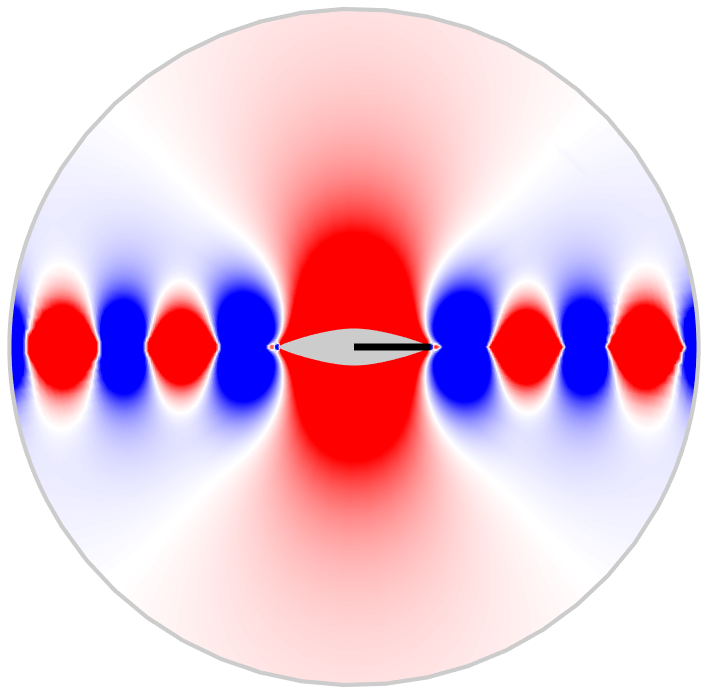} \end{center}
	\caption{$\phasediff = \pi$} 
	\label{Fig:sumThrustHeave180}
	\end{subfigure}\hfill
	\begin{subfigure}[h]{.1\textwidth}
	\begin{center}
\myCBarSum
\vspace{.7cm}
\end{center}
	\end{subfigure}
	\caption{The dependence of the average thrust $(T_1+T_0)/2$ on the relative positions of a pair of heaving foils at the indicated phase difference $\phasediff$ with $\sigma = 2$. One plate is indicated by the black line and has thrust $T_0$. Points in the plane correspond to the leading edge of the other plate, which has thrust $T_1$. 
	}	
	\label{Fig:sumThrustHeave}
	\begin{subfigure}[h]{.28\textwidth} \begin{center}
	\includegraphics[width = \linewidth]{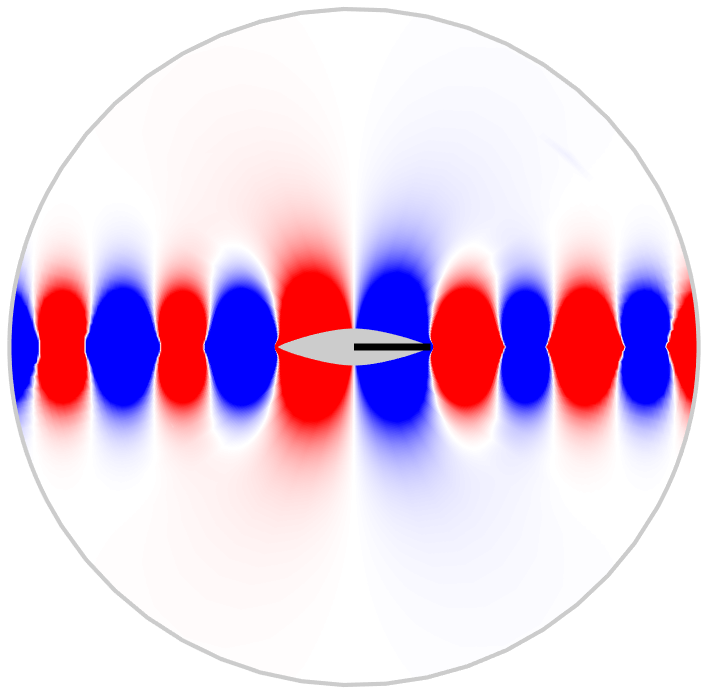} \end{center}
	\caption{$\phasediff = 0$}
	\label{Fig:diffThrustHeave0}
	\end{subfigure} \hfill
	\begin{subfigure}[h]{.28\textwidth} \begin{center}
	\includegraphics[width = \linewidth]{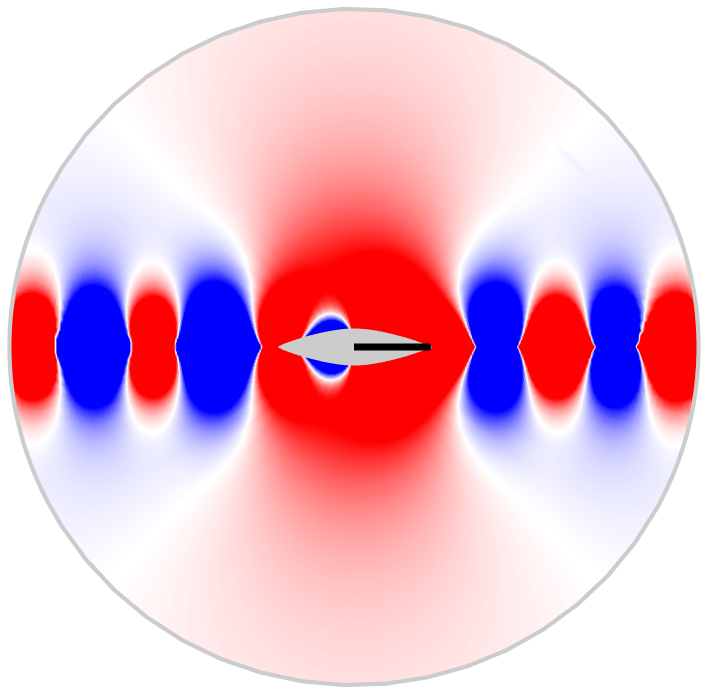} \end{center}
	\caption{$\phasediff = \pi/2$} 
	\label{Fig:diffThrustHeave90}
	\end{subfigure} \hfill
	\begin{subfigure}[h]{.28\textwidth} \begin{center}
	\includegraphics[width = \linewidth]{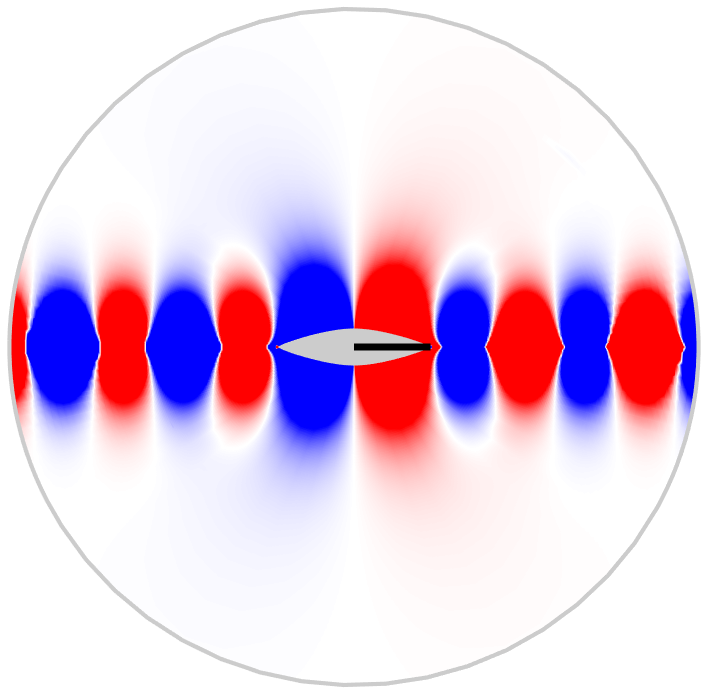} \end{center}
	\caption{$\phasediff= \pi$} 
	\label{Fig:diffThrustHeave180}
	\end{subfigure}
	\hfill
		\begin{subfigure}[h]{.1\textwidth}
	\begin{center}
\myCBar{0.1}{$T_1-T_0$}
\vspace{.7cm}
\end{center}
	\end{subfigure}
	\caption{The dependence of the thrust difference $T_1-T_0$ on the relative positions of a pair of heaving foils at the indicated phase difference $\phasediff$ with $\sigma = 2$. 
	}
	\label{Fig:diffThrustHeave}
\end{figure}

Unlike previous investigations, the present theory does not constrain the swimmers to the inline configuration. Thus, Figures~\ref{Fig:sumThrustHeave}--\ref{Fig:diffThrustHeave} expand the discussion of the preceding section to interacting swimmers having arbitrary lateral positions. In these figures, the position of one foil is fixed at the origin, and points in the plane correspond to the leading edge of the second foil. The average (Fig.~\ref{Fig:sumThrustHeave}) and difference (Fig.~\ref{Fig:diffThrustHeave}) of the thrusts are shown in color for three different phase differences, $\phasediff = 0, \pi/2, \pi$. 
Hence, each point in the plane of Figs.~\ref{Fig:sumThrustHeave}--\ref{Fig:diffThrustHeave} corresponds to the computation of a different swimmer-pair interaction, with different points representing  different configurations. Altogether, roughly 21,000 swimmer-pair interactions are represented in each of these figures. The narrow gray region surrounding the origin is excluded from consideration, since, here, the close proximity of the foils degrades the numerical accuracy of the solutions.

Figure~\ref{Fig:sumThrustHeave} shows the average thrust, $(T_1+T_0)/2$, generated by the pair of heaving foils. Viewing Fig.~\ref{Fig:sumThrustHeave0} along the positive $x$-axis recovers the inline results from Fig.~\ref{fig:ThrustA}, and viewing the figure off axis generalizes those results. In particular, picking a point on the $y$-axis of Fig.~\ref{Fig:sumThrustHeave0} corresponds to a side-by-side configuration. For this arrangement, Fig.~\ref{Fig:sumThrustHeave0} shows that in-phase heaving reduces the net thrust, while Fig.~\ref{Fig:sumThrustHeave180} shows that anti-phase heaving increases the net thrust. Both results are consistent with the physical intuition gained from the pressure fields in Figs.~
\ref{fig:HeaveInPhase2}--\ref{fig:HeaveOutPhase2}. That is, in-phase heaving causes the fluid between the plates to move cohesively thereby reducing net thrust, whereas the squeezing effect of anti-phase heaving creates a high-pressure region in between which enhances net thrust.

Figure \ref{Fig:diffThrustHeave} shows the difference in thrusts, $T_1-T_0$, where $T_0$ corresponds to the foil at the origin and $T_1$ to the second foil located arbitrarily. Thus, red (blue) indicates that the second foil generates greater (less) thrust than the one located at the origin. Of particular importance are the zero-contours of $T_1-T_0$, along which two free-swimmers generate equal thrust and would therefore maintain their relative positions. These contours  generalize the predictions for inline equilibria (\S\ref{SSec:Inline}) to arbitrary arrangements not considered in previous models~\cite{ramananarivo2016flow, Newbolt2019}.

Interestingly, Figs.~\ref{Fig:diffThrustHeave0}--\ref{Fig:diffThrustHeave180} show several bounded contours of $T_1-T_0=0$, and a handful of unbounded contours. The unbounded contours, in particular, imply that equilibrium configurations with arbitrarily large lateral separation are possible. That is, the swimmers do not necessarily need to be tightly clustered in order to achieve an equilibrium configuration.
More specifically, for both in-phase (Fig.~\ref{Fig:diffThrustHeave0}) and anti-phase (Fig.~\ref{Fig:diffThrustHeave180}) heaving, an unbounded contour extends vertically along the $y$-axis, indicating that the side-by-side configuration is an equilibrium. Moreover, examining the surrounding sign of $T_1-T_0$ (i.e.~red or blue on the figure) indicates this configuration to be stable to streamwise perturbations if the foils heave out of phase (Fig.~\ref{Fig:diffThrustHeave0}), and unstable if they heave in phase (Fig.~\ref{Fig:diffThrustHeave180}).

Perhaps the most surprising results come from the case of a 90-degree phase offset in kinematics, $\phasediff=\pi/2$, as shown in Fig.~\ref{Fig:diffThrustHeave90}. Here, the side-by-side arrangement is not an equilibrium, and instead the unbounded contours of $T_1-T_0=0$ extend outwards at an angle. Hence, equilibrium is achieved by a staggered arrangement. By evaluating the sign of $T_1-T_0$, we find the rightward-extending branches to be unstable and the leftward ones to be stable. The physical consequence is that the kinematic phase $\phasediff$ pre-selects the swimmer that will ultimately emerge as the positional leader. That is, the swimmer that leads in phase will ultimately lead the configuration, provided that the initial arrangement lies within the basin of attraction of the stable contour. Otherwise, the horizontal separation between the swimmers will grow indefinitely.

\begin{figure}
	\begin{subfigure}[h]{.28\textwidth}\begin{center}
	\includegraphics[width = \linewidth]{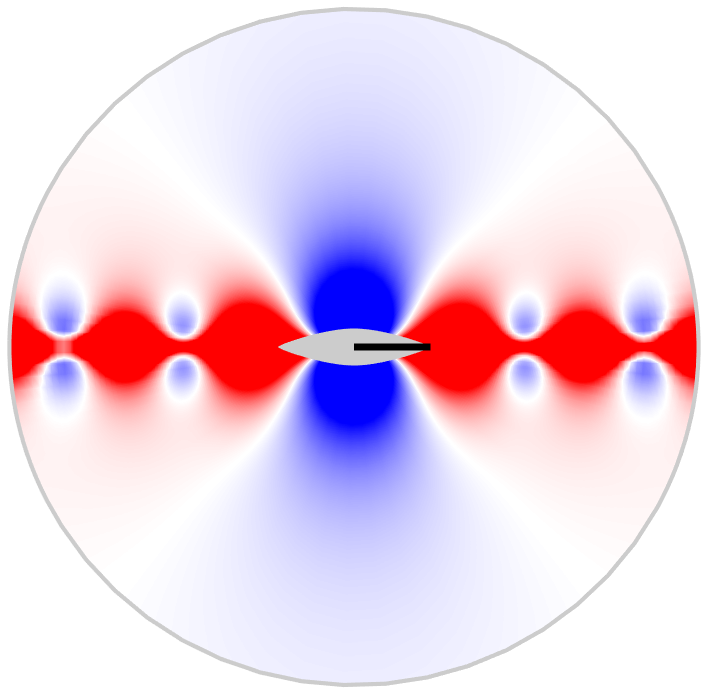}\end{center}
	\caption{$\phasediff = 0$} 
	\label{Fig:sumThrustPitch0}
	\end{subfigure} \hfill
	\begin{subfigure}[h]{.28\textwidth}\begin{center}
	\includegraphics[width = \linewidth]{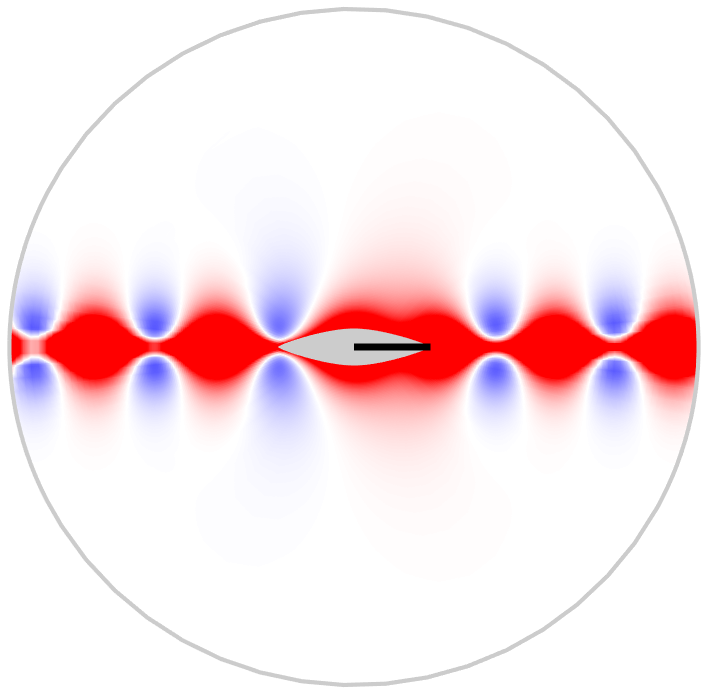}\end{center}
	\caption{$\phasediff = \pi/2$}
	\label{Fig:sumThrustPitch90}
	\end{subfigure}\hfill
	\begin{subfigure}[h]{.28\textwidth}\begin{center}
	\includegraphics[width = \linewidth]{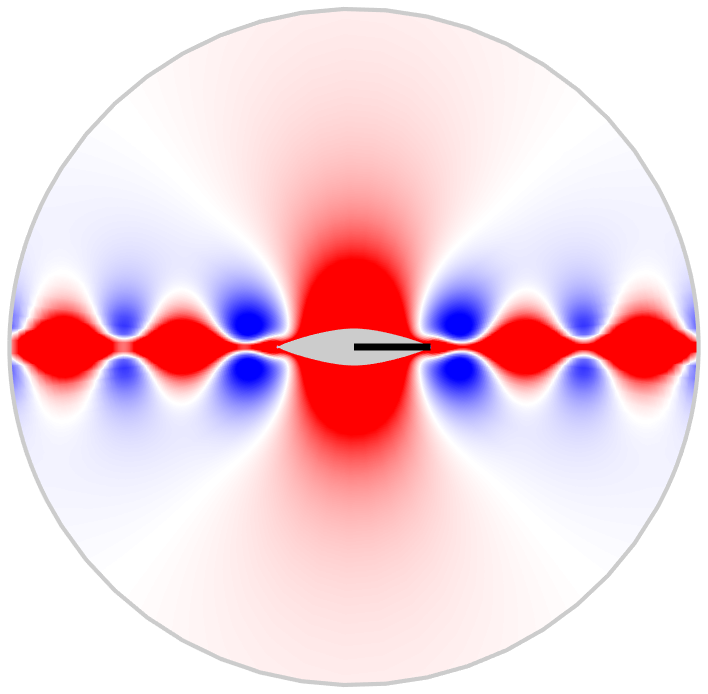}\end{center}
	\caption{$\phasediff = \pi$}
	\label{Fig:sumThrustPitch180}
	\end{subfigure} \hfill
	\begin{subfigure}[h]{.1\textwidth}
	\begin{center}
    \myCBarSum
    \vspace{.7cm}
    \end{center}
	\end{subfigure}
	\caption{The dependence of the average thrust $(T_1+T_0)/2$ on the relative positions of a pair of pitching foils at the indicated phase difference $\phasediff$ with $\sigma = 2$.
	}	
	\label{Fig:sumThrustPitch}
	\begin{subfigure}[h]{.28\textwidth}\begin{center}
	\includegraphics[width = \linewidth]{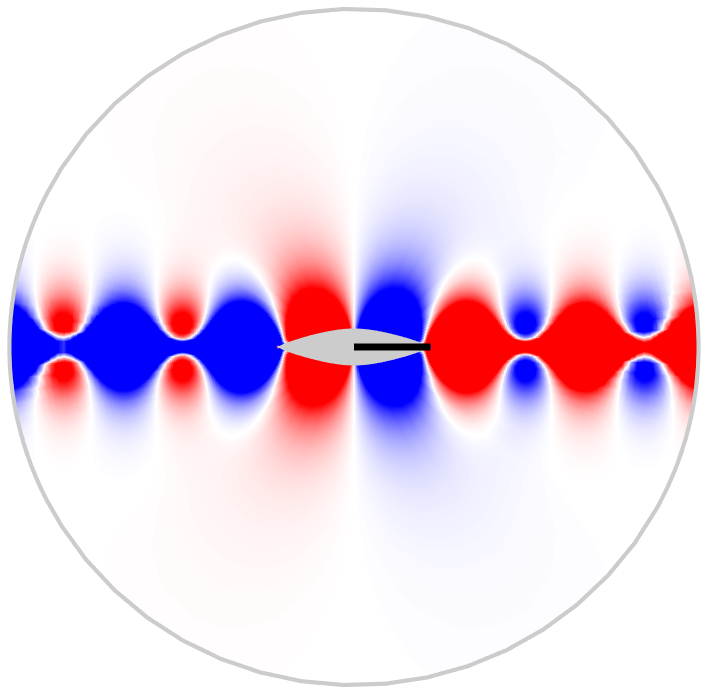}\end{center}
	\caption{$\phasediff = 0$}
	\label{Fig:diffThrustPitch0}
	\end{subfigure}\hfill
	\begin{subfigure}[h]{.28\textwidth}\begin{center}
	\includegraphics[width = \linewidth]{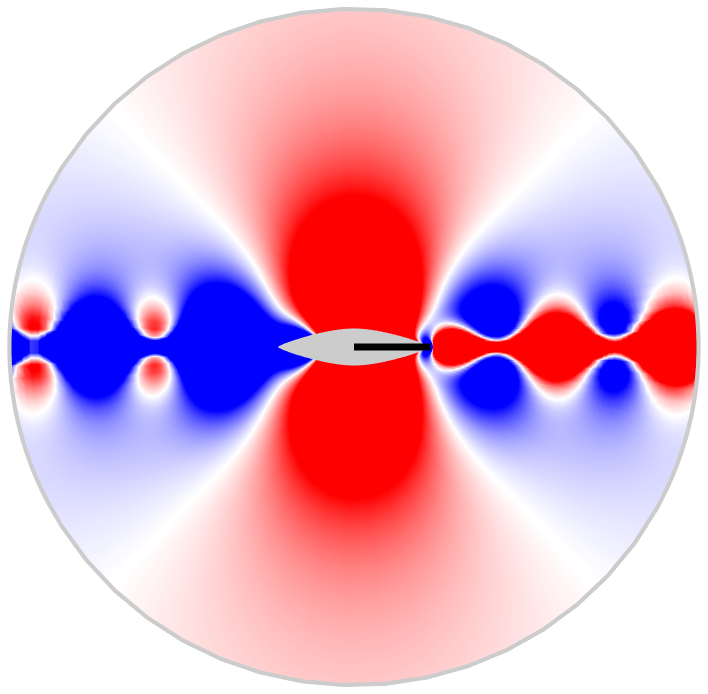}\end{center}
	\caption{$\phasediff = \pi/2$}
	\label{Fig:diffThrustPitch90}
	\end{subfigure}\hfill
	\begin{subfigure}[h]{.28\textwidth}\begin{center}
	\includegraphics[width = \linewidth]{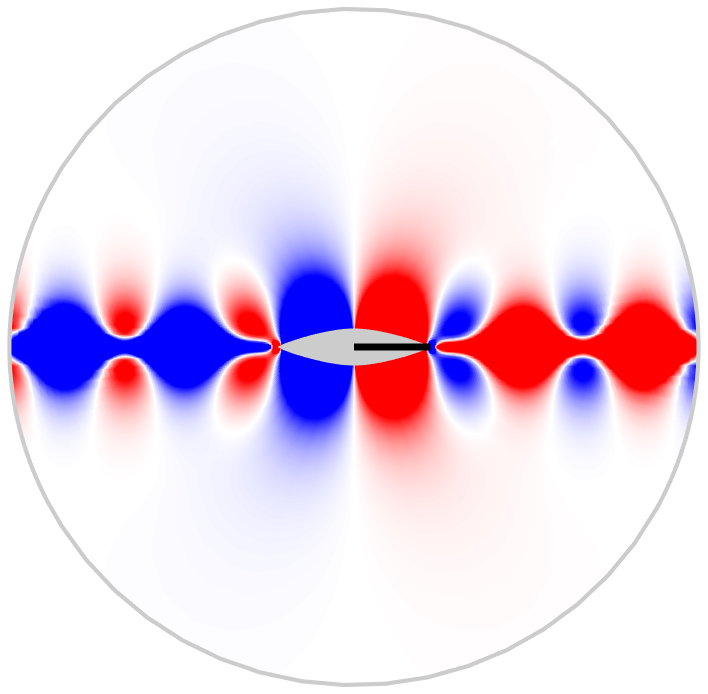}\end{center}
	\caption{$\phasediff = \pi$}
	\label{Fig:diffThrustPitch180}
	\end{subfigure} \hfill
	\begin{subfigure}[h]{.1\textwidth}
	\begin{center}
    \myCBar{0.1}{$T_1-T_0$}
    \vspace{.7cm}
    \end{center}
	\end{subfigure}
	\caption{The dependence of the thrust difference $T_1-T_0$ on the relative positions of a pair of pitching foils at the indicated phase difference $\phasediff$ with $\sigma = 2$.}	
	\label{Fig:diffThrustPitch}
\end{figure}

How are these results affected by the mode of locomotion? We briefly address this question in Figs.~\ref{Fig:sumThrustPitch}--\ref{Fig:diffThrustPitch} by showing analogous plots for {\em pitching} foils. By comparing the figures to the corresponding plots for heaving kinematics, we observe that the average (Fig.~\ref{Fig:sumThrustHeave} vs. Fig.~\ref{Fig:sumThrustPitch}) and difference (Fig.~\ref{Fig:diffThrustHeave} vs. Fig.~\ref{Fig:diffThrustPitch}) of thrusts have a qualitatively similar structure along the vertical midplane ($y$-axis). Specifically, panel A (panel C) in both Fig.~\ref{Fig:sumThrustHeave} and Fig.~\ref{Fig:sumThrustPitch} show that the hydrodynamic thrust is relatively low (high) for in-phase (antiphase) flapping. Moreover, the vertical contour of $T_1-T_0=0$ indicates that the side-by-side arrangement is an equilibrium for both in-phase (Fig.~\ref{Fig:diffThrustPitch0}) and anti-phase (Fig.~\ref{Fig:diffThrustPitch180}) pitching, as was also observed for heaving foils.

However, there are some notable differences between heaving and pitching kinematics. For in-line configurations of heaving foils, the average thrust oscillates between values greater and less than unity as the distance from the leader is progressively increased (Fig.~\ref{fig:Thrust} and Fig.~\ref{Fig:sumThrustHeave}), but it is always greater than unity for pitching foils (Fig.~\ref{Fig:sumThrustPitch}). Moreover, Fig.~\ref{Fig:diffThrustPitch0} suggests that for in-phase inline pitching foils, the follower always generates greater thrust and thus would tend to catch up with the leader. The zero-contour of $T_1-T_0$ lies off axis and therefore an equilibrium arrangement requires the swimmers to be offset vertically. The same is nearly true when the follower has a nonzero phase offset (Fig.~\ref{Fig:diffThrustPitch90} and Fig.~\ref{Fig:diffThrustPitch180}), except for a small region near the trailing edge of the leader where the thrust difference is negative, which leads to a single (stable) equilibrium point for which the foils are nearly touching. We note that the presence of either zero (Fig.~\ref{Fig:diffThrustPitch0}) or one (Fig.~\ref{Fig:diffThrustPitch90} and Fig.~\ref{Fig:diffThrustPitch180}) in-line equilibrium configuration is in contrast to the case of heaving kinematics (Fig.~\ref{Fig:diffThrustHeave}), for which there exists a discrete set of equilibria along the horizontal midplane. Our predictions in Fig.~\ref{Fig:diffThrustPitch180} are qualitatively consistent with recent experiments on pitching hydrofoils in a water tank with an imposed free stream flow~\cite{Kurt2021}, in which a single in-line equilibrium with closely-spaced foils was observed.

\begin{figure}
	\begin{subfigure}[h]{.28\textwidth}\begin{center}
	\includegraphics[width = \linewidth]{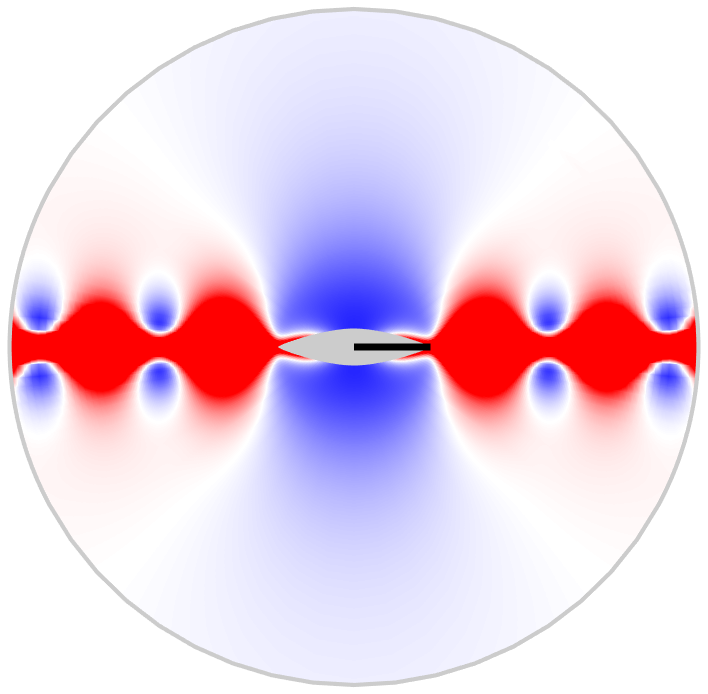}\end{center}
	\caption{$\phasediff = 0$} 
	\label{Fig:sumThrustWave0}
	\end{subfigure} \hfill
	\begin{subfigure}[h]{.28\textwidth}\begin{center}
	\includegraphics[width = \linewidth]{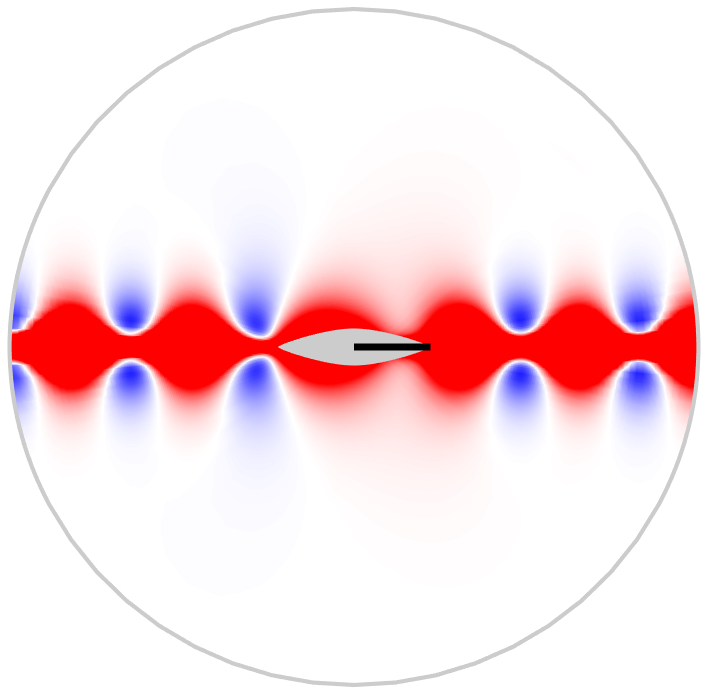}\end{center}
	\caption{$\phasediff = \pi/2$}
	\label{Fig:sumThrustWave90}
	\end{subfigure}\hfill
	\begin{subfigure}[h]{.28\textwidth}\begin{center}
	\includegraphics[width = \linewidth]{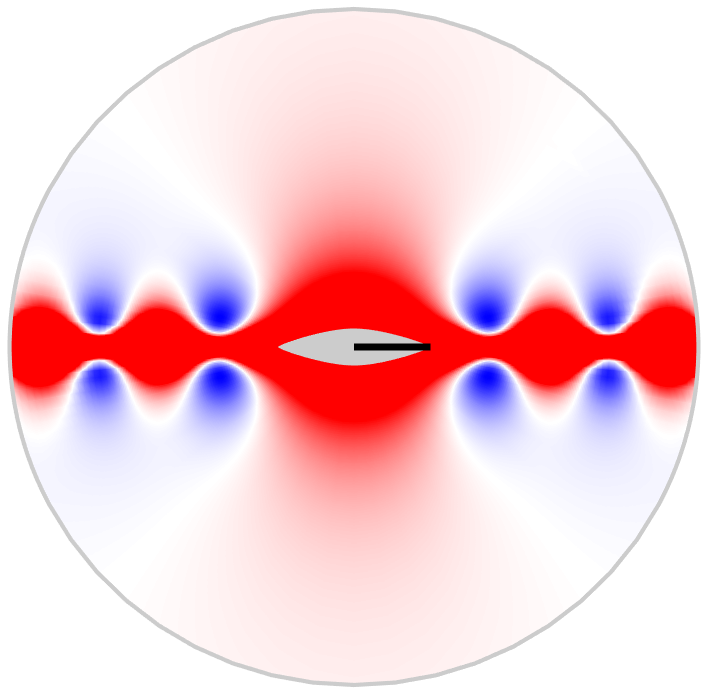}\end{center}
	\caption{$\phasediff = \pi$}
	\label{Fig:sumThrustWave180}
	\end{subfigure} \hfill
	\begin{subfigure}[h]{.1\textwidth}
	\begin{center}
    \myCBarThree
    \vspace{.7cm}
    \end{center}
	\end{subfigure}
	\caption{The dependence of the average thrust $(T_1+T_0)/2$ on the relative positions of a pair of waving, or undulating, foils at the indicated phase difference $\phasediff$ with $\sigma = 2$.}	
	\label{Fig:sumThrustWave}
	\begin{subfigure}[h]{.28\textwidth}\begin{center}
	\includegraphics[width = \linewidth]{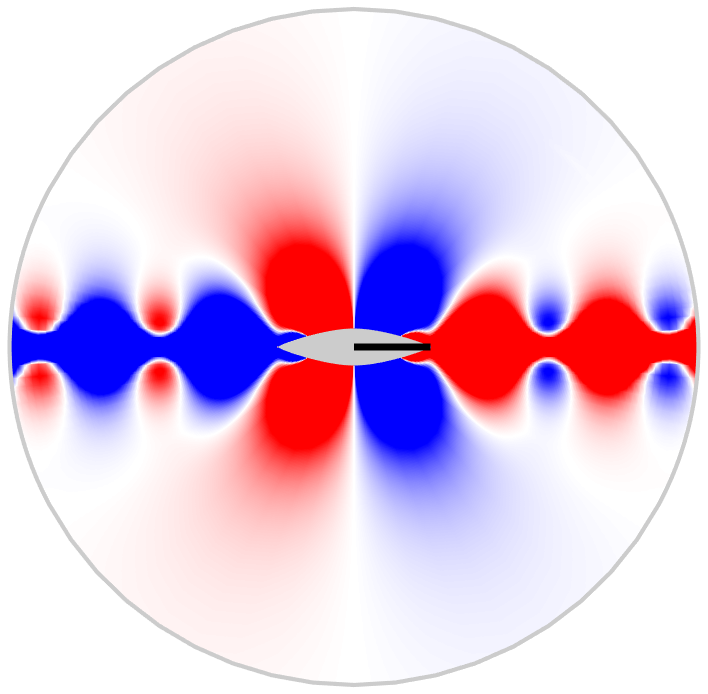}\end{center}
	\caption{$\phasediff = 0$}
	\label{Fig:diffThrustWave0}
	\end{subfigure}\hfill
	\begin{subfigure}[h]{.28\textwidth}\begin{center}
	\includegraphics[width = \linewidth]{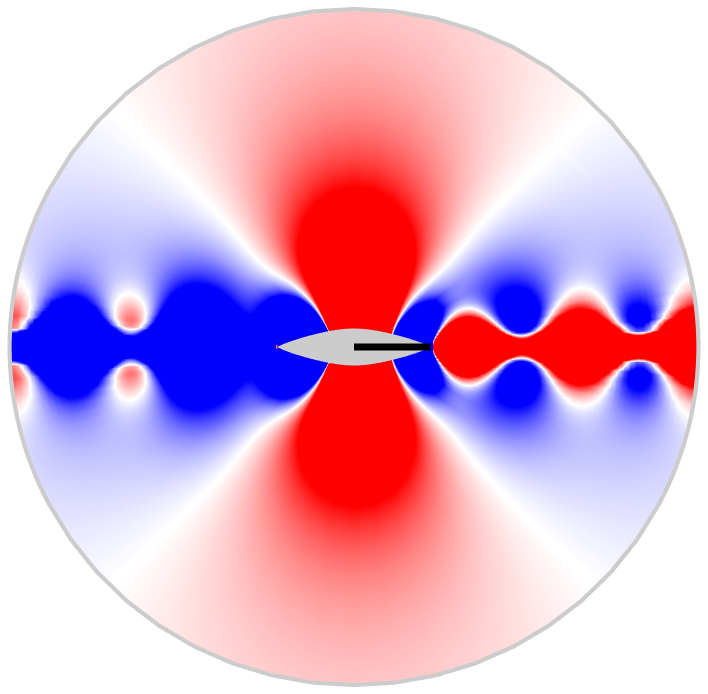}\end{center}
	\caption{$\phasediff = \pi/2$}
	\label{Fig:diffThrustWave90}
	\end{subfigure}\hfill
	\begin{subfigure}[h]{.28\textwidth}\begin{center}
	\includegraphics[width = \linewidth]{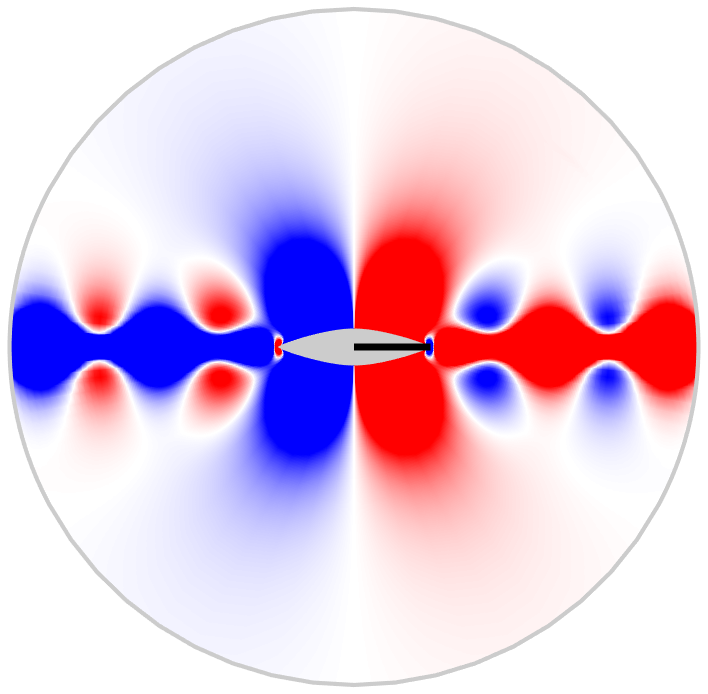}\end{center}
	\caption{$\phasediff = \pi$}
	\label{Fig:diffThrustWave180}
	\end{subfigure} \hfill
	\begin{subfigure}[h]{.1\textwidth}
	\begin{center}
    \myCBarFour
    \vspace{.7cm}
    \end{center}
	\end{subfigure}
	\caption{The dependence of the thrust difference $T_1-T_0$ on the relative positions of a pair of undulating foils at the indicated phase difference $\phasediff$ with $\sigma = 2$.}	
	\label{Fig:diffThrustWave}
\end{figure}

In addition to heaving and pitching, our theory can easily handle more intricate kinematics, such as the so-called waving motions originally studied in the single-body case by Wu~\cite{wu1961swimming}. To illustrate, Figs.~\ref{Fig:sumThrustWave}--\ref{Fig:diffThrustWave} show thrust values associated with the undulatory kinematics  {$h_m(x,t)=(1/2)(x-x_m^-)\cos(2(x-x_m^--1)+2\pi t+m\phasediff)$} where $m=0$ represents the foil located at the origin and $m=1$ the one located arbitrarily. The results for the average thrust (Fig.~\ref{Fig:sumThrustWave}) are qualitatively similar to those for pitching foils (Fig.~\ref{Fig:sumThrustPitch}) for all phase differences considered. The results for the thrust difference (Fig.~\ref{Fig:diffThrustWave}) are also similar to the pitching case (Fig.~\ref{Fig:diffThrustPitch}). The most salient difference is observed for antiphase kinematics, for which the zero-contour of $T_1-T_0$ just downstream of the foil shown consists of a single connected component for the pitching configuration (Fig.~\ref{Fig:diffThrustPitch180}) but three separate components for the undulatory one (Fig.~\ref{Fig:diffThrustWave180}).

Some of the physical results in this section have been corroborated by numerical simulations of the Navier--Stokes equations and experiments on pairs of flapping foils in a water tank. For the inline configuration, numerical~\cite{Akhtar2007, Deng2007, Broering2012b, Muscutt2017} and experimental~\cite{Warkentin2007, ramananarivo2016flow, Boschitsch2014, Gong2015, Lua2016} studies have found that the hydrodynamic thrust oscillates with the separation distance and/or the phase offset $\phasediff$, as we observed in Fig.~\ref{fig:exptB}. Meanwhile, for the side-by-side configuration, Figs.~\ref{Fig:sumThrustHeave0} and~\ref{Fig:sumThrustPitch0} show that in-phase oscillations lead to a decrease in the net thrust, as observed in experiments~\cite{Dewey2014}. By contrast, Figs.~\ref{Fig:sumThrustHeave180} and~\ref{Fig:sumThrustPitch180} show that anti-phase oscillations cause a dramatic increase in the thrust, a finding that has been observed in experiments~\cite{Dewey2014} and simulations~\cite{Dong2007,Bao2017}. Furthermore, for the case of anti-phase pitching (Fig.~\ref{Fig:diffThrustPitch180}), our prediction of a single stable inline equilibrium is consistent with recent experiments on pitching foils in a water tank~\cite{Kurt2021}.
 \section{Conclusions}
\label{Sec:Conclusions}

In this paper, we have generalized classical waving-plate theory to handle multiple swimming bodies. Our analysis relies on conformally mapping multiply-connected domains via the prime function, a special function whose application to varied physical problems has only recently been explored. In our theory, flow-mediated interactions between swimmers are handled directly via the acceleration potential (or pressure field), with vortex shedding accounted for by the singularity structure of the solution. Importantly, the analysis introduces no new ad-hoc assumptions regarding the separation distance between swimmers or the direction of influence (e.g.~a leader/follower relationship), in contrast with other recently developed models.

For the case of two interacting swimmers, we have presented an efficient method for evaluating the solutions and compared against existing experimental results. The theory recovers a range of observed behaviors, including the dependence of thrust on swimmer configuration and relative phase, and the emergence of integer-valued ``schooling modes'' for free swimmers.
Generally, Fig.~\ref{fig:exptB} suggests that the dominant physics of the interaction between inline heaving wings is controlled by the phasing between the follower's leading edge and the leader's wake. This physical picture is consistent with that proposed in 2D~\cite{Maertens2017} and 3D~\cite{Li2019} numerical simulations of pairs of swimming bodies with realistic fish-like kinematics.

We remark that most of the two-body effects mentioned above are traditionally thought of as wake-induced interactions, a perspective that we do not dispute. However, it is remarkable that our method accurately captures these effects {\em without ever explicitly resolving vortex wakes}. Wakes are certainly present in the model and accounted for by the singularity structure of the acceleration potential, but they do not need to be computed explicitly in order to determine the hydrodynamic forces on the swimmers. Rather the pressure field serves as the intermediary.

Because our solutions are primarily analytical in nature, the numerical requirements are light (mainly evaluation of special functions and quadrature), which enables extremely efficient sweeps through parameter space. Figure~\ref{Fig:sumThrustHeave} for example contains roughly 21,000 solution evaluations, each one of which represents the simulated interaction between a pair of swimmers in a different configuration. The creation of this entire figure required roughly five minutes of CPU time, whereas other simulation methods might require significant resources to obtain just a handful of data points on this plot.

We have invested the effort in extending the waving-plate formulation of Wu to multiply-connected domains because we see distinct advantages in working with the acceleration potential, or pressure field, as the main state variable. First, unlike a velocity potential, the acceleration potential enjoys analyticity throughout the entire fluid domain, even crossing vortex-sheet wakes where the velocity potential suffers a discontinuity. This feature provides great mathematical convenience. Second, compared to a method that explicitly resolves vortex sheets (e.g.~through solving integral equations), the pressure field offers a more direct description of hydrodynamic interactions. The vorticity formulation, while it remains an indispensable theoretical tool, introduces a few degrees of separation between the main state variable and the physics of interest. That is, once the vortex sheets are computed, the Biot--Savart law must be invoked to obtain the velocity field, and then the unsteady Bernoulli equation gives the pressure distribution, after which it is finally possible to calculate the hydrodynamic forces acting on each surface. This indirect path might call into the question the ability to draw meaningful conclusions regarding hydrodynamic interactions from visualizations of vorticity. The current approach, however, reframes the mathematical problem in terms of the pressure field, thus directly linking the primary state variable to the physics of interest.

At the same time, the newly developed theory is not without limitations. In particular, the flat structure of vortex wakes implied by the theory precludes discussion of more exotic phenomena, such as vortex sheet roll-up or wake deflection \cite{Bao2017, Lagopoulos2020, Gungor2020, Kurt2020}, which may play a role in altering hydrodynamic forces under certain circumstances. It might be possible to extend the theory to higher order to capture such effects, and such an extension is a worthwhile endeavor for future investigations.

Lastly, since our theory does not constrain the relative location of the swimmers or their actuation, it offers a number of new predictions. One example is the off-axis equilibrium configurations anticipated by Figs.~\ref{Fig:diffThrustHeave}, \ref{Fig:diffThrustPitch}, and \ref{Fig:diffThrustWave} for heaving, pitching, and undulatory kinematics respectively. Most previous laboratory experiments constrained the vertical positions of swimmers, and thus only observed the inline equilibria, but this constraint could be removed to search for staggered arrangements (e.g.~\cite{Kurt2021}). For biological systems, the vertical constraint is completely artificial, so staggered arrangements would be expected to be just as likely as inline ones. In a similar vein, our theory easily handles swimmers actuated at different frequencies (see Appendix \ref{Ap:Fourier}). Thoroughly exploring the range of new predictions and seeking corroboration from laboratory experiments is an exciting avenue for future work.

\subsection*{Acknowledgements}

The authors would like to thank the Isaac Newton Institute for Mathematical Sciences for support and hospitality during the programme  ``Complex analysis: techniques, applications and computations'' (supported by EPSRC grant no. \mbox{EP/R014604/1}) where the ideas for this paper were first generated. NJM thanks the NSF (grant no.~DMS-2012560) and the Simons Foundation (grant no.~524259) for their support. AUO's visit to the programme was partially supported by a travel award from the NSF (grant no. DMS-1933403), and he also acknowledges the support of the Simons Foundation (Collaboration Grant for Mathematicians, grant no. 587006).

\begin{appendices}

\section{General harmonic motions}
\newcommand{\mfreq}{\omega}
\newcommand{\freqMat}{\boldsymbol{V}}
\newcommand{\numFreq}{M_{\mfreq}}
\newcommand{\bSigma}{\boldsymbol{\Sigma}}

\label{Ap:Fourier}
The calculations presented in the main text may readily be generalized to the case of plates flapping with different frequencies.
Suppose that the $m$-th plate executes motions with frequency $\mfreq_m$ of the form
\begin{align}
	h_{m}(x,t) = h_{m}^c(x) \cos(\mfreq_m t) +  h_{m}^s(x) \sin(\mfreq_m t).
\end{align}
There will be a subset $\{\tilde{\mfreq}_m \} \subseteq \{\mfreq_m \}$ of $\tilde{M}$ distinct frequencies.
Since the problem is linear, the acceleration potential and velocity fields will be composed of linear responses to these distinct frequencies:
\begin{align}
\aPot(z,t) &= \sum_{m=0}^{\tilde{M}} \left[\aPot^c_m(z) \cos(\tilde{\mfreq}_m t) +\aPot^s_m(z) \sin(\tilde{\mfreq}_m t)\right],\\
w(z,t) &= \sum_{m=0}^{\tilde{M}} \left[w^c_m(z) \cos(\tilde{\mfreq}_m t) + w^s_m(z) \sin(\tilde{\mfreq}_m t)\right].  
\end{align}
We therefore introduce the vectorized quantities
\begin{align}
	\baPot(z) &= \begin{bmatrix} \aPot^c_0(z)& \,\aPot^s_0(z)& \cdots &
	\:\aPot^c_{\tilde{M}}(z) & \,\,\aPot^s_{\tilde{M}}(z)\end{bmatrix}^T,\\
	\bw(z) &= \begin{bmatrix} w^c_0(z)& w^s_0(z)& \cdots &
	w^c_{\tilde{M}}(z) & w^s_{\tilde{M}}(z)\end{bmatrix}^T.
\end{align}
In this notation, \eqref{Eq:velPres} becomes
\begin{align}
	\partial_z \baPot(z) =\freqMat \bw(z)  + U \partial_z \bw(z)
	\label{Eq:pdeVarFreq}
\end{align}
where $\freqMat\in \mathbb{R}^{2\tilde{M} \times 2 \tilde{M}}$ is a block diagonal matrix:
\begin{align}
	\freqMat =
	\begin{bmatrix}
 \tilde{\mfreq}_0 \bJ &  & \boldsymbol{0}\\
 & \ddots  & 	\\
 \boldsymbol{0} & & \tilde{\mfreq}_{\tilde{M}} \bJ
	\end{bmatrix}=
	\textrm{diag} \left(
	\begin{bmatrix}\tilde{\mfreq}_0 \bJ & \tilde{\mfreq}_1 \bJ &\cdots &
	\tilde{\mfreq}_{\tilde{M}} \bJ\end{bmatrix}
	\right).
\end{align}
Equation \eqref{Eq:pdeVarFreq} can be inverted to obtain
\begin{align}
	\bw(z) = \frac{1}{U} \e^{-\bSigma z}
	\int_{-\infty}^z \e^{\bSigma z^\prime}
	\partial_z \baPot(z^\prime) \d z^\prime \label{Eq:matSol2}
\end{align}
where $\bSigma = \bV/U$.
Since the exponential of a block diagonal matrix is the diagonal of the exponential of the
blocks, we have $\e^{\bSigma z} =\textrm{diag} \left(
	\begin{bmatrix}\e^{\sigma_0 \bJ}&\cdots &
	\e^{\sigma_{\tilde{M}} \bJ}\end{bmatrix}\right)$
	where $\sigma_k = \tilde{\mfreq}_k/U$.
	Now, all the other calculations of the paper follow with $\sigma \bJ$ replaced with $\bSigma$.

Another perspective is that the motion of each plate can be represented as a superposition of contributions from all frequencies in the system. For example, consider two plates flapping with frequencies $\mfreq_0$ and $\mfreq_1$, respectively. 
Then, the problem is equivalent to solving two problems, each of which consists of a single plate flapping with a particular frequency ($\mfreq_0$ or $\mfreq_1$) while the other plate is stationary.
\section{Comparison to Wu's solution}
\label{Sec:wuComp}
Here we show that our single plate solution in \S\ref{Sec:singlePlate} is equivalent to that derived by Wu \cite{wu1961swimming}.

In the single plate case ($M=0$), the canonical circular domain is the unit disc and the mapping is the Joukowski map:  $\map(\zeta) = (\zeta + \zeta^{-1})/2$.
The associated prime function is simply $\omega(\zeta, \alpha) = \zeta - \alpha$.
Wu's solution is expressed in the circular domain as the Laurent series
\begin{align}
    L(\zeta) = \i U^2\sum_{n=1}^\infty {a_n \zeta^n} \label{Eq:wuSol}
\end{align}
where the coefficients $a_n$ are given by the recurrence relation
\begin{align}
    a_n = \lambda_n + \frac{\j \sigma}{2 n} \left(\lambda_{n-1} - \lambda_{n+1} \right).
\end{align}
The $\lambda_n$ are the coefficients of the Chebyshev expansion of the material derivative of the plate motion:
\begin{align}
    (\j \sigma + \partial_x)h_0(x) = -\frac{\lambda_0}{2} - \sum_{n=1}^\infty \lambda_n T_n(x), 
\end{align}
where $T_n$ denotes the Chebyshev polynomials of the first kind.
Applying the orthogonality property of Chebyshev polynomials yields
    \begin{align*}
    \lambda_n = 
    \frac{-2}{\pi}
    \int_{-1}^1 \frac{T_n(x)}{\sqrt{1-x^2}} 
     (\j \sigma + \partial_x)h_0(x) \d x.
    \end{align*}
As discussed in \S\ref{Sec:harmonic}, Wu uses the complex `$\j$' notation
where the real part of the solution is assumed with respect to the imaginary unit $\j$.
For ease of comparison, we also adopt this notation in this section.
Wu also treats $h_0$ and $\partial_x h_0$ with similar Chebyshev expansions:
\begin{align}
    h_0(x) = \frac{\beta_0}{2}+ \sum_{n=1}^\infty \beta_n T_n(x), \qquad \qquad
    \partial_x h_0(x) = \frac{\gamma_0}{2} + \sum_{n=1}^\infty \gamma_n T_n(x)
\end{align}
so that $\lambda_n = -(\gamma_n + \j \sigma \beta_n)$. 
The coefficients $\gamma_n$ and $\beta_n$ are
connected by the recurrence relation $\gamma_{n-1} - \gamma_{n+1} = 2 n \beta_n$ for $n>0$.

To achieve correspondence with Wu's solution, we also expand our boundary data \eqref{Eq:PsiDefn} in Chebyshev polynomials.
The integral formula
\begin{align}
    \int^x T_n(x^\prime) \d x^\prime = \begin{dcases} T_1(x), &n= 0,\\
    \frac{1}{4} T_2(x),& n = 1,\\ 
    \frac{1}{2} \left(\frac{T_{n+1}(x)}{n+1} - \frac{T_{n-1}(x)}{n-1} \right), & n>1,
    \end{dcases}
\end{align}
allows us to write
\begin{align}
    \Psi_{0}(\zeta(x)) 
    &=\textnormal{const.} - U^2\left[ \sum_{n=1}^{\infty} (\gamma_n +2 \j \sigma \beta_n) T_n(x) \right.
    \notag \\
    &\left.- \sigma^2 \frac{\beta_0}{2} T_1(x)- \sigma^2 \frac{\beta_1}{4} T_2(x) - \frac{\sigma^2}{2} \sum_{n=2}^\infty \beta_n  \left(\frac{T_{n+1}(x)}{n+1} - \frac{T_{n-1}(x)}{n-1} \right)\right].
\end{align}
Expanding the first bracketed term and rearranging the term on the second line yields
\begin{align}
    \Psi_0(\zeta(x))
    &=\textnormal{const.} + U^2 \sum_{n=1}^{\infty} \left(\lambda_n - \j \sigma \beta_n + \frac{\sigma^2}{2 n} \left( \beta_{n-1} - \beta_{n+1} \right) 
    \right) T_n(x).
\end{align}
Finally, one can show that the bracketed term above is simply equal to $a_n$, so,
\begin{align}
    \Psi_{0}(\zeta(x))
    &= U^2 \sum_{n=1}^\infty a_n T_n(x), \label{Eq:hCheb}
\end{align}
where we have set the constant of integration to be zero without loss of generality.

We may now insert the Chebyshev expansion \eqref{Eq:hCheb} into the Poisson integral formula \eqref{Eq:Poisson} to verify that we obtain the same regular solution~\eqref{Eq:wuSol} as Wu. Since $|\zeta|< 1$, \eqref{Eq:Poisson} may be written as
\begin{align}
    L(\zeta) = \textnormal{const.} +
   \sum_{n=1}^\infty  \frac{\zeta^n}{\pi} \oint_{C_{0}} \Psi_{0}(\zeta^\prime) 
\frac{\d \zeta^\prime}{(\zeta^{\prime})^{n+1}},
\label{Eq:WuComp2}
\end{align}
where we have interchanged the order of summation and integration.
The constant term in~\eqref{Eq:WuComp2} is purely imaginary and may thus be ignored. For the Joukowski map~\eqref{Jouk}, $x=\cos\vartheta=\map(\e^{\i\vartheta})$, so substituting~\eqref{Eq:hCheb} into~\eqref{Eq:WuComp2} and using 
$T_k(\cos(\vartheta)) = \cos(k \vartheta)$ produces
\begin{align}
    L(\zeta) =  \frac{U^2 \i}{\pi} \sum_{n=1}^\infty  \zeta^n\sum_{k=1}^\infty a_k \int_0^{2\pi}
 \cos(k \vartheta)\e^{-\i n \vartheta} \d \vartheta.
\end{align}
The remaining integral equals $\pi \delta_{k,n}$ so we finally have
\begin{align}
    L(\zeta) = \i U^2 \sum_{n=1}^\infty a_n \zeta^n 
\end{align}
which is identical to the regular part of Wu's solution \eqref{Eq:wuSol}.

The singular part of Wu's solution is also recovered by our analysis.
Detailing every step would involve tedious algebra so we will only provide an overview of the main steps.
In particular, we will explicate how Bessel functions and the celebrated Theodorsen function arise.
Wu expresses the singular part of the pressure field as
\begin{align}
\frac{\i U^2 a_0}{\zeta + 1}
\end{align}
which, in the notation of the present paper, is
\begin{align}
  a_0 U^2 \hat{\K}(\zeta,\zeta^-_0)
\end{align}
where $\zeta^-_0=-1$ is the preimage of the leading edge. 
In the simply connected case, the integrals in \S\ref{Sec:suctions} can be computed analytically in terms of Bessel functions so an analytical formula for ${a}_0$ can be found.
The integral for the matrix $\boldsymbol{D}$ given in~\eqref{Eq:DMatrix}, which has just a single entry, becomes
\begin{align}
   U^2 D &= -U
	\Im_{\i} \left[
	\int_{-\infty}^{z_0}
	\e^{\sigma \j (z - z_0)}
\frac{\d}{\d z} \left(\frac{\i}{\zeta(z)+1} \right)
\d z \right]
\end{align}
where {$z_0$ is an arbitrary point on the plate,} and we are using $\Im_\i$ to represent the imaginary part with respect to $\i$ specifically.
Applying integration by parts then yields
\begin{align}
   U^2 D &= -U
	\Im_{\i} \left[\frac{\i}{\zeta(z_0)+1}- \sigma \i \j
	\int_{-\infty}^{z_{0}}
\frac{\e^{\sigma \j (z- z_{0})}}{\zeta(z)+1}
\d z \right].
\label{Eq:Dsing}
\end{align}
The remaining integral can be expressed using Bessel functions:
by inverting the Joukowski map~\eqref{Jouk}, we obtain
\begin{align}
	\frac{1}{\zeta(z)+1} =
	\frac{1}{1 + z + \sqrt{z^2-1}}  &=  
\frac{1}{2} 
	\left(\frac{\d }{\d z} \left( z - \sqrt{z^2 - 1} \right)  
	- \frac{1}{\sqrt{z^2 -1}} \right).
	\label{Eq:mapInvert}
\end{align}
On the other hand, the modified Bessel functions of the first kind can be expressed as 
\begin{align}
	K_0(\j \sigma) =\int_{-\infty}^{-1} \frac{\e^{\j \sigma z} }{\sqrt{z^2 -1}} \d z,\qquad
	K_1(\j \sigma) = \j \sigma \int_{-\infty}^{-1} \e^{\j \sigma z} \sqrt{z^2 -1} \,\d z.
		\label{Eq:bessel}
\end{align}
Thus, choosing the integration endpoint as $z_0=-1 + \epsilon \i$, applying \eqref{Eq:mapInvert} and \eqref{Eq:bessel} and taking the limit $\epsilon \rightarrow 0$ generates
\begin{align}
\sigma \j	\int_{-\infty}^{z_0} \frac{\e^{\j \sigma (z - z_0)}}{\zeta(z)+1} \d z
	&= \frac{1}{2} \left[ 1
		- \j \sigma \e^{\j \sigma}(K_1(\j \sigma) + K_0(\j \sigma))
	\right].
	\label{Eq:int3}
\end{align}
Substituting \eqref{Eq:int3} into \eqref{Eq:Dsing}  yields
\begin{align}
    U^2 D &= -{\j \sigma U} (K_1(\j \sigma) + K_0(\j \sigma)).
\end{align}
While we do not go into the details, similar arguments can be applied to express the integral \eqref{Eq:qDef} in terms of Bessel functions.
Thus all the terms in the linear system \eqref{Eq:linear} -- which here is just a single complex scalar equation -- can be expressed purely in terms of the motion of the plate and these Bessel functions.
Solving this scalar equation yields
\begin{align*}
    a_0 = (\lambda_0 + \lambda_1) C(\sigma) - \lambda_1
\end{align*}
where
\begin{align*}
    C(\sigma) = \frac{K_1(\j \sigma)}{K_1(\j \sigma) + K_0(\j \sigma)}
\end{align*}
is the Theodorsen function.
Thus, our solution in \S\ref{Sec:singlePlate} is identical to that derived by Wu \cite{wu1961swimming}.
\section{Degrees of freedom in the map}
\label{Sec:dof}     
It is worth pointing out that the number of parameters matches the number of constraints in the map $\map(\zeta)$ specified by~\eqref{phimap}. In the physical domain, there are $3(M+1)$ constraints, since
each of the $M+1$ slits is determined by three parameters: two for position and one for length.
As for the parameters, in the circular domain there are three parameters for each of the $M$ excised circles:
two for position and one for radius.
Additionally, the complex parameters $A$, $B$ and $\beta$ contribute $2$ further parameters each. There are three real degrees of freedom (two for translation and one for rotation) available in the Riemann Mapping Theorem. We insist that $A$ is real, which fixes one of these. In general, we can pick $\beta$ as we like, which fixes the other two. 
As such, there are $3M + 2 + 2 + 2 - 3= 3(M+1)$ parameters remaining, which matches the number of constraints. We note that, for the two-plate case (\S\ref{Sec:TwoSwim}), we instead fixed $\delta_1=0$ for the convenience of $D_{\zeta}$ being a concentric annulus, and had to find $\beta$. 
\section{Explicit expressions for the regular {part of the} solution}
\label{Ap:exact}
In certain simple situations it is possible to write down explicit solutions to the regular part of the solution $\boldsymbol{L}(\zeta)$ without using the integral formula \eqref{Eq:schwarzSol}.
For example, the motion of $(M+1)$ plates executing identical, in-phase heaving motions may be described by
\begin{align}
	h_{m}^c(x) = h,\quad h_{m}^s(x) = 0, \qquad m = 0,\dots,M.
\end{align}
Thus, the sine contribution $L_s$ to the regular part is zero, whereas the cosine contribution $L_c$ depends on $h$.
$L_c$ solves the boundary value problem \eqref{Eq:schwarz3}, where the streamfunction takes values
\begin{align}
	\Psi_{m}^c(\zeta) =
		4 \pi^2 h x(\zeta).
	\label{Eq:hPitch}
\end{align}
To determine $L_c$ we must find a function that is analytic in $D_\zeta$ and takes imaginary part equal to $4 \pi^2 h x$ on the boundary as specified by \eqref{Eq:schwarz3}.
Since the boundaries in the physical domain are horizontal slits, the conformal map satisfies \mbox{$\Im[\i \map(\zeta)] = x(\zeta)$} when $\zeta \in \partial D_\zeta$.
An ansatz for the regular solution is thus
\begin{align}
    L_c(\zeta) &=
		4 \pi^2 \i h \map(\zeta).
	\label{Eq:ansatz1}
\end{align}
The ansatz~\eqref{Eq:ansatz1} satisfies the boundary data specified by~\eqref{Eq:schwarz3}.
However, this ansatz 
is not a valid solution to the boundary value problem because it is not analytic everywhere, since $\map(\zeta)$ has a simple pole at $\zeta = \beta$.
In particular,
\begin{align}
    L_c(\zeta) \sim 
\frac{4A\pi^2 \i h}{\zeta - \beta} \qquad \qquad \text{as}\quad\zeta \rightarrow \beta.
\label{Eq:exactPole}
\end{align}
Hence, to obtain a valid solution to the boundary value problem we must remove the pole of the ansatz \eqref{Eq:ansatz1} without disturbing the imaginary part on the boundary $\partial D_{\zeta}$.
To achieve this, we combine the ansatz \eqref{Eq:ansatz1} with a function that has a simple pole of the correct magnitude and constant imaginary part on the boundaries (different constants on different boundaries).
Crowdy \cite{Crowdy2010} showed that
\begin{align*}
    \mathcal{U}(\zeta)
    =-\i\left(
    \frac{1}{\overline{\beta}}
    \mathcal{K}\left(\zeta,1/\overline{\beta} \right)
    +    \frac{1}{{\beta}}
\mathcal{K}(\zeta, \beta)\right)
\end{align*}
takes constant imaginary values on $\partial D_\zeta$ and
\begin{align*}
\mathcal{U}(\zeta) \sim \frac{\i}{\zeta - \beta} \qquad \qquad \text{as}\quad\zeta \rightarrow \beta. 
\end{align*}
Thus, the new ansatz
 $   L_c(\zeta) = 4 \pi^2
    h (\i \map(\zeta)-A\mathcal{U}(\zeta))$
takes the correct boundary values and is analytic in $D_\zeta$. 
Expanding $\map$ and $\mathcal{U}$ then gives
\begin{align}
    L_c(\zeta) =8 \pi^2\i h\frac{ A}{\overline{\beta}} \K(\zeta,1/\overline{\beta}).
\end{align}
The above expression is the regular part of the acceleration potential for any number of plates in any configuration provided that they are heaving with the same amplitude and phase.
Similar expressions for pitching motions can be derived, though the algebra becomes quite complicated.
\section{Fast computation of {integrals comprising the singular part of the solution}}
\label{Ap:integrals}
We present a computationally efficient method for evaluating the integrals that determine the unknown coefficients $a_{m}$ in the singular part of the solution (\S\ref{Sec:suctions}).
The integrals to be computed in \eqref{Eq:sucInts} generally take the form
\begin{align}
	\int_{-\infty}^{z_{m}}
	\e^{\sigma \bJ (z - z_{m})}
\frac{\d \boldsymbol{r}}{\d z}\,\, \d z
\label{Eq:intFirst}
\end{align}
for some vectorised function $\boldsymbol{r}$ whose components are analytic in $D_z$.
Additionally, the components of $\boldsymbol{r}$ are bounded as $|z| \rightarrow \infty$.
Recalling the expression~\eqref{Eq:eigen} for $\e^{\sigma \bJ z}$ means that to compute \eqref{Eq:intFirst} it is sufficient to compute
\begin{align}
	I^\pm = \int_{-\infty}^{z_{m}}
	\e^{\pm\i \sigma (z - z_{m})}
\frac{\d r}{\d z}\,\, \d z
\label{Eq:int2}
\end{align}
where $r$ is any component of $\boldsymbol{r}$. However, an expression for $r(z)$ is not readily available, so we rewrite the integral in $D_{\zeta}$:
\begin{align}
I^\pm = 
\int_{\beta^-}^{\zeta_m}
\exp\left[\pm\i \sigma \left(\map(\zeta) - z_m\right)\right]
\frac{\d R }{\d \zeta}(\zeta)\d \zeta,\label{Eq:intzeta}
\end{align}
where $R(\zeta) = r(z)$ and $\beta^-$ is the preimage of $-\infty$, i.e. $\map(\beta^-)=-\infty$. 

By \eqref{phimap}, we can express the conformal map as
\begin{align}
\map(\zeta) = \frac{A}{\zeta - \beta} + \tilde{\map}(\zeta),  
\label{Eq:residue}
\end{align}
where $\tilde{\map}$ is analytic in $D_\zeta$ and finite at $\beta$. 
The residue of $\map$ at $\beta$ is given by $A\in\mathbb{R}$.
There is an essential singularity at $\zeta = \beta$
and we cannot integrate through this singularity.

\begin{figure}[tpb]
\begin{center}
	\resizebox{.8\linewidth}{!}{
\input{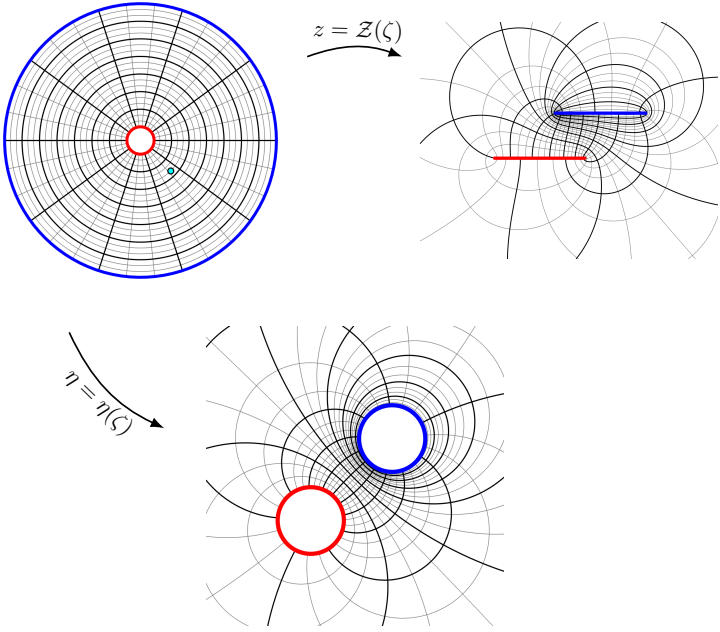}
}
\end{center}
\caption{Illustration of the mappings between  $D_\zeta$ (top left), $D_z$ (top right) and $D_\eta$ (bottom).}
\label{Fig:etaDiagram}
\end{figure}

We now introduce the substitution
\begin{align}
\eta(\zeta) = \frac{\zeta_m - \zeta}{\zeta - \beta} \cdot \frac{1}{\zeta_m -\beta},
\qquad \qquad
\zeta(\eta) = \frac{\eta \beta(\beta - \zeta_m) - \zeta_m }
{(\beta -\zeta_m) \eta - 1},
\label{Eq:eta}
\end{align}
where $\zeta_m$ is the pre-image of the endpoint of the integral: $z_m = \map(\zeta_m)$.
This is a M\"obius map so the boundary circles are mapped to boundary circles.
Note that the point $\zeta = \beta^-$ is mapped to $\eta = -\infty$, so $D_\eta$ is exterior to the boundary circles.
This domain is similar to the initial physical domain, except the flat plate boundaries are now circles;
the arrangement is illustrated in figure \ref{Fig:etaDiagram}.

The exponential term in the integrals can now be manipulated using~\eqref{Eq:residue} into a form amenable to Gauss--Laguerre quadrature: 
\begin{align}
\exp \left[ \pm \i \sigma \left(\map(\zeta) - z_m \right) \right] &=
\exp \left[\pm \i \sigma A \eta \right]
\exp \left[ \pm \i \sigma \left( \tilde{\map}(\zeta(\eta))+ \frac{A}{\zeta_m - \beta} - \map(\zeta_m) \right)  \right].
\end{align}
The integrals~\eqref{Eq:intzeta} may thus be written in $D_\eta$:
\begin{align}
	I^\pm &= 
\exp \left[ \pm \i \sigma \left(
\frac{A}{\zeta_m - \beta} - \map(\zeta_m) \right)  \right]\notag \\
	      &\times \int_{-\infty}^{0} \exp \left[\pm \i \sigma A \eta \right]
	      \exp \left[ \pm \i \sigma \tilde{\map}(\zeta(\eta)) \right]
	      	\frac{\d \zeta}{\d \eta} (\eta)
	  \frac{\d R }{\d \zeta} (\zeta(\eta))
\d \eta. 
\end{align}
Also, note that
\begin{align}
\frac{\d \zeta}{\d \eta}(\eta) = -\frac{ ( \zeta_m - \beta)^2}
{\left( (\zeta_m - \beta) \eta +1 \right)^2}
= \mathcal{O}\left( \frac{1}{\eta^2}  \right)\qquad\text{as}\quad |\eta| \rightarrow \infty.
\end{align}
Since $\d R/\d \zeta(\beta)$ and $\tilde{\map}(\beta)$ are bounded, the integrand of $I^\pm$ decays like $o(1/\eta)$ for $\text{Im}(\eta)\gtrless 0$, so the limit of integration of $I^\pm$ can be deformed to $\pm \i \infty$:
\begin{align}
	I^\pm &= 
-	\exp \left[ \pm\i \sigma \left(\frac{A}{\zeta_m - \beta}  - \map(\zeta_m)\right)\right]\notag\\
	      &\times \int_{0}^{\pm \i \infty} \exp \left[\pm \i \sigma A \eta\right]
		\exp \left[ \i \sigma \tilde{\map}(\zeta(\eta)) \right]
	\frac{\d \zeta}{\d \eta}(\eta)\frac{\d R }{\d \zeta}(\zeta(\eta)) \d \eta.
\end{align}
The advantage of this approach is that the oscillatory exponential term is now a decaying exponential term.
Now the integrand decays rapidly as $\eta \rightarrow \textrm{sign}(A) \times \i \infty$ and 
can be evaluated with a Gauss--Laguerre quadrature rule.

\begin{figure}[t]
	\begin{subfigure}{.45\textwidth}
	\begin{center}
		\begin{tikzpicture}[scale = 0.65]
	\node at (2,-1) {$D_\eta$};
	\skCircBlankBThick{(0,0)}{1cm}{blue!50!white}
	\skCircBlankBThick{(1,3)}{1cm}{red}
	\draw[->-=0.5,ultra thick] (-5,.75)--(-.75,.75);
	\draw[green,->-=0.5,ultra thick] (-.75,.75)--(-.75,5) node [black, midway, left]
	{$\e^{+\i A \sigma \eta}$};
	\draw[purple,->-=0.5,ultra thick] (-.75,-.7)--(-.75,-4) node [black, midway, left]
	{$\e^{-\i A \sigma \eta}$};
	\draw[purple,ultra thick] (-.75,-.7) arc (-135:-225:1);
	\fill[black] (-.75,.75) circle (0.15) node [above left] {$0$};
	\draw[ultra thick,purple,->-=0.5] ([shift=(180:2cm)]-3,-1) arc (180:270:2cm);
	\draw[ultra thick,green,->-=0.5] ([shift=(180:2cm)]-3,2) arc (180:90:2cm);
\end{tikzpicture}	
	\end{center}
	\caption{}
	\label{Fig:deformA}
	\end{subfigure}
\hfill
\begin{subfigure}{.45\textwidth}
	\begin{center}
		\begin{tikzpicture}[scale = 0.65]
	\skCircBlankBThick{(0,0)}{1cm}{blue}
	\skCircBlankBThick{(1,3)}{1cm}{red!30!white}
	\draw[->-=0.5,ultra thick] (-5,2.5)--(.1,2.5);
	\draw[ultra thick,green] ([shift=(210:1cm)]1,3) arc (210:135:1cm);
	\draw[ultra thick,purple] ([shift=(210:1cm)]1,3) arc (210:315:1cm);
	\draw[->-=0.5,ultra thick,green] (.3,3.7)--(.3,6) node [black, midway, left]
	{$\e^{+\i A \sigma \eta}$};
	\draw[->-=0.5,ultra thick,purple] (1.75,2.35)--(1.75,-3) node [black, midway, right]
	{$\e^{-\i A \sigma \eta}$};
	\fill[black] (.1,2.5) circle (0.15) node [above left] {$0$};
	\draw[ultra thick,purple,->-=0.5] ([shift=(180:2cm)]-3,1) arc (180:270:2cm);
	\draw[ultra thick,green,->-=0.5] ([shift=(180:2cm)]-3,4) arc (180:90:2cm);
	\draw[ultra thick,purple,->-=0.5] ([shift=(0:1.2cm)]0,0) arc (0:360:1.2cm);
\end{tikzpicture}
	\end{center}
	\caption{}
	\label{Fig:deformB}
	\end{subfigure}
	\caption{An example of the deformed contours for $A>0$.
	The black contour can be deformed to the green contour (for $I^+$) or the red contour (for $I^-$).
	In case (a) the endpoint is on the blue circle whereas in case (b) the endpoint is on the red circle.
	}
\label{Fig:deform}
\end{figure}
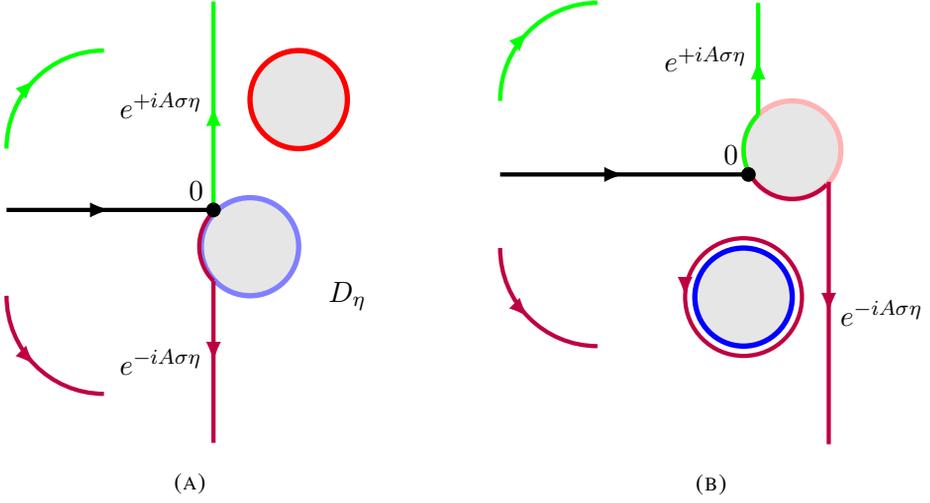

It is important to ensure that this contour deformation does not cross any poles, or their residue contribution should be included.
For example, $R$ could have a simple pole at the pre-image of the leading edge, and the new contour should not enclose this point in $D_\eta$.
Some typical contour deformations are illustrated in figure \ref{Fig:deform}.
If $A >0$, then the contour for $I^+$ should be deformed
into the upper half-plane (as indicated by the green contours), and 
the contour for $I^-$ into the lower half-plane (purple contours).
If $A<0$, then the contours for $I^\pm$ should be deformed into the lower/upper half-plane.

The deformed contour typically must include a circular arc section to ensure that it does not pass through any circles, since the integrand is not defined inside the circles.
For example, in figure \ref{Fig:deformA} there is a circular arc integral around part of the blue circle.
Additionally, sometimes the deformed contour encloses a boundary circle, as in Fig.~\ref{Fig:deformB}.
In this case, the residue around the enclosed circle must also be included. The specific choice of contour deformation described here is not unique, but experience suggests that it is the simplest to implement and automate when studying a large number of configurations, as we do in \S\ref{sec:results}.

\section{Expressions for forces}
\label{Ap:forces}
Here we present expressions for the lift and thrust forces exerted on each plate. {While the expression for the lift is not used in the main text and is presented for the sake of completeness, the thrust is used in \S\ref{SSec:Inline} and \S\ref{SSec:Staggered}.}

The instantaneous lift exerted on each plate by the fluid is
\begin{align*}
    L_m(t) = \int_{x_m^-}^{x_m^+}
    \Delta p_m(x,t)
    \d x
\end{align*}
where $\Delta p_m$ is the jump in pressure across the $m$-th plate.
Transforming the integral to $D_\zeta$ yields
\begin{align}
    L_m(t) = -\rho \mathcal{X}_m \oint_{C_m}
    \phi(\zeta,t) \frac{\d \map}{\d \zeta}(\zeta)
    \d \zeta,
    \label{Eq:liftInt}
\end{align}
where 
\begin{align*}
    \mathcal{X}_m = \begin{cases}
    -1 & \textrm{for }m =0,\\
    +1 & \textrm{for }m \neq 0
    \end{cases}
\end{align*}
accounts for the change in orientation when integrating around $C_0$.
Analytical expressions for the lift force can be derived by applying the residue theorem to \eqref{Eq:liftInt}.
However, we find this approach unwieldy and instead recommend evaluating the integrals  using the trapezoidal rule, which converges exponentially
fast in this context \cite{Trefethen2014}.

The instantaneous thrust on the $m$-th plate may be decomposed as
\begin{align}
    T_m(t)&= T_m^{(p)}(t) + T_m^{(l)}(t)
\end{align}
where 
\begin{align*}
    T_m^{(p)}(t) = \int_{z_m^-}^{z_m^+}
    \Delta p_m(x,t) \frac{\partial h_m}{\partial x}(x,t)
    \d x
\end{align*}
is the contribution to the thrust from the pressure jump across the plate
and $T_m^{(l)}$ is the leading-edge suction force.
Again, $T_m^{(p)}$ can be transformed into $D_\zeta$ as
\begin{align*}
    T_m^{(p)}(t) = \mathcal{X}_m \rho \oint_{C_m}
    \phi(\zeta,t) \frac{\partial h_m}{\partial x}(x(\zeta),t)
    \frac{\d \map}{\d \zeta}(\zeta)
    \d \zeta.\end{align*}

The leading-edge suction forces $T_m^{(l)}$ are obtained by considering the dominant terms in
the Bernoulli equation near the leading edge.
Since the unsteady velocity potential and its time derivative are bounded, the dominant contribution comes from the quadratic interaction of the singular velocity.
(Further details of this argument can be found on page 328 of \cite{wu1961swimming}.)
Thus, the leading-edge suction force is
\begin{align}
    T_m^{(l)}(t) =\frac{\i \rho}{2} \oint_{\mathcal{B}_m}
    w(z,t)^2 \d z
    \label{Eq:thrust1}
\end{align}
where $\mathcal{B}_m$ is an $\epsilon$-circle centered at $z_m^-$.

Transforming the integral into $D_\zeta$ produces
\begin{align}
    T_m^{(l)}(t) =\frac{ \rho}{2}\i  \mathcal{X}_m  \oint_{ \mathcal{S}_m} 
    w(\map(\zeta),t)^2 \frac{\d \map}{\d \zeta}(\zeta) \d \zeta
    \label{Eq:thrust2}
\end{align}
where $\mathcal{S}_m$ is an epsilon-semicircle centered
at $\zeta_{m}^-$.
Since the image of each boundary circle is a horizontal slit, the derivative of the mapping function has the Taylor expansion
\mbox{$\tfrac{\d \map}{\d \zeta}(\zeta) \sim (\zeta - \zeta_m^-)\tfrac{\d^2 \map}{\d \zeta^2}(\zeta_m^-)$} at  $\zeta_m^-$. Additionally, by \eqref{Eq:velPres} we have $w(\map(\zeta), t) \sim U \APot(\zeta,t)$.
Combining these behaviours with \eqref{Eq:Kasymp}, \eqref{Eq:hatKdef} and \eqref{Eq:singularityExt} indicates that the integrand in \eqref{Eq:thrust2} has a simple pole of the form
\begin{align*}
    w(\map(\zeta),t)^2  \frac{\d \map}{\d \zeta} (\zeta)
    \sim
    -\dfrac{U^2  a_{m}(t)^2 q_m^2 \e^{2 \i \vartheta^{-}_{m}}}{\zeta - \zeta_{m}^-}
    \frac{\d^2 \map}{\d \zeta^2}(\zeta_m^-).
\end{align*}
Applying the residue theorem to \eqref{Eq:thrust2} yields the final expression for the leading edge suction force:
\begin{align*}
  T_m^{(l)}(t) = \frac{\rho}{2}\mathcal{X}_m \pi U^2 a_{m}(t)^2 q_m^2 \e^{2 \i \vartheta^{-}_{m}}
    \frac{\d^2 \map}{\d \zeta^2}(\zeta_m^-).
\end{align*}
We recover the known leading-edge suction force for a single plate as {$T_0^{(l)} = \pi \rho U^2 a_0(t)^2/2$}
when the map $\map$ is taken to be the Joukowski map and the leading edge is
$\zeta_0^- = \e^{2 \i \vartheta_0^-} = -1$ \cite{wu1961swimming,moore2017fast}.

For harmonic motions we can similarly decompose the thrust averaged over a single period as $\left< T_m \right> = \left< T_m^{(p)} \right> + \left< T_m^{(l)} \right>$ where
\begin{align*}
    \left< T_m^{(p)} \right> &=
   \frac{\rho}{2} \mathcal{X}_m \oint_{C_m}
    \left(\phi^c(\zeta) \frac{\d h_m^c}{\d x}(x(\zeta))
    \frac{\d \map}{\d \zeta}(\zeta)+ 
    \phi^s(\zeta) \frac{\d h_m^s}{\d x}(x(\zeta))
    \frac{\d \map}{\d \zeta}(\zeta)\right) \d \zeta,
    \\
    \left< T_m^{(l)} \right> &= \frac{\rho}{4}
    \mathcal{X}_m \pi U^2 \left((a_m^c)^2 + (a_m^s)^2 \right)
    q_m^2 \e^{2 \i \vartheta^{-}_{m}}
    \frac{\d^2 \map}{\d \zeta^2}(\zeta_m^-).
\end{align*}
Similar expressions can be derived for the moment, power, and kinetic energy of each plate.
\end{appendices}

\bibliographystyle{plain}
\bibliography{bibliography}
\end{document}